\documentclass[dvipdfm,brazil,12pt]{article} 
\usepackage[ansinew]{inputenc}
\usepackage{amssymb,amsthm,amsfonts,amsmath,natbib}
\usepackage[pdftex]{graphicx}   
\usepackage{epsfig} 
\usepackage{multicol}
\usepackage{multirow}   
\usepackage{scalefnt} 
\usepackage{epstopdf}	
\usepackage{rotating} 
\usepackage{float}   
\usepackage{subfigure}
\usepackage{bbm}
\usepackage{booktabs}
\DeclareMathOperator*{\argmax}{argmax}

\usepackage{wrapfig}
\usepackage{appendix}
\usepackage{fancyhdr}
\usepackage{url}
\graphicspath{{./Figuras/}} 

\usepackage[paperwidth=215.9mm,paperheight=279.4mm,hmargin={25mm,25mm},vmargin={25mm,25mm}]{geometry}

\renewcommand{\arraystretch}{3.0}



\begin{document}


\begin{center}
{\Large\bf MCEM and SAEM Algorithms for Geostatistical Models under Preferential Sampling}
\end{center}
\centerline{\bf Douglas Mateus da Silva\footnotemark[1], Lourdes C. Contreras Montenegro}\centerline\\ 
\hspace{35mm}{\footnotesize{Department of Statistics, Universidade Federal de Minas Gerais, Brazil.}}\\ \vskip 0.5cm
\footnotetext[1]{Department of Statistics, Universidade Federal de Minas Gerais, Brazil\\
\text{\hspace{7mm}}E-mail: douglas\_est@yahoo.com.br}

\begin{abstract}

The problem of preferential sampling in geostatistics arises when the choise of location to be sampled is made with information about the phenomena in the study. 
The geostatistical model under preferential sampling deals with this problem, but parameter estimation is challenging because the likelihood function has no closed form. We developed an MCEM and an SAEM algorithm for finding the maximum likelihood estimators of parameters of the model and compared our methodology with the existing ones: Monte Carlo likelihood approximation and Laplace approximation. Simulated studies were realized to assess the quality of the proposed methods and showed good parameter estimation and prediction in preferential sampling. Finally, we illustrate our findings on the well known moss data from Galicia.


\end{abstract}

 
\section{Introduction}

Geostatistics is the field of Statistics that deals with spatially correlated data in a continuous domain. In the traditional geostatistical model, the variable of interest is modeled through the sum of three components: one deterministic that represents the mean of the process, one random and spacially correlated and one that represents a random noise. Since we observe data with error, the component that models the correlation structure is also called of latent process. The traditional methodology of geostatistics supposes that the sampling design is constructed independently of the phenomena in the study and we denominate this sampling process as non preferential. A random choice of the locals or a regular grid in the region are examples of non preferential sampling. When the sampling design is constructed with information about the variable of interest, some locals have more chance of being selected in comparision to others. We denominate this kind of sampling as preferential. For example, suppose we have the problem of allocating monitoring stations to analyze pollution in a city. We probably want to choose locals near of pollution sources, like factories, and thus we use the information of the problem to construct our sampling design. In other words, locals are more likely to be chosen than others and, in this context, the traditional geostatistical model is not adequated.

\cite{diggle2010} presents a methodology to deal with preferential sampling in geostatistics. They introduced a point process with intensity function depending on the latent process to model the sampling design and show that ignoring this dependence may lead to biased parameter estimates and biased prediction. Since the prediction of the variable of interest at unobserved locations is the principal aim of geostatistics, biased predictions are several problem. In this way, the joint distribution of the point process and the latent process must be considered. Based on their work, some studies have arisen in the field. \cite{pati2011} presents theoric properties of the model in the Bayesian approach, \cite{gelfand2012} analyzed the impact of the preferential sampling in prediction and, \cite{ferreiragam} studied the effect of preferential sampling in optimal design.

In classical statistics, we must find the maximum likelihood estimator (MLE) of the parameters of the model for estimation. 
Since the likelihood of the geostatistical model under preferential sampling has no closed form, obtaining those estimates may be a dificult task. In this way, \cite{diggle2010} used a Monte Carlo likelihood approximation approach and importance sampling for estimating the parameters of the model. As an alternative, a Laplace approximation approach is presented in \cite{dinsdale2019}. The EM algorithm \citep{dempster1977} is a natural tool for finding the MLE of this model, since we do not observe the latent process. To the best of the authors' knowledge, there is no EM algorithm for parameter estimation for this model presented in the literature before. 
As we will see, there is no closed form of the terms in the E-step and a simulation approach is necessary. Thus, we used the MCEM \citep{weitanner1990} and SAEM \citep{delyon1999} algorithms. In this way, the contribution of this paper is the development of an MCEM and an SAEM algorithmS for parameter estimation of the geostatistical model under preferential sampling. 

The paper is organized as follows. We present a review of the traditional geostatistical model and some results of the spatial point process in Section 2. In Section 3 we discuss the geostatistical model under preferential sampling, the methods of parameter estimation presented in \cite{diggle2010} and \cite{dinsdale2019} and we develop our proposed MCEM and SAEM algorithms. Section 4 presents a simulation study and we analyze a real dataset in Section 5. Finally, Section 6 presents our final remarks.

\section{Preliminaries}
\subsection{Geostatistical model}

Let $B$ be a compact region in $\mathbb{R}^d$, $d \geq 1$, the region of study. A point in the $B$ region will be denoted by $\boldsymbol{x} = (b_1,b_2,...,b_d)$, the spatial coordinates of the point $\boldsymbol{x}$.
We define the following geostatistical model
\begin{equation}
Y_i = \mu(\boldsymbol{x_i}) + S(\boldsymbol{x_i}) + \epsilon_i, \quad i=1,...n, \label{eq:geo_model}
\end{equation}
where $Y_i$ is the observed value at location $\boldsymbol{x_i}$, $\mu(\boldsymbol{x_i})$ is the deterministic component that represents the expected value of $Y$ at $\boldsymbol{x_i}$, $S(.)$ is a stationary Gaussian process with zero mean, $\sigma^2$ variance and isotropic correlation function $\rho(\boldsymbol{x},\boldsymbol{x}')$ and $\epsilon_i$ is the error component. We assume that $\epsilon_i's$ are independent Gaussian variables with zero mean and $\tau^2$ variance and are independent of $S$.

The process $S$ is responsible for modeling the spatial dependence and is a latent process, not measured directly, while the error component measures the non-spatial variability. Note that $Y - \mu$ variable is a noise process of $S$. If there's no spatial structure, the model (\ref{eq:geo_model}) reduces to a traditional regression model. 

The correlation function must be positive definite and checking this condition is not easy. In this way, there are many parametric functions that attends this condition and \cite{diggle1998} shows various of them. Since we assume the correlation function is isotropic, it depends only on the Euclidian distance of locations $\boldsymbol{x}$ and $\boldsymbol{x}'$, denoted by $h$. The most common family of correlation functions is the Matérn family, given by
\begin{equation}
\rho(h) = \{2^{\kappa-1}\Gamma(\kappa)\}^{-1}(h/\phi)^{\kappa}K_{\kappa}(h/\phi), \label{eq:matern_corr} \nonumber
\end{equation}
in which $K_{\kappa}(.)$ denotes a modified Bessel function of order $\kappa$, $\phi>0$ is a scale parameter with the dimensions of distance, $\kappa>0$ is a shape parameter which determines the smoothness of the process $S$. 
In this article, we assume a special case of the Matérn family when $\kappa$ equals to 0.5: the exponential correlation function, given by
\begin{equation}
\rho(h) = \exp\{-h/\phi\}. \label{eq:exp_corr}
\end{equation}

We can incorporate covariates in geostatistical model assuming that, for some known functions of $\boldsymbol{x}_i$, $d_1(\boldsymbol{x}_i), d_2(\boldsymbol{x}_i),...,d_p(\boldsymbol{x}_i)$, the deterministc component is expressed by
\begin{equation}
\mu(\boldsymbol{x}_i) = \eta_0 + \sum_{j=1}^{p}\eta_j d_j(\boldsymbol{x_i}), \nonumber
\end{equation}
in which $\eta_0,\eta_1,...,\eta_p$ are regression parameters to be estimated. Model (\ref{eq:geo_model}) can be expressed in matrix notation by
\begin{equation}
\boldsymbol{Y} = \boldsymbol{D\eta} + \boldsymbol{S} + \boldsymbol{\epsilon},
\end{equation}
where $\boldsymbol{Y} = (Y_1,...,Y_n)'$, $\boldsymbol{D}$ is a $n\times(p+1)$ matrix with $i$th row as $(1,d_1(\boldsymbol{x}_i),...,d_p(\boldsymbol{x}_i))$, $\boldsymbol{\eta} = (\eta_0,\eta_1,...,\eta_p)'$, $\boldsymbol{S} = (S(\boldsymbol{x}_1),...,S(\boldsymbol{x}_n))'$ and $\boldsymbol{\epsilon} = (\epsilon_1,..,\epsilon_n)'$. Note that $\boldsymbol{Y}$ has a $n$-variate Gaussian distributioin with $\boldsymbol{D\eta}$ mean and $\boldsymbol{\Sigma}$ covariance matrix, where $\boldsymbol{\Sigma} = \tau^2\boldsymbol{I}_n + \sigma^2 \boldsymbol{R}(\phi)$, $\boldsymbol{I}_n$ is the $n$-dimensional identity matrix and $\boldsymbol{R} = \boldsymbol{R}(\phi)$ is the correlation matrix.

\subsection{Spatial Point Process}

Spatial Point Process (PP) are used to model point patterns in which the points usually are locals in a bi or tridimensional region (in fact, dimension $d>1$). We can see it's applicability in many fields, such as population studies, forestry, epidemiology, agriculture and material science \citep{mollerw}. Statistical models for these data sets are commonly given by densities with respect to a Poisson point process. With the advance of computational technology, the use of methods like Markov Chain Monte Carlo (MCMC) has grown and has allowed the evolve of the study of spatial process.

A spatial point process $X$ is a random countable subset of a space $B \subset \mathbb{R}^d$, where usualy $d = 1$, 2 or 3. A realization of $X$ is denoted by $\boldsymbol{x} = (\boldsymbol{x}_1,...,\boldsymbol{x}_n)$, in which $n$ represents the number of points of $X$ and $\boldsymbol{x_i} = (b_{1i},...,b_{di})$, where $b_{1i},...,b_{di}$ are the spatial coordinates of point $\boldsymbol{x_i}$. Each $\boldsymbol{x}_i$ is called a event. From \cite{mollerw}, we have the following definition: a point process $X$ on $B$ is a Poisson point process with intensity function $\lambda$ if for any $B' \subset B$ with $m(B') = \int_{B'}\lambda(\xi)d\xi <\infty$, (i) the number of points on $B'$ has Poisson distribution width mean $m(B')$ and (ii) for any $n \in \mathbb{N}$, conditional on the number of points $n$, $X_{B'} \sim Binomial(B',n,f)$, with $f(\xi) = \lambda(\xi)/m(B')$. Heuristically, $\lambda(\xi)d\xi$ is the probability for the occurrence of a point in an infinitesimally small ball with centre $\xi$ and volume $d(\xi)$.

In many situations, Poisson point process is to simple for analyze real data. A natural extension of the Poisson point process is the Cox process, obtained by considering the intensity function of the Poisson process as a realisation of a random field. Let $S = {S(\xi): \xi \in B}$ be a nonnegative random field. If the conditional distribution of $X$ given $S$ is a Poisson process on $B$ with intensity function $S$, then $X$ is said to be a Cox process driven by $S$. Additionaly, if $\lambda(\boldsymbol{x}) = \exp\{S(x)\}$ and $S$ is a stationary Gaussian random field, so $X$ is said to be a log-Gaussian Cox process (LGCP).

The likelihood function associated with LGCP, given the number of observations $n$, is written as
\begin{equation}
L(\lambda|\boldsymbol{x},n) = \left(\int_{B}\exp\{S(\xi)\}d\xi\right)^{-1}\left[\prod_{i=1}^{n}\exp\{S(\boldsymbol{x_i})\}\right]. \label{eq:cox_process_likel}
\end{equation}
This function is not analytically tractable and approximated methods have to be used. \cite{mollerw} use a fine grid on $B$ to approximate the Gaussian process $S$ and make inference feasible. Although this method make approximations relatively simple of the likelihood function (\ref{eq:cox_process_likel}), the computational cost increase very fast when we grow the refination of grid, since simulation of Gaussian process is of $O(n^3)$. So, a trade-off of computational cost and refination of the grid has to be considered.

\section{Geostatistics under preferential sampling}
In geostatistical literature, it is common to assume that no information about the fenomena under study is used in the choice of sampled locations. This type of sampling is called \textit{non preferential} and inference can be made using the traditional model (\ref{eq:geo_model}). However, there are some situations that the sampling design is constructed making use of information about the fenomena, such as allocating air pollutant monitoring stations close to possible sources of pollution. In this case, we have \textit{preferential sampling}, since locals close to source of pollution are more likely to be chosen than locals far from those sources.

Here, we assume that $X = (\boldsymbol{x_1},...,\boldsymbol{x}_n)$ is a pontual process responsible to specify the sampling design. In this context, the complete model needs to specify the joint distribution of $S$, $X$ and $Y$. In non preferential context, $X$ is independent of $S$ and $f(s,x,y) = f(s)f(x)f(y|s(x))$, which means that we can ignore $X$ in the inference step. But, in the preferential context, $X$ and $S$ are not independent and the functional $S(X)$ can not be ignored, althoug we have mislead inference \citep{diggle2010}.

In the pioneer article \cite{diggle2010}, the authors proposed the following class of models for geostatistical with preferential sampling:
\begin{enumerate}
	\item $S$ is a stationary Gaussian process on $B$ with zero mean, variance $\sigma^2$ and correlation function $\rho(h,\phi)$, in which $h$ is the Euclidian distance between $\boldsymbol{x}$ and $\boldsymbol{x}'$ and $\phi$ is a parameter (or a vector of parameters) of the correlation function $\rho(.)$.
	\item Conditional on $S$, $X$ is a Poisson point process with intensity function given by
	\begin{equation}
	\lambda(\boldsymbol{x}) = \exp\{\alpha + \beta S(\boldsymbol{x})\},
	\end{equation}
	\item Conditional on $S$ and $X$, $Y$ is a set of mutually independent Gaussian variates with $Y_i \sim N(\mu + S(\boldsymbol{x}_i),\tau^2)$.
\end{enumerate}

\begin{figure}[tb!]
	\centering
	\subfigure[]{\includegraphics[scale=0.25]{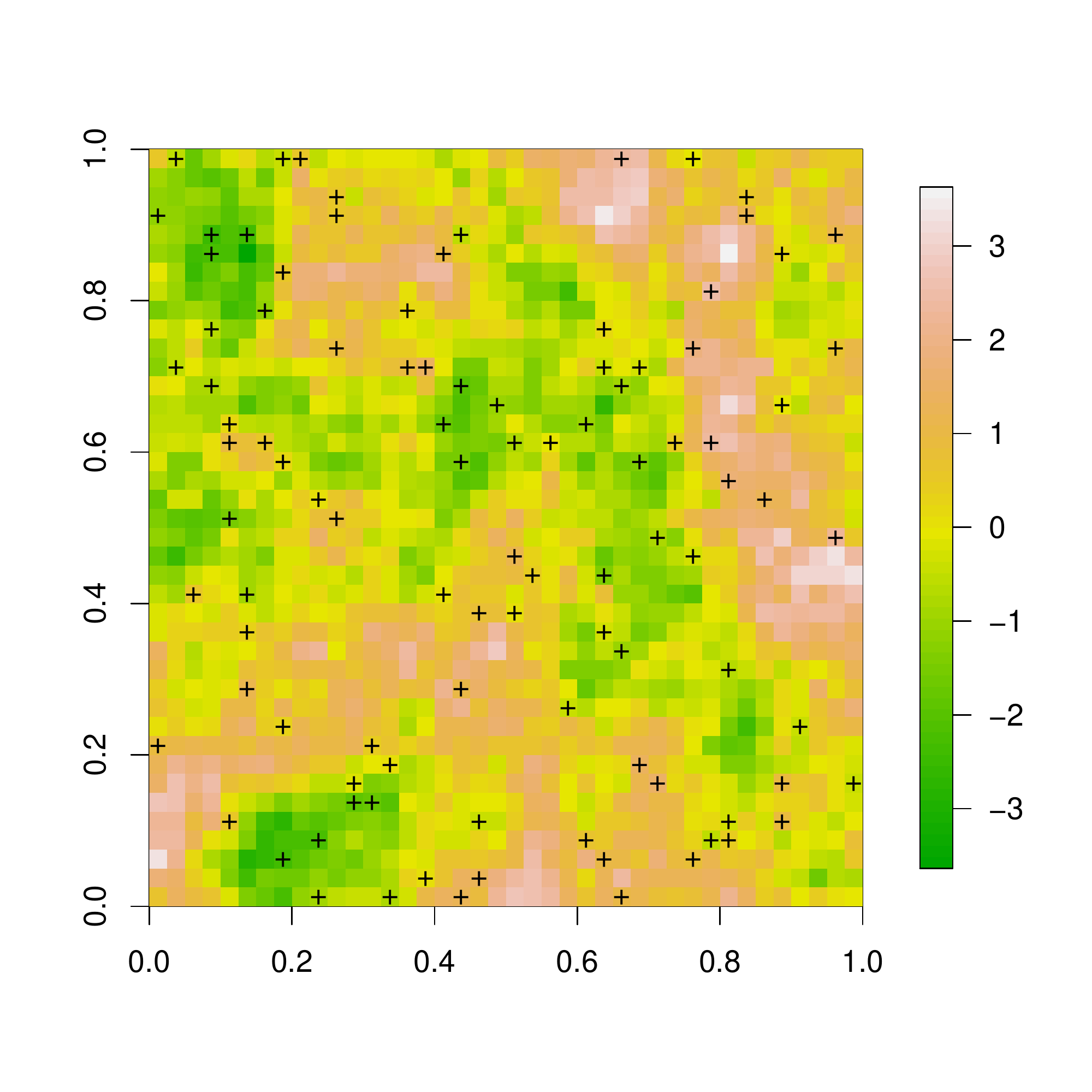}}
	\subfigure[]{\includegraphics[scale=0.25]{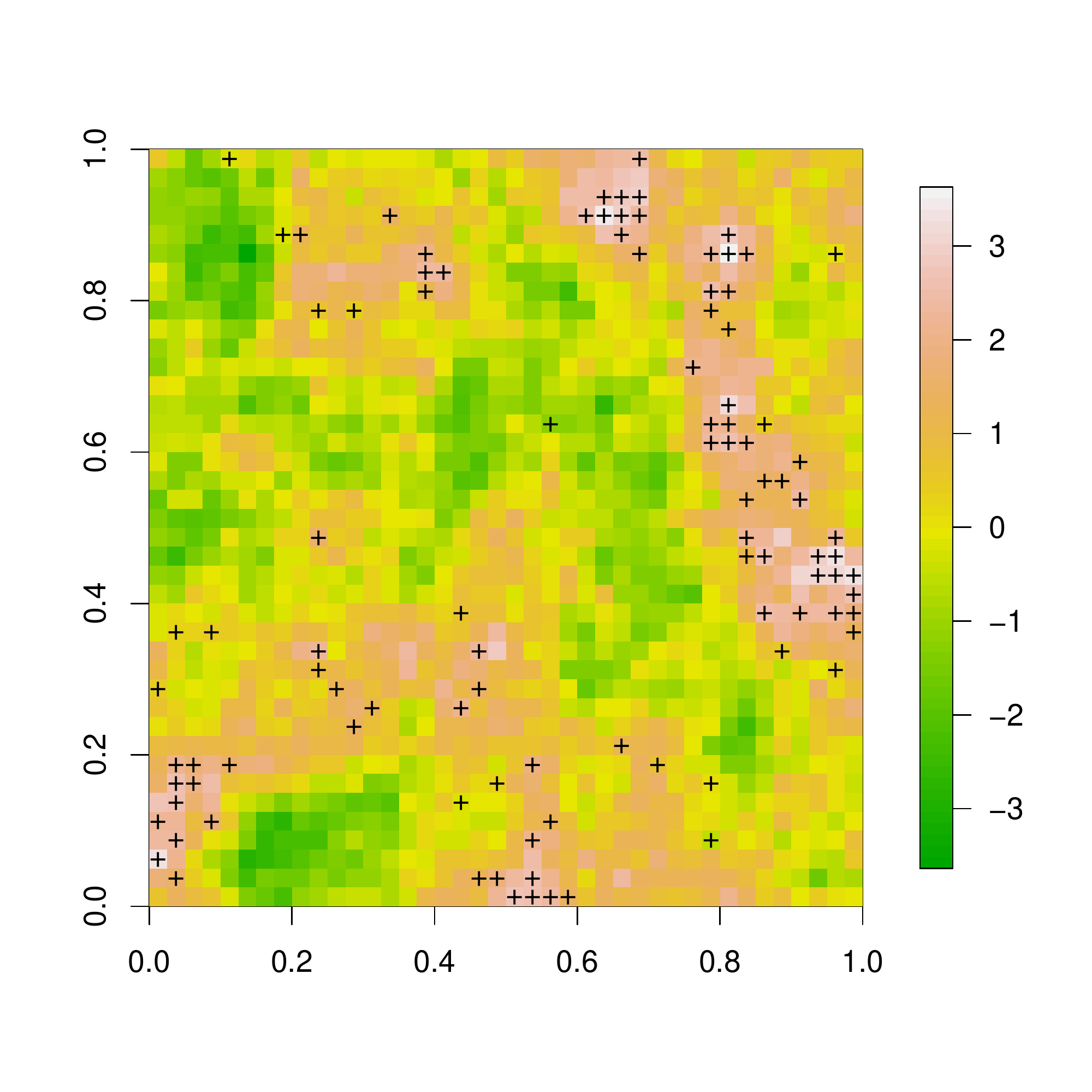}}
	\subfigure[]{\includegraphics[scale=0.25]{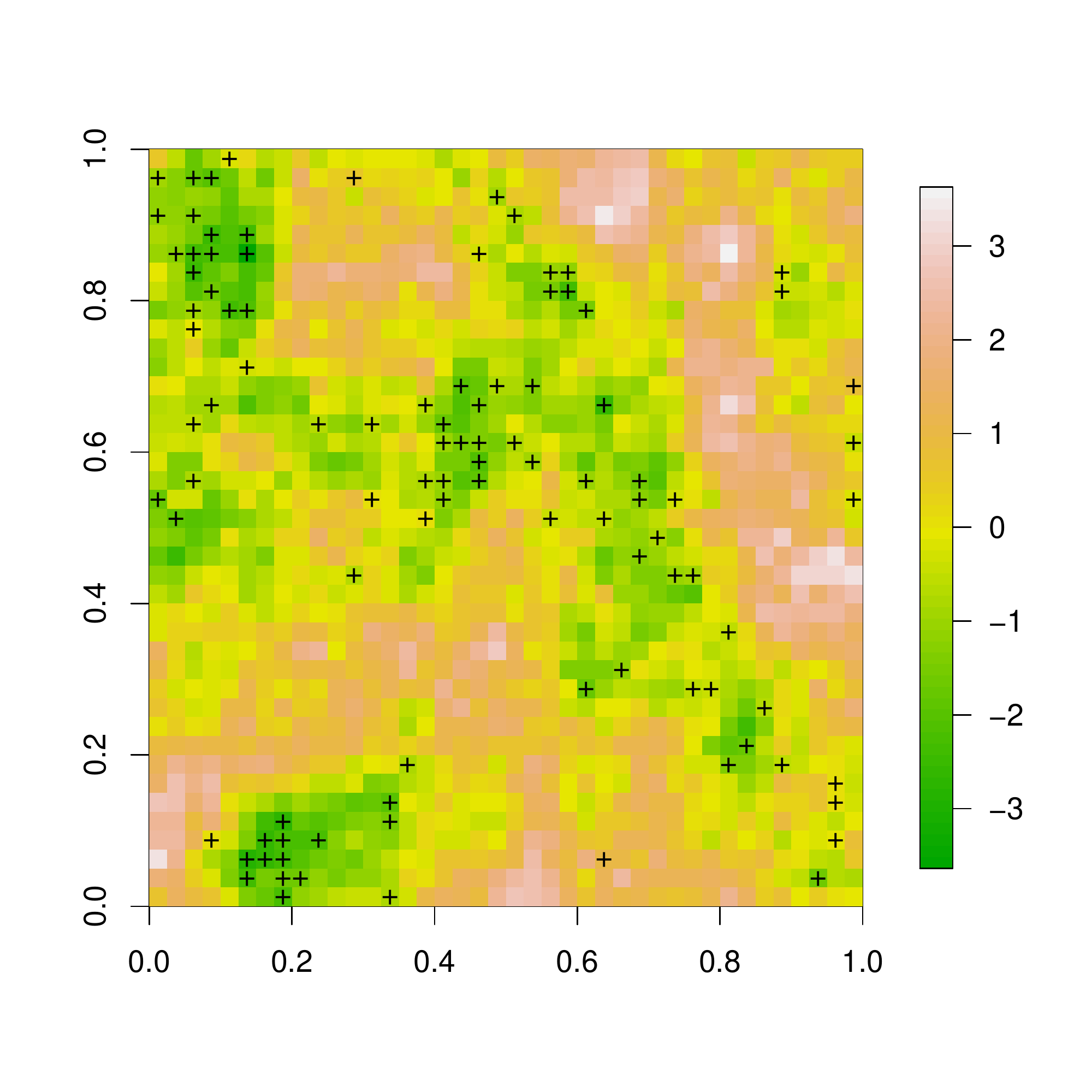}}
	\caption{Examples of samples obtained considering $\beta = 0$, $\beta = 1,5$ e $\beta = -1,5$ from the geostatistical model under preferential sampling. \label{fig:map:exemplogauss}}
\end{figure}

From assumption (1) and (2), marginally, $X$ is a LGCP \citep{mollerw}. The parameter $\beta$ controls the degree of preferability on the sampling design. If $\beta$ is equal to zero, $X$ do not depend on $S$ and we fall to the case of non preferential sampling. A positive value of $\beta$ indicates that $S$ and $X$ are positive associated, which means that regions of larger values of $S$ are more likely to be chosen. On the other hand, a negative value of $\beta$ indicates a negative association of $S$ and $X$ in the sense that lower values of $S$ are preferential sampling. Figure \ref{fig:map:exemplogauss} shows the effect of the preferability parameter.  Note that, if $\beta$ is not zero, the marginal distribution of $Y$ has no closed form. 

The density function of $X|S,n$ is given by
\begin{equation}
f(X|S,n) = \prod_{i=1}^n\dfrac{\text{exp}\{\alpha + \beta S(x_i)\}}{\int_B\text{exp}\{\alpha + \beta S(\xi)\}d\xi} = \Big(\int_B\text{exp}\{\beta S(\xi)\}d\xi\Big)^{-n} \prod_{i=1}^n \text{exp}\{\beta S(x_i)\}. \label{eq:distX:s_n}
\end{equation}
Note the above density does not depend on $\alpha$ parameter, since the number of points is fixed and it is the context that \cite{diggle2010} work. We follow this idea. The integral above has no closed form, so consider a fine grid over the region $B$ and approximate this integral by
\begin{equation}
\int_{B}\text{exp}\{\beta S(\xi)\}d\xi \simeq \sum_{i=1}^{N}\Delta_i\text{exp}\{\beta S(x_i)\} \label{eq:intaprox}
\end{equation}
in which $N$ is the number of cells and $\Delta_i$ is the area (or volume) of the $i$th cell of the grid. If we use a regular grid, then $\Delta_i = \Delta$, for all $i = 1,...,N$. Substituting (\ref{eq:intaprox}) in (\ref{eq:distX:s_n}), we have the following approximation
\begin{equation}\label{eq:pxs:app}
f(X|S) \simeq \left(\sum_{i=1}^{N}\Delta_i\text{exp}\{\beta s(x_i)\} \right)^{-n} \left[\prod_{i=1}^N \text{exp}\{\beta S(x_i)\}^{n_i}\right],
\end{equation}
where $n_i$ is the number of points in the $i$th cell of the grid and $\sum_{i=1}^{N} n_i = n$. Note that if there are more than one observed point in a specific cell, say cell $j$, these observed values brings information for just one value of $S$ at centroid of cell $j$. If the area $\Delta_i$ is reduced such that the maximum number of points expected in each cell is one point, this aproximation corresponds to that used in \cite{diggle2010}.

The likelihood function of the geostatistical model under preferential sampling is given by
\begin{eqnarray}
L(\boldsymbol{\theta};\boldsymbol{y},\boldsymbol{x}) &=& f(\boldsymbol{y},\boldsymbol{x}|\boldsymbol{\theta}) = \int_{B}f(\boldsymbol{y}|\boldsymbol{x},S,\boldsymbol{\theta}) f(\boldsymbol{x}|S,\boldsymbol{\theta})f(S|\boldsymbol{\theta})dS, \label{eq:likel:pref}
\end{eqnarray}
in which $\boldsymbol{\theta} = (\mu,\tau^2,\sigma^2,\phi,\beta)$. For parameter estimation, we must find the maximum likelihood estimator (MLE) of vector parameter $\boldsymbol{\theta}$, but, with no analytical form for likelihood function, this task may be dificult. Some alternatives has arised in the literature: \cite{diggle2010} use Monte Carlo likelihood approximation (MCLA) and \cite{dinsdale2019} make use of a Laplace approximation. We propose a SAEM algorithm for estimation parameter.

\subsection{Parameter estimation}
We first discuss the methods of parameter estimations proposed by \cite{diggle2010} and \cite{dinsdale2019} and, then, show our proposed MCEM and SAEM algorithms.
\subsubsection{Monte Carlo likelihood approximation}
The likelihood function (\ref{eq:likel:pref}) can be written as
\begin{eqnarray}
L(\boldsymbol{\theta};\boldsymbol{y},\boldsymbol{x}) &=& E_S[f(y|x,s)f(x|s)] \label{eq:veross1}
\end{eqnarray}
and we can approximate it using sampling techniques as Monte Carlo. The problem of this strategy being applied direct in (\ref{eq:veross1}) is that the simulation of samples of $S$ (unconditional) will not be compatible with the observed values of $Y$ when $\tau^2$ is equal to zero or is very small \citep{diggle2010}. Then the authors use importance sampling with $S|Y$ as the importance distribuition. 

Let $S = \{S_0,S_1\}$, where $S_0$ denotes the Gaussian process on the observed locations and $S_1$, the Gaussian process on the rest of points of the grid. We rewrite the likelihood function (\ref{eq:likel:pref}) as
\begin{eqnarray}
L(\boldsymbol{\theta};\boldsymbol{y},\boldsymbol{x}) &=& \int f(y|x,s)f(x|s)\dfrac{f(s|y)}{f(s|y)}f(s)ds  = E_{S|Y}\left[ f(x|s)\dfrac{f(y|s_0)}{f(s_0|y)}f(s_0) \right]. \label{eq:veromc}
\end{eqnarray}
Note that
\begin{eqnarray}
\dfrac{f(y|s_{0j})}{f(s_{0j}|y)}f(s_{0j}) = \dfrac{f(y,s_0)}{f(s_0|y)} = f(y), \nonumber
\end{eqnarray}
and (\ref{eq:veromc}) turns to
\begin{equation}
L(\boldsymbol{\theta};\boldsymbol{y},\boldsymbol{x}) = E_{S|Y}\left[ f(x|s)f(y) \right]
\end{equation}
where $f(y)$ is the marginal density of $Y$ in the non preferential sampling context. The MCLA is given by
\begin{eqnarray}\label{eq:verossmc3}
L_{MC}(\boldsymbol{\theta};\boldsymbol{y},\boldsymbol{x}) &=& \dfrac{1}{m}f(y)\sum_{j=1}^{m}f(x|s_j),
\end{eqnarray}
where $s_j$ are simulations of $S$ conditional on $Y$. To simulate a realization of the $S|Y$ distribution, the authors utilize a result from \cite{rueheld2005} that shows that the variable $S + \Sigma C'(C\sigma^2 \boldsymbol{R}(\phi) C' + \tau^2\textbf{I}_n)^{-1}(y-\mu+Z-CS)$ has the desired distribution $S|Y = y$, where $Z\sim N(0,\tau^2)$, $S \sim N_N(0,\sigma^2 \boldsymbol{R}(\phi))$ and $C$ is a $n \times N$ matrix whose $i$th row consists of $N-1$ zeros and a single one to identify the position of $\boldsymbol{x}_i$ within the $N$ cells of the grid. 

Obtained $m$ samples of $S|X$, the estimate of $\boldsymbol{\theta}$ is given by
\begin{equation}
\hat{\boldsymbol{\theta}} = \argmax_{\boldsymbol{\theta}} \big( L_{MC}(\boldsymbol{\theta}) \big). \nonumber
\end{equation}

Since this approximation is constructed considering the distribution of $Y$ in the non preferential context, this is the main problem that \cite{dinsdale2019} pointed out on their article about this method, showing that this is a good approximation only in the non preferential case.
Then, they proposed to use a Laplace approximation for parameter estimation.

\subsubsection{Laplace approximation}
As an alternative method to the Monte Carlo simulation technique, \cite{dinsdale2019} suggest to use Laplace approximation for parameter estimation. We can write the likelihood function (\ref{eq:likel:pref}) as 
\begin{eqnarray}
L(\boldsymbol{\theta}) = \int_B f(y,x,s)ds = \int_B \exp\{f(S,\boldsymbol{\theta})\}ds, \label{eq:verolaplace}
\end{eqnarray}
in which $f(S,\boldsymbol{\theta}) = \log(f(y,x,s))$. Considering a fine grid on $B$, the Laplace approximation is given by
\begin{eqnarray}
\hat{L}(\boldsymbol{\theta}) = (2\pi)^{N/2} |\boldsymbol{H}(\boldsymbol{\theta})|^{1/2} \exp\{f(\hat{\boldsymbol{S}}(\boldsymbol{\theta}),\boldsymbol{\theta})\}, \label{eq:gapg:laplace}
\end{eqnarray}
where $N$ is the number of cells of the approximation grid, $\hat{\boldsymbol{S}}(\boldsymbol{\theta})$ is the $\boldsymbol{S}$ value that maximizes $f(\boldsymbol{S},\boldsymbol{\theta})$,
$$\hat{\boldsymbol{S}}(\boldsymbol{\theta}) = \argmax_{\boldsymbol{S}} f(\boldsymbol{S},\boldsymbol{\theta}),$$
and $\boldsymbol{H}(\boldsymbol{\theta})$ is the Hessian matrix of $f(\boldsymbol{S},\boldsymbol{\theta})$ with respect to $\boldsymbol{S}$ applied to $\hat{\boldsymbol{S}}(\boldsymbol{\theta})$,
$$\boldsymbol{H}(\boldsymbol{\theta}) = \left. \left[ -\dfrac{d}{d\boldsymbol{S}d\boldsymbol{S}'}f(\boldsymbol{S},\boldsymbol{\theta}) \right]^{-1} \right|_{\boldsymbol{S} = \hat{\boldsymbol{S}}(\boldsymbol{\theta})}. $$

Parameter estimation of vector $\boldsymbol{\theta}$ is given by maximizing the approximated likelihood function (\ref{eq:gapg:laplace}). To compute the Hessian matrix, we need the second derivatives of $f(\boldsymbol{S},\boldsymbol{\theta})$ with respect to $\boldsymbol{S}$. Although we can compute those derivatives analytically, we can use automatic differentiation methods (AD) to do this task. The R package \textit{Template Model Buider} (\texttt{TMB}) proposed by \cite{kristensen2016}, utilizes automatic differentiation of a Laplace approximation to the marginal likelihood to maximize the likelihood efficiently with respect to the parameters of interest. 

To use the package \texttt{TMB} regarding the variational approximation and the Laplace approximation methods, it is necessary to write the function $f(\boldsymbol{S},\boldsymbol{\theta})$ in a C ++ script. The package then integrates the $\boldsymbol{S}$ random effect and evaluates the function in (\ref{eq:likel:pref}) approximately.

\subsubsection{EM, MCEM and SAEM algorithms}

The Expectation-Maximization (EM) algorithm (Dempster et al. 1977) is an iterative method for the computation of the maximize the log-likelihood function in the missing data context. One of the advantages of the EM algorithm  is that the M-step involves only complete data ML estimation, which is often computationally simple. 

For the model under study we considered the latent process $S$ as missing data. Denoting $\boldsymbol{S}$ as the approximated Gaussian process, the complete log-likelihood function is given by
\begin{equation}\label{eq:verocomp}
l_c(\theta) = \text{log}f(y,x,s) = \text{log}   \left[ f(y|x,s)f(x|s)f(s) \right] = \text{log}f(y|x,s) + \text{log}f(x|s) + \text{log}f(s). \nonumber
\end{equation}

For the geostatistical model under preferential sampling and using the approximation of the integral in (\ref{eq:intaprox}), the complete log-likelihood function turns to
\begin{eqnarray}
l_c(\theta) &=& -\dfrac{n}{2} \text{log}(\tau^2) -\dfrac{1}{2\tau^2} \sum_{i=1}^{n}(y_i - \mu - s(x_i))^2 +  \beta \sum_{j=1}^{N} n_j S(x_j)  - n\text{log}\Big(\sum_{j=1}^{N}\Delta\text{exp}\{\beta S(x_j)\}\Big)  \nonumber\\
& &  -\dfrac{N}{2} \text{log}(\sigma^2) -\dfrac{1}{2} \text{log}(|\textbf{R}|) -\dfrac{1}{2\sigma^2} S'\textbf{R}^{-1}S +c^*,  \label{eq:verocompform2}
\end{eqnarray}
in which $n_j$ is the number of sample locations inside the $j$ht cell of the grid and $c^*$ is a constant with respect to the parameters.

In the first step of EM algorithm, we must compute the conditional expectation of the complete log-likelihood function, the $Q$ function $Q(\boldsymbol{\theta}|\boldsymbol{\theta}^{(k)}) =  E[l_c(\theta)|y,x,\theta^{(k)}]$, where $\boldsymbol{\theta}^{(k)}$ is the estimate of the parameter vector $\boldsymbol{\theta}$ in the $k$th iteration. For the model under study, this step is given by
\begin{eqnarray}
Q(\theta|\theta^{(k)}) &=& -\dfrac{n}{2} \text{log}(\tau^2) -\dfrac{1}{2\tau^2} \sum_{i=1}^{n}E[(y_i - \mu - S(x_i))^2|y,x,\theta^{(k)}] + \nonumber\\
&& + \beta \sum_{j=1}^{N} n_j E[S(x_j)|y,x,\theta^{(k)}] - nE\Big[\text{log}\Big(\sum_{j=1}^{N}\Delta\text{exp}\{\beta S(x_j)\}\Big)\Big|y,x,\theta^{(k)}\Big] \nonumber\\
&&-\dfrac{N}{2} \text{log}\sigma^2 -\dfrac{1}{2} \text{log}(|\textbf{R}|) - \dfrac{1}{2\sigma^2} E[S'\textbf{R}^{-1}S|y,x,\theta^{(k)}] +c^*. 
\end{eqnarray} 

Note that the terms of conditional expectation have no closed form. Thus, we will approximate those terms by stochastic simulation. This is the aim of SAEM algorithm (\textit{Stochastic approximation of EM algorithm}), proposed by \cite{delyon1999}. The $Q$ function is approximated in the following way:
\begin{eqnarray}\label{eq:SAEM}
\hat{Q}(\boldsymbol{\theta}|\boldsymbol{\theta}^{(k)}) &=& \hat{Q}(\boldsymbol{\theta}|\boldsymbol{\theta}^{(k-1)}) + \gamma_k \left( \dfrac{1}{L}\sum_{i=1}^{L} l_c(\boldsymbol{\theta};S^{(l)},x,y) - \hat{Q}(\boldsymbol{\theta}|\boldsymbol{\theta}^{(k-1)}) \right), \nonumber
\end{eqnarray}
where $S^{(l)}$ is a simulation from the condicional distribution of $S|\boldsymbol{X},\boldsymbol{Y}$ and $\{\gamma_j\}_{j \geq 1}$ is a positive sequence of weigths such that $\sum_{j=1}^{\infty}\gamma_j = \infty$ and $\sum_{j=1}^{\infty}\gamma_j^2 < \infty$. In this way, the $Q$ function is approximated by incrementing a new value for the function, the average of the complete log-likelihood function given $L$ simulations of $S|\boldsymbol{X},\boldsymbol{Y}$.

The weights $\gamma_j$'s control the importance of the simulated values of $l$th iteration. If $\gamma_j = 1$ for all $j$, the algorithm has no memory (does not use information about past iterations) and it is equivalent to MCEM algorithm, proposed by \cite{weitanner1990}. The SAEM with no memory will converge quickly in distribution to a solution neighborhood. On the other hand, the SAEM with memory will converge slowly almost surely to the MLE \cite{delyon1999}. As in \cite{galarza2017}, we use the following choice of weiths
\begin{equation*}		
\gamma_j = \left \{ \renewcommand\arraystretch{1.5} \begin{matrix} 1, & \mbox{se} \; 1 \leq j \leq cW;\\
\frac{1}{j - cW} & \mbox{se} \; cW+1 \leq j \leq W;
\end{matrix} \right.
\end{equation*}	
in which $W$ is the maximum number of iterations and $c$ is the proportion of inicial iterations with no memory. For example, if $c=0$, the algorithm will have memory for all iterations and will converge slowly and $W$ must be large. If $c = 1$, the algorithm will converge quickly (in distribution) to a solution neighborhood. A number between 0 and 1 of $c$ will allow the algorithm converge quickly in distribution to a solution neighborhood for the initial iterations and converge almost sure for the rest of the iterations. In this way, we mix the SAEM with no memory and with memory to achieve the convergence to MLE faster then using only SAEM with memory.

Note that we need to sample from the distribution of $S|X,Y$ that has no closed form. As this distribution is the predictive distribution, the method used to sample from it is described in the next session. 

For the model under study, the $\hat{Q}$ function is given by
	\begin{eqnarray}
\hat{Q}(\boldsymbol{\theta}|\boldsymbol{\theta}^{(k)}) &=&  -\dfrac{n}{2} \text{log}(\tau^2) -\dfrac{1}{2\tau^2}  \sum_{i=1}^{n} \Big((y_i - \mu - S(x_i))^2\Big)_{SA}^{(k)} + \nonumber\\
&& + \beta  \sum_{j=1}^{N} n_j \big(S(x_j) \big)_{SA}^{(k)} - n \bigg(\text{log}\Big(\sum_{j=1}^{N}\Delta\text{exp}\{\beta S(x_j)\}\Big) \bigg)_{SA}^{(k)} \nonumber\\
&&-\dfrac{N}{2} \text{log}\sigma^2 -\dfrac{1}{2} \text{log}(|\textbf{R}|) - \dfrac{1}{2\sigma^2} \Big( S'\textbf{R}^{-1}S\Big)_{SA}^{(k)} +c^*. \label{eq:saem:gauss}
\end{eqnarray}
where
\begin{eqnarray}
\big(S(x_j) \big)_{SA}^{(k)} &=& \gamma_k \dfrac{1}{L}\sum_{l=1}^{L} S_l(x_j) + (1-\gamma_k) \big(S(x_j) \big)_{SA}^{(k-1)},\nonumber\\
\Big((y_i - \mu - S(x_i))^2\Big)_{SA}^{(k)} &=& \gamma_k \dfrac{1}{L}\sum_{l=1}^{L} (y_i - \mu - S_l(x_i))^2 + (1-\gamma_k) \Big((y_i - \mu - S(x_i))^2\Big)_{SA}^{(k-1)}, \nonumber\\
\bigg(\text{log}\Big(\sum_{j=1}^{N}\Delta\text{exp}\{\beta S(x_j)\}\Big) \bigg)_{SA}^{(k)} &=& \gamma_k \dfrac{1}{L}\sum_{l=1}^{L}\text{log}\Big(\sum_{j=1}^{N}\Delta\text{exp}\{\beta S_l(x_j)\}\Big) + \nonumber\\
&& (1-\gamma_k) \bigg(\text{log}\Big(\sum_{j=1}^{N}\Delta\text{exp}\{\beta S(x_j)\}\Big) \bigg)_{SA}^{(k-1)}, \nonumber\\		
\Big( S'\textbf{R}^{-1}S\Big)_{SA}^{(k)} &=& \gamma_k \dfrac{1}{L}\sum_{l=1}^{L}S_l'\textbf{R}^{-1}S_l + (1-\gamma_k) \Big( S'\textbf{R}^{-1}S\Big)_{SA}^{(k-1)}. \nonumber
\end{eqnarray}

The second step is the Maximization step, which is the same for EM and SAEM algorithm: maximize the $Q$ function with respect to the parameter vector $\boldsymbol{\theta}$ to obtain a new estimate $\hat{\boldsymbol{\theta}}^{(k+1)}$. Maximizing (\ref{eq:saem:gauss}) with respect to $\boldsymbol{\theta}$ leads to the following estimators:
\begin{eqnarray}
\hat{\mu}^{(k+1)} &=& \frac{1}{n}\sum_i\Big(y_i - \big(S(x_i) \big)_{SA}^{(k)}\Big), \nonumber\\
\hat{\tau^2}^{(k+1)} &=& \frac{1}{n}\sum_i \Big((y_i - \mu - S(x_i))^2\Big)_{SA}^{(k)}, \nonumber\\
\hat{\sigma^2}^{(k+1)} &=& \frac{1}{N}\big( S'\textbf{R}^{-1}S\big)_{SA}^{(k)}. \nonumber
\end{eqnarray}
There is no closed form of estimator parameters for $\phi$ and $\beta$, so they must be estimated numerically, 
\begin{equation}
(\hat{\phi}^{(k+1)},\hat{\beta}^{(k+1)}) = \argmax_{(\phi,\beta)} \big( \hat{Q}(\boldsymbol{\theta}|\boldsymbol{\theta}^{(k)}) \big).\nonumber
\end{equation}

The two steps of SAEM algorithm are iterated until we observe convergenge, which can be made graphically with a plot of estimate of each parameter versus iteration. The convergence also can be observed by verifing if some distance of two successive evaluations of parameter estimation (or the log-likelihood fuction) becomes small enough. 

\subsection{Prediction}\label{sec:prediction}
In geostatistic, predictions of the response variable at unobserved locations are usually made by \textit{kriging}. This method is widely used in non preferential context, where the response variable $Y$ is normaly distributed. Let $Y_n$ be the response variable at observed locations and $Y_o$ at unobserved locations. Thus, $(Y_n,Y_o)$ has multivariate Gaussian distribution with vector mean $\boldsymbol{\mu}$ and covariance matrix $\boldsymbol{\Sigma}$. The condicional distribution of $Y_o$ given the observer values $Y_n$ is also normally distributed with mean and covariance matrix given by
\begin{eqnarray}
E[\boldsymbol{Y}_o|\boldsymbol{Y}_n] &=& \boldsymbol{\mu}_o + \boldsymbol{\Sigma}_{12}\boldsymbol{\Sigma}_{22}^{-1} (\boldsymbol{Y}_n - \boldsymbol{\mu}), \label{eq:Eyo_y}\\
Var[\boldsymbol{Y}_o|\boldsymbol{Y}_n] &=& \boldsymbol{\Sigma}_{11} - \boldsymbol{\Sigma}_{12} \boldsymbol{\Sigma}_{22}^{-1} \boldsymbol{\Sigma}_{21}. \label{eq:Varyo_y} 
\end{eqnarray}
in which $\boldsymbol{\mu}_o$ is the mean vector of $\boldsymbol{Y}_o$, $\boldsymbol{\mu}_n$ is the mean vector of $\boldsymbol{Y}_n$, $\boldsymbol{\Sigma}_{11}$ corresponds to the covariance matrix of $\boldsymbol{Y}_o$, $\boldsymbol{\Sigma}_{22}$ corresponds to the covariance matrix of $\boldsymbol{Y}_n$, $\boldsymbol{\Sigma}_{12}$ corresponds to the covariance matrix of $\boldsymbol{Y}_o$ e  $\boldsymbol{Y}_n$ e $\boldsymbol{\Sigma}_{21} = \boldsymbol{\Sigma}_{12}'$. The constrution of $\Sigma$ matrix depends on the choise of covariance function.

Note that \textit{kriging} does not take account the sampling design, in other words, does not use the information about the pontual process $X$. This can lead to biased predictors in preferential sampling context \citep{dinsdale2019} and others approachs must be considered. Also note that, in non preferential sampling, we can work directly with the $Y$ process, since it's marginal distribution has closed form. In preferential sampling context, we must predict the Gaussian process $S$ first, then use those values to predict the $Y$ process since the distribution of $Y|X,S$ has closed form.

An alternative approuch is to take the mode of $f(S,\boldsymbol{\theta})$ in (\ref{eq:verolaplace}) given the estimated parameter vector $\hat{\boldsymbol{\theta}}$. Thus, $\hat{\boldsymbol{S}}(\hat{\boldsymbol{\theta}})$ is the value of $S$ that maximizes $f(S,\boldsymbol{\theta})$ and can be viewed as a predictor of $S$ \citep{dinsdale2019}. \texttt{TMB} package computes both $\hat{\boldsymbol{\theta}}$ and  $\hat{\boldsymbol{S}}(\hat{\boldsymbol{\theta}})$.

As an alternative for the previous methods, we can use Monte Carlo technique to sample from the condicional distribution of $S|X,Y$. This method is used to construct predict surfaces of the $S$ process, for example, but also is used as a tool in the E-step of SAEM algorithm described in the previous session. As presented in \cite{ferreiragam}, we construct a Metropolis Hasting algorithm to sample from the predictive distritution of $S$.

The predictive distribution $S|X,Y$ is given by
\begin{equation}\label{eq:funcpred}
f(S|X,Y) = \dfrac{f(S,X,Y)}{f(X,Y)} = \dfrac{f(Y|X,S)f(X|S)f(S)}{\int_{B}f(Y|X,S)f(X|S)f(S)dS} 
\end{equation}
We do not recognize any known distribution in (\ref{eq:funcpred}), so approximated method must be considered. For the Gaussian model under preferential sampling, we have
\begin{eqnarray}
f(S|X,Y) &\propto& f(Y|X,S)f(X|S)f(S) \nonumber\\
&\propto& \text{exp}\left\{-\dfrac{1}{2} (y-\mu-S(x))'(\tau^2I)^{-1}(y-\mu-S(x)) \right\} \times \left[ \prod_{i=1}^{n} \dfrac{\text{exp}\{\beta S(x_i)\}}{\int_{B}\text{exp}\{\beta S(\xi)\}d\xi} \right] \times \nonumber\\
& & \text{exp}\left\{ -\dfrac{1}{2}S'(\sigma^2R(\phi))^{-1}S \right\}. \label{eq:fpredg}
\end{eqnarray}
We use (\ref{eq:intaprox}) to approximate the integral term on the right-hand of the equation above.

Since we do not have closed form for the distribution (\ref{eq:fpredg}), we choose to use MH. Note that the dimension of $\boldsymbol{S}$ depends on the number of cell of the approximation grid, the acceptance probability will be very slow and the algorithm turns extremely inefficient. \cite{dinsdale2019} propose to sample from each condicional distribution of $S_i|\boldsymbol{S}_{(-i)}$, where $\boldsymbol{S}_{(-i)}$ is the vector $\boldsymbol{S}$ without the $i$th position. To improve the computational cost, we propose to sample the vector $\boldsymbol{S}$ by blocks and, in the simulation session, we show that the time needed for the chain of log-likelihood function achieves convergence is lower than the way sampling each component of the vector $\boldsymbol{S}$.
	
Let $g$ be the number of blocks of vector $\boldsymbol{S}$ and denote $G_c$ as the $c$th block, $c = 1,...,g$, the distribution of $S_{G_c}|S_{-G_c},X,Y$, in which $S_{-G_c} = S\backslash S_{G_c}$, is given by
\begin{eqnarray}
f(S_{G_c}|S_{-G_c},X,Y) &\propto& \text{exp}\left\{-\dfrac{1}{2\tau^2} \sum_{i=1}^n\big[(y_i-\mu-S(x_i))^2 \mathbbm{1}\{S(x_i) \in G_c\} \big] \right\} \times \nonumber\\
&&\left[ \prod_{j=1}^{N} \big(\text{exp}\big\{\beta S(x_j) \mathbbm{1}\{S(x_j) \in G_c\}\big\}\big)^{n_i} \right]  \left[\sum_{j=1}^N\text{exp}\{\beta S(x_j)\} \right] ^{-n} \times \nonumber\\
&&\text{exp}\left\{ -\dfrac{1}{2}S'(\sigma^2R(\phi))^{-1}S \right\}. \label{eq:fpredg_cond}
\end{eqnarray}

Thus, we sample a proposed value for each element of $S_{G_c}$ from a univatiate normal distribution centered in the element of $S_{G_c}$ of previous iteration and $\delta_\sigma$ variance. The acceptance probability of block $G_c$ is
\begin{eqnarray}
p_{S_G} &\propto& \text{exp}\left\{-\dfrac{1}{2\tau^2} \sum_{i=1}^n\Big[\big((y_i-\mu-S_p(x_i))^2 -  (y_i-\mu-S_c(x_i))^2\big)\mathbbm{1}\{S(x_i) \in G_c\} \Big] \right\} \times \nonumber\\
&&\left[ \prod_{j=1}^{N} \big(\text{exp}\big\{\beta (S_p(x_j)-S_c(x_j)) \mathbbm{1}\{S(x_j) \in G_c\}\big\}\big)^{n_i} \right]  \left[  \dfrac{\sum_{j=1}^N\text{exp}\{\beta S_p(x_j)\}}{\sum_{j=1}^N\text{exp}\{\beta S_c(x_j)\}} \right] ^{-n} \times \nonumber\\
&&\text{exp}\left\{ -\dfrac{1}{2\sigma^2}(S_p'R^{-1}S_p-S_c'R^{-1}S_c) \right\}. \nonumber
\end{eqnarray}

Given the vector of predictions of $\boldsymbol{S}$, we can predict $\boldsymbol{Y}$ by using the fact that the distribution of $\boldsymbol{Y}|\boldsymbol{X},\boldsymbol{S}$ has expected value $\boldsymbol{\mu} + S(\boldsymbol{x})$. Thus, prediction values for $\boldsymbol{Y}$ can be obtained by pluging-in the estimated value of mean parameter and the predicited vector of $\boldsymbol{S}$.

\section{Simulation study}
We performed a simulation study to illustrate the performance of our MCEM and SAEM algorithms. Then, we compare it with the others algorithms previous described.

\subsection{Predictive distribution of $S$}
The predictive distribution of $S$ in preferential sampling context is obtained empirically through sampling methods. We use the MH algorithm for sampling from the distribution of $S|X,Y$, as described in Section \ref{sec:prediction}. Since we use a MCMC method for sampling, we need to verify the convergence of the elements sampled. The dimension of vector $\boldsymbol{S}$ depends on the number of cells of the grid considered and usually is large (more then 225 which correspond a $15 \times 15$ grid). analyze the chain of each element of $\boldsymbol{S}$ is impracticable, so we choose to work with the log-density of $\boldsymbol{S}$.

Our data were generated following model (\ref{eq:likel:pref}), assuming the true values of parameters $(\mu,\tau^2,\sigma^2,\phi,\beta) = (4,0.1,1.5,0.15,2)$. The $S$ process had exponential correlation function and was evaluated over a $50 \times 50$ regular grid over the unit square. Conditionally on each realization, we obtained 100 sampling locations of $X$ generated by using it's density (\ref{eq:distX:s_n}). Conditionally on $S$ and $X$ observations, we simulated the process $Y$.

\begin{figure}[h!]
	\centering
	\includegraphics[scale=0.4]{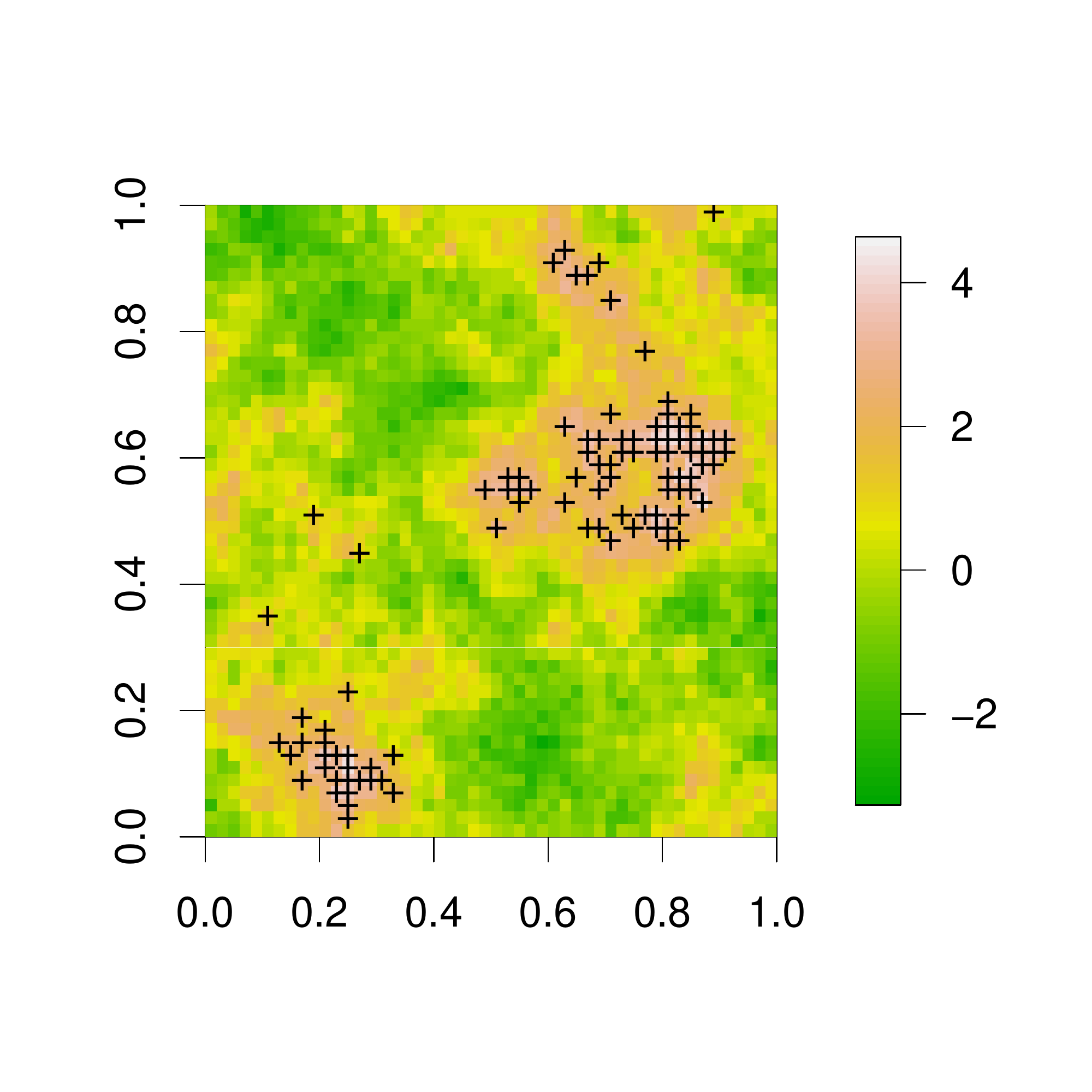}
	\caption{Map of the simulated process $S$ and the pontual process $X$ for the Gaussian model under preferential sampling. \label{fig:map:gauss625}}
\end{figure}

For prediction, we considered regular grids for size 225, 400, 625 and 900 over the unit square. The size of blocks for sampling the vector $\boldsymbol{S}$ were 1, 5 and 10. In this way, we can verify the effect of blocking vector $\boldsymbol{S}$ on the behavior of it's chains. Note that blocks of size 1 means sampling $\boldsymbol{S}$ element by element. As initial vector for $\boldsymbol{S}$, we simulate a value of the standard normal distribution for each $S_i$ and we made 1000 iterations of MH algorithm.
Figure \ref{fig:map:gauss625} shows the map of simulated process $S$ and the $n = 100$ locations sampled. As we expected, those locations are related to higher values of $S$, since we chose a positive value for $\beta$. 

\begin{figure}[tb!]
	\centering
	\subfigure[]{\includegraphics[scale=0.4]{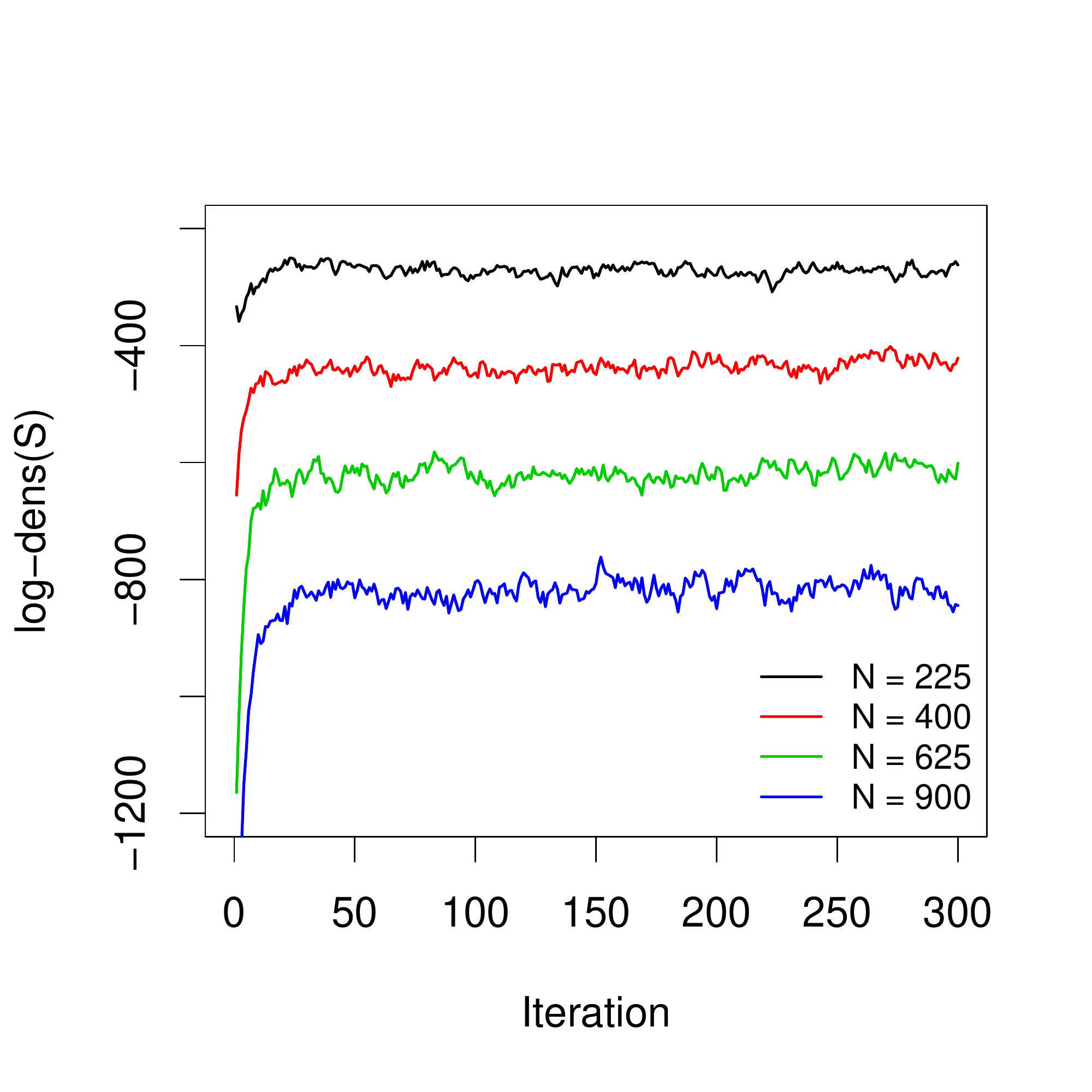}}
	\subfigure[]{\includegraphics[scale=0.4]{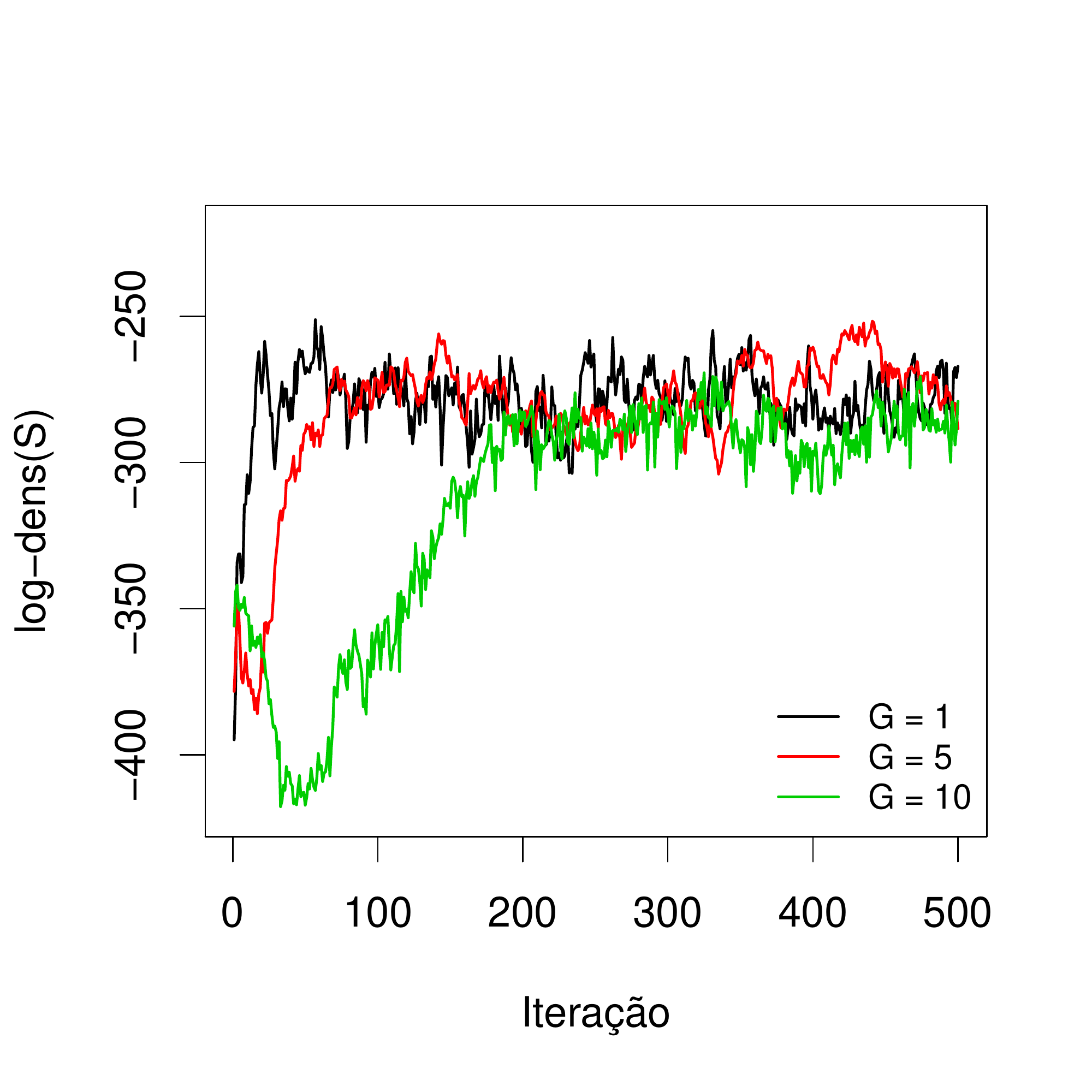}}
	\caption{Markov chains of log-density of $\boldsymbol{S}$ (a) for the values of grid size considered and block size 1 and (b) for the size of blocks considered and grid size 225. \label{fig:cad:gaussconj}}
\end{figure}

Figure \ref{fig:cad:gaussconj}(a) shows the chain of log-density of $\boldsymbol{S}$ for each grid size considered and block size equal 1. We can observe the chains converge very fast and need less than 100 iterations to achieve convergence. 
As we can see in Figure \ref{fig:cad:gaussconj}(b), the chains achieve convergence to the same region for each block size considered, although it takes more iteration to converge when we increase the block size (results presented for grid size 225). Table \ref{tab:tempo:predicao_blocos} presents the time taken to made 1000 iterations of MH algorithm for the block sizes considered. As expected, increase the block size reduces the time considerably, increasing the performance of the algorithm. In both grid sizes, the time was reduced to 10\% when we pass the block size 1 (element by element) to 10.
\begin{table}[hbt!]
	\caption{Time (in secs) to make 1000 iterations of MH algorithm for sampling from the predictive distribution of $S$ in preferential sampling.}\label{tab:tempo:predicao_blocos}
	\centering \renewcommand\arraystretch{1.1}
	\begin{tabular}{lrr}
		\hline
		Block size&\multicolumn{2}{c}{Grid size}\\
		 & 225 & 900 \\ \hline
		1 & 17.14&1082.15\\
		5 & 3.76&210.62\\
		10 &1.94&105.82\\
		\hline
	\end{tabular}
\end{table}

\begin{figure}[t!]
	\centering
	\subfigure[]{
		\includegraphics[scale=0.27]{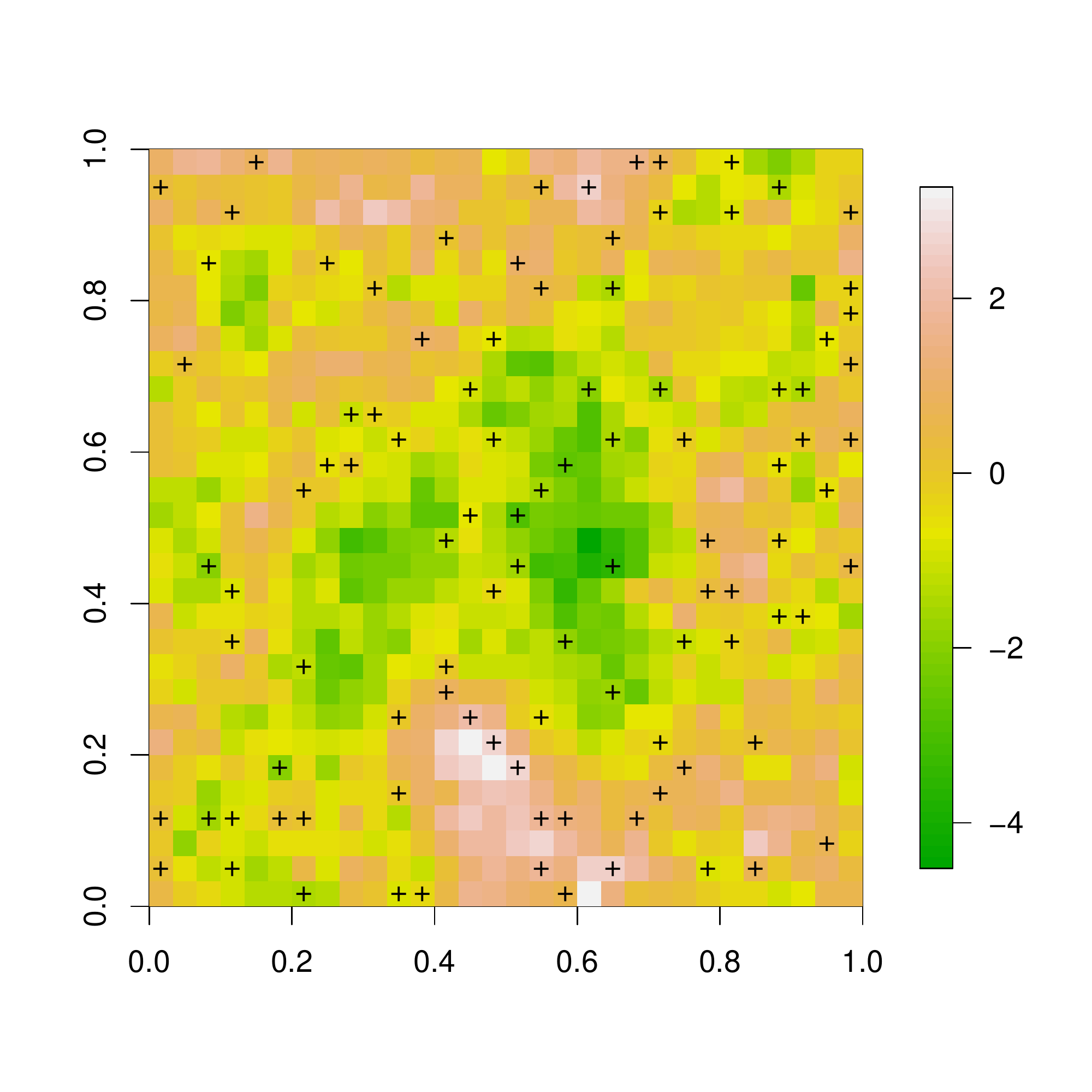}
		\includegraphics[scale=0.27]{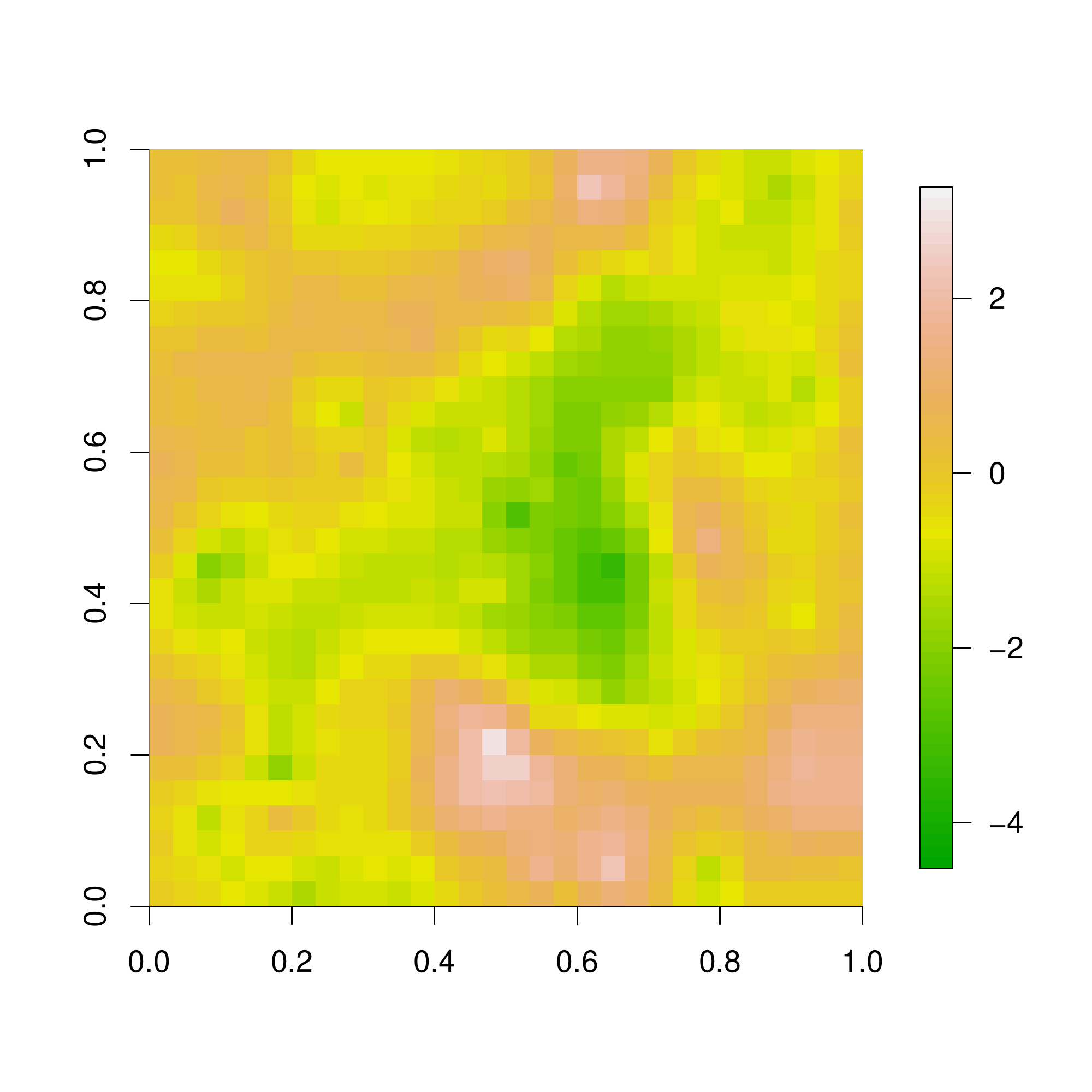}
		\includegraphics[scale=0.27]{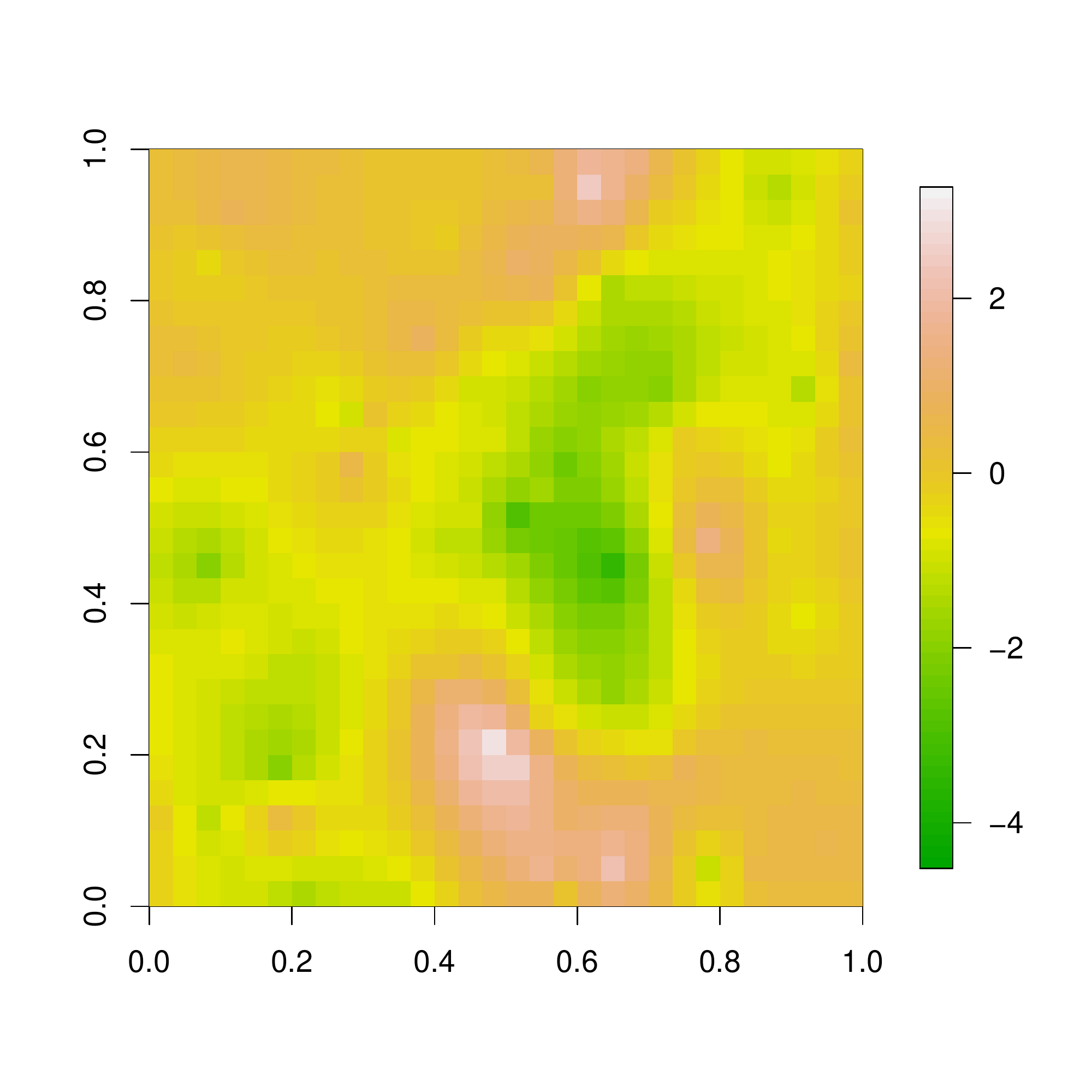}
	}\\
	\subfigure[]{
		\includegraphics[scale=0.27]{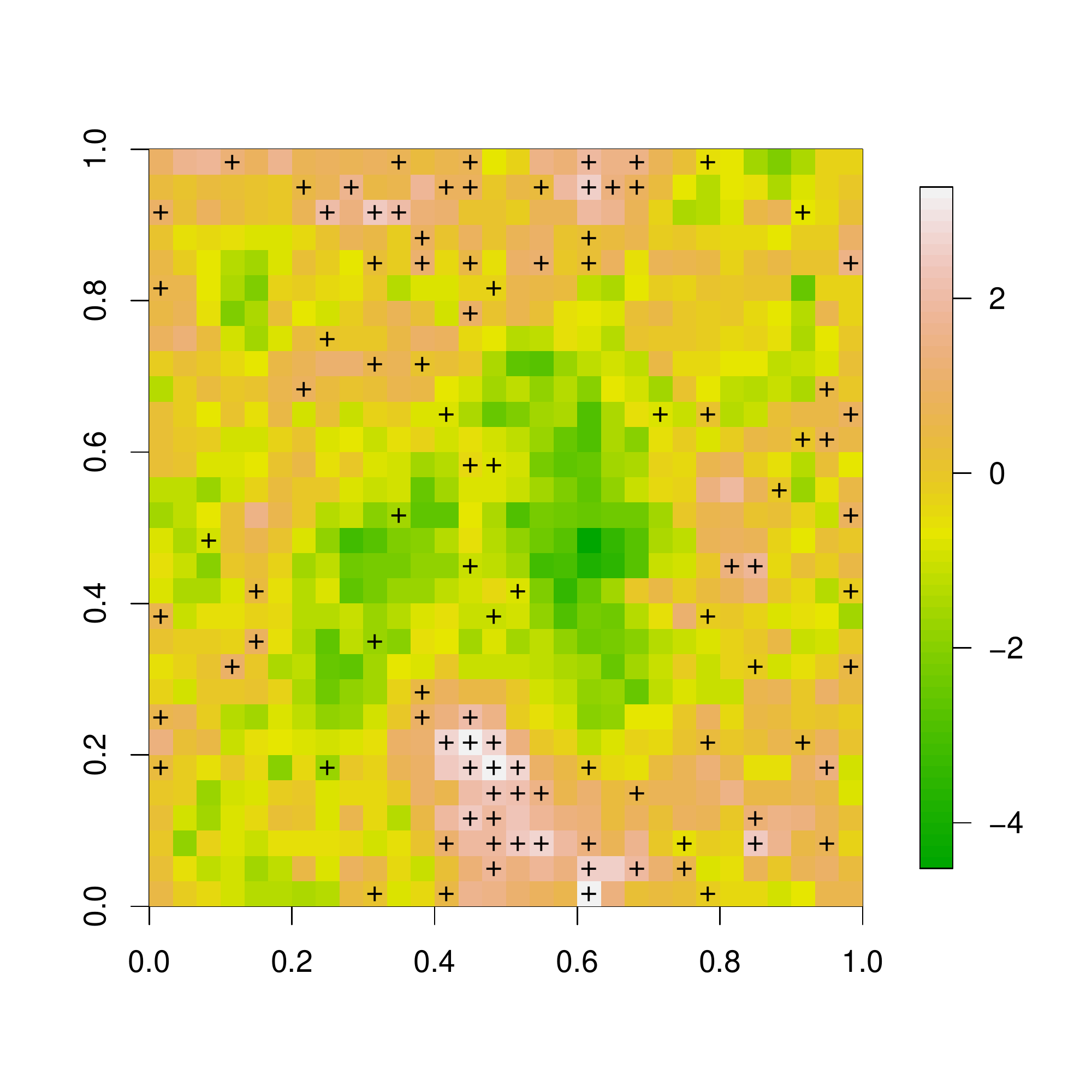}
		\includegraphics[scale=0.27]{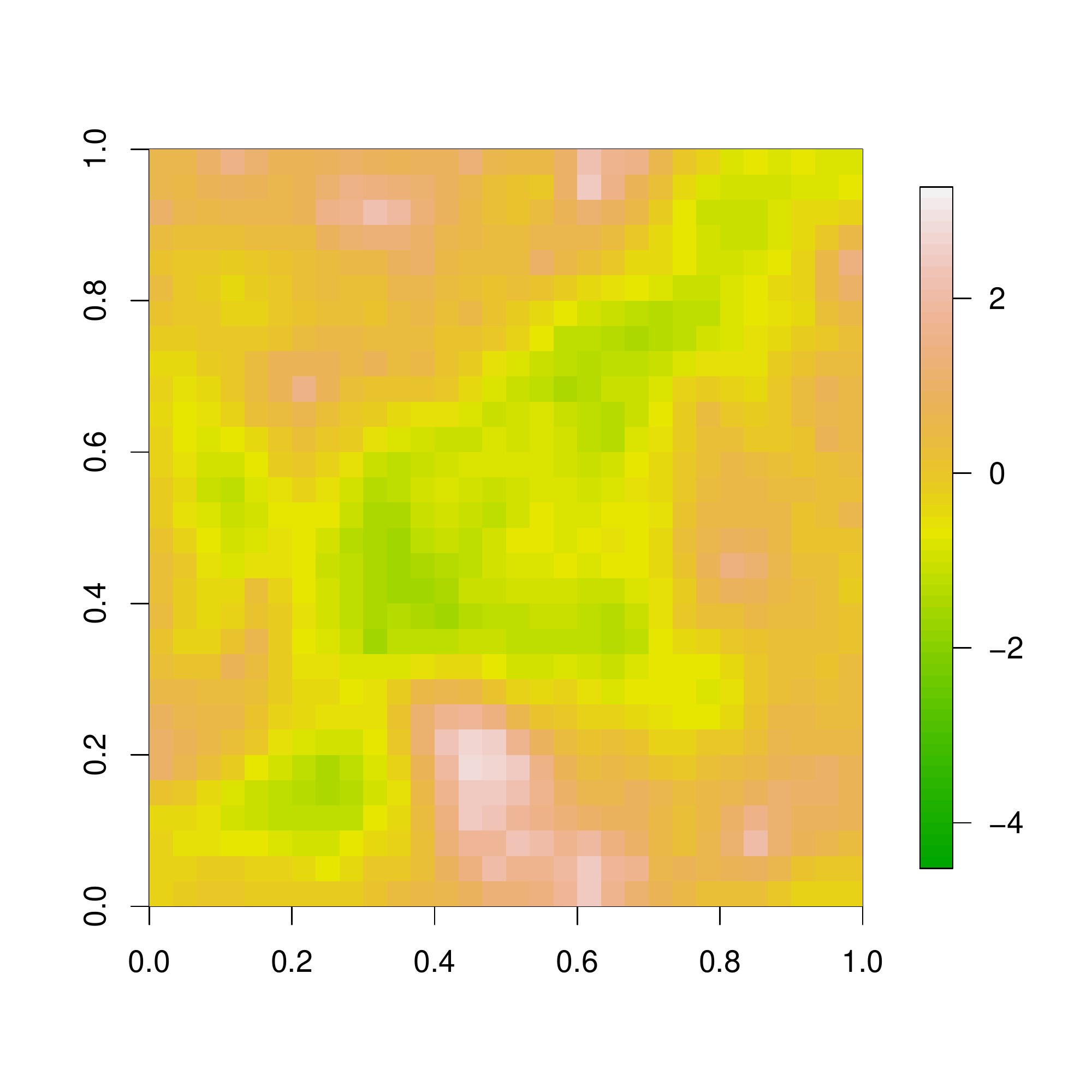}
		\includegraphics[scale=0.27]{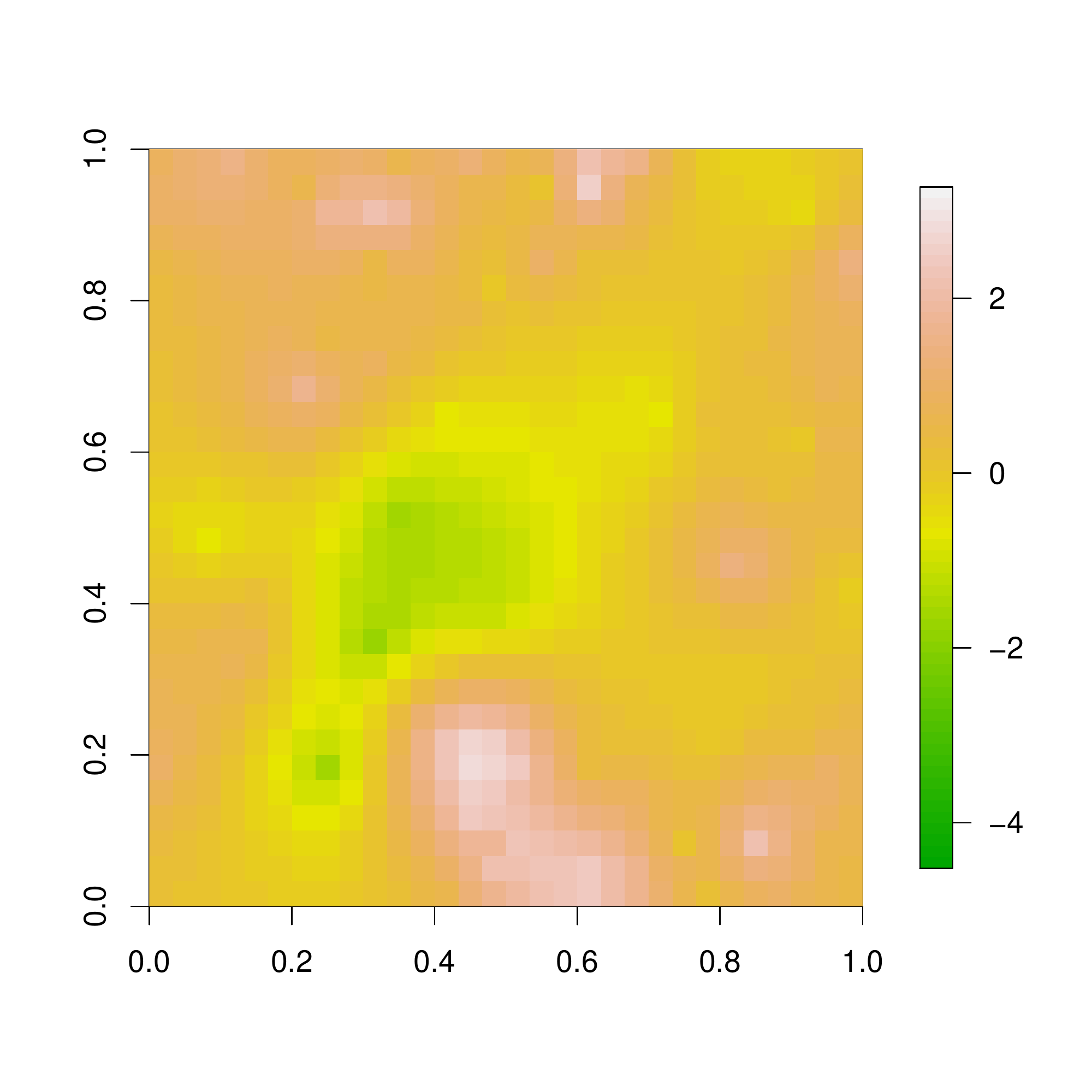}
	}\\
	\subfigure[]{
		\includegraphics[scale=0.27]{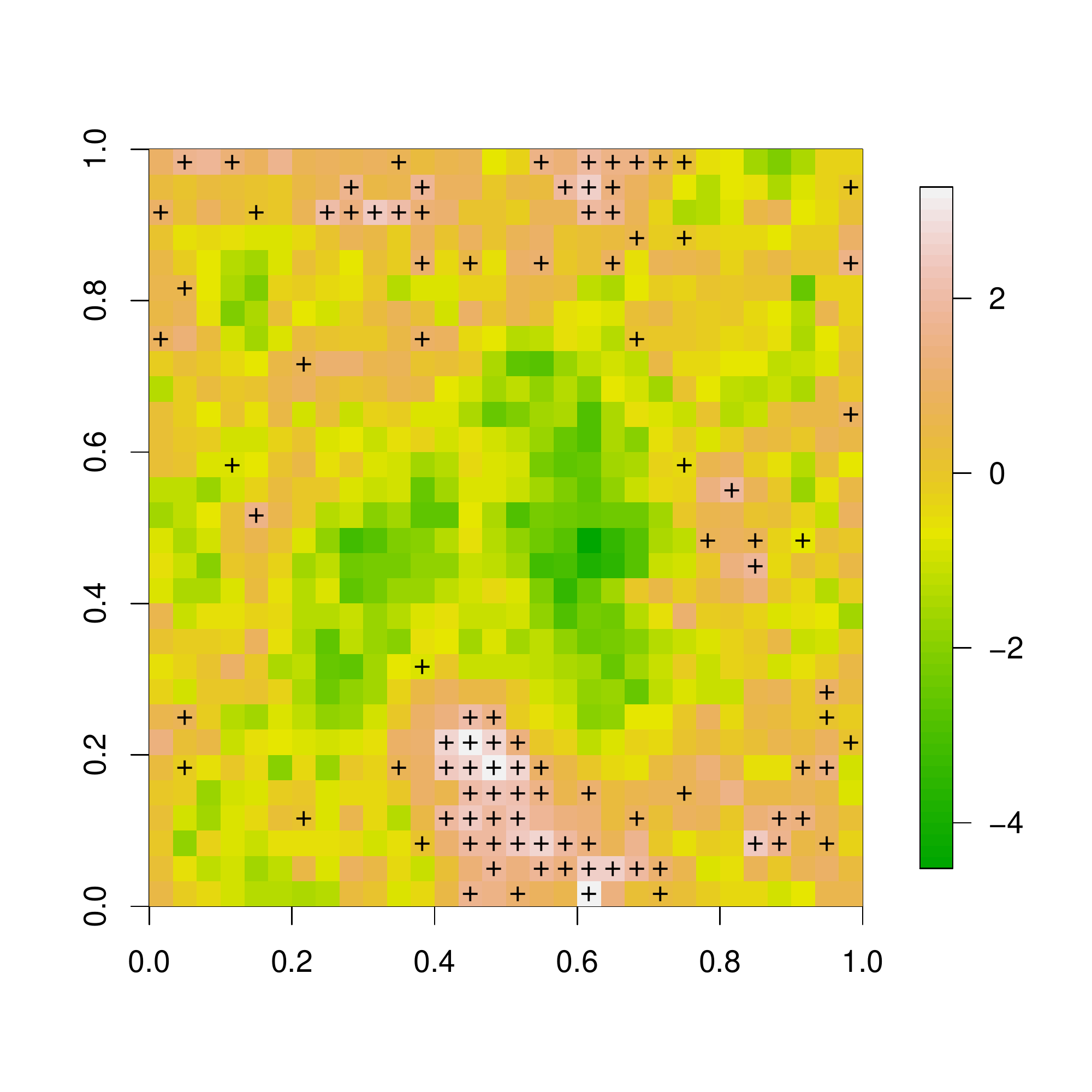}
		\includegraphics[scale=0.27]{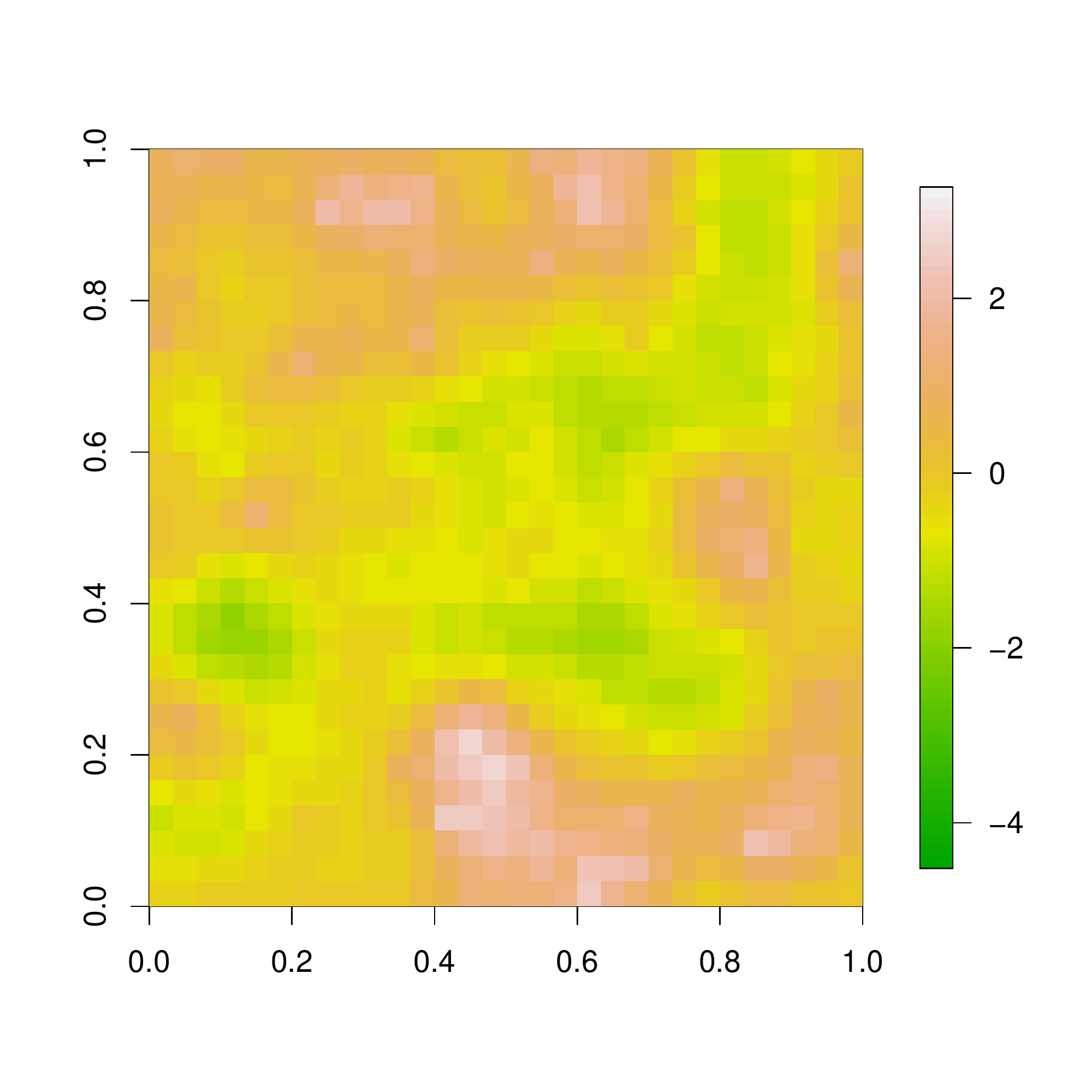}
		\includegraphics[scale=0.27]{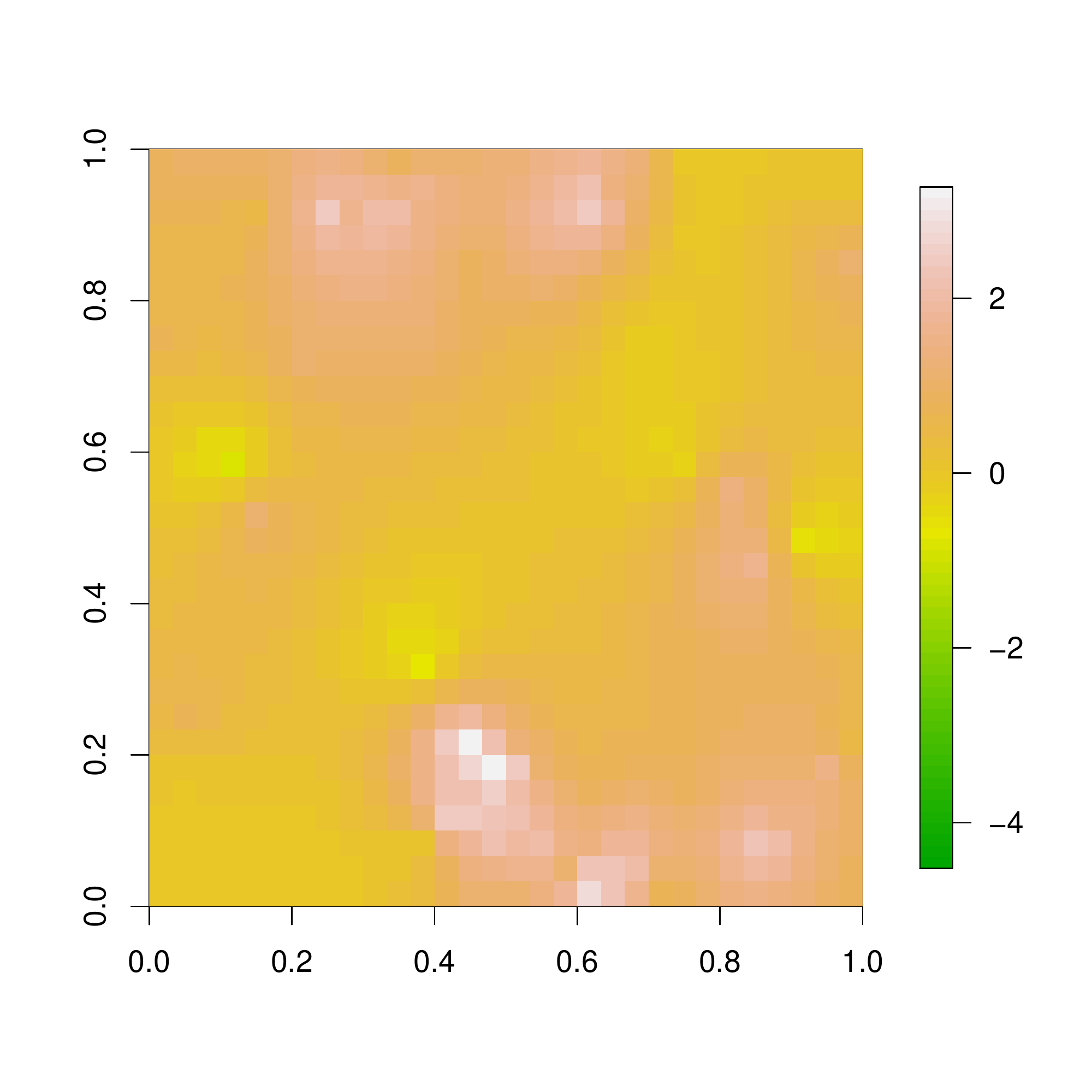}
	}
	\caption{Original map from the simulated $S$ process (column 1), predicted map from the $S|X,Y$ distribution (column 2) and predicted map from \textit{kriging} (column 3) for $\beta$ values of (a) 0, (b) 1 and (c) 2. \label{fig:mapa:compgauss}}
\end{figure}

After we observe the convergence of the chain of log-density of $\boldsymbol{S}$, the last value sampled is a realization of the predictive distribution $S|X,Y$. Thus, we need to assess the quality of this prediction. We compare it to the prediction obtained from \textit{kriging}, which does not take account the information of $X$. Figure \ref{fig:mapa:compgauss} presents the original map from the simulated $S$ process, predicted map from the $S|X,Y$ distribution and predicted map from \textit{kriging}. We considered grid size 900, 300 iterations of the MH algorithm and burn in of size 100. Thus, the predict values were calculated as the mean of the 200 remaining iterations. We note that the prediction are similar in non preferential sampling context. As we increase the value of $\beta$, \textit{kriging} provides biased predictions since we have preferential sampling and this method can not afford this. On the other hand, we observe good results using the correct distribution of $S|X,Y$, even to $\beta$ = 2. 

In order to assess the quality of those predictions, we used two measures: the mean absolute error (MAE) and the root of mean square error (RMSE), defined by
\begin{eqnarray}
MAE &=& \dfrac{1}{N}\sum_{i=1}^{N}|\hat{S}_i - S_i|,\\
RMSE &=& \sqrt{ \dfrac{1}{N}(\hat{S} - S)'(\hat{S} - S)}
\end{eqnarray}
in which $\hat{S}_i$ is the estimate of $S$ on the $i$th cell of the grid and $\hat{S}$ is the estimate of vector $\boldsymbol{S}$. As we can see in Table \ref{tab:pred:quali_pred_gauss}, we obtain better predict values when we use the correct predictive distribution in preferential sampling. Note that in the case of $\beta$ = 2, we have a great distance between quality measures of both methods of prediction, which agrees with the result we analyzed in Figure \ref{fig:mapa:compgauss}. In the non preferential case ($\beta=0$), \textit{kriging} provides better results since the method uses the exact correct predictive distribution.
\begin{table}[hbt!]
	\caption{Quality measures for the predict values of the model.}\label{tab:pred:quali_pred_gauss}
	\centering \renewcommand\arraystretch{1.2}
	\begin{tabular}{lrrrr}
		\hline
		&\multicolumn{2}{c}{MAE}&\multicolumn{2}{c}{RMSE}\\
		$\beta$ & $S|X,Y$ & \textit{Kriging} & $S|X,Y$ & \textit{Kriging}\\ \hline
		0 & 0,635 & 0,583 & 0,826 & 0,754\\
		1 & 0,613 & 0,744 & 0,793 & 0,971\\
		2 & 0,623 & 1,007 & 0,817 & 1,279\\
		\hline
	\end{tabular}
\end{table}

\subsection{Parameter estimation}
We compared the parameter estimates that we obtained by using our proposed method MCEM and SAEM for the model and obtained by the methods from \cite{diggle2010} (MCLA) and \cite{dinsdale2019} (TMB). 

Again, we assumed the true values of parameters $(\mu,\tau^2,\sigma^2,\phi,\beta) = (4,0.1,1.5,0.15,2)$ and generated data following model (\ref{eq:likel:pref}), considering a $50 \times 50$ grid and sampling size 100. We generated 200 samples for estimate parameters from MCLA and TMB methods and 50 samples for MCEM and SAEM. The reason for this is the computacional time to make a Monte Carlo simulation for SAEM algorithm in preferential sampling context. We considered blocks of size 15, burn in of size 500 and used 20 samples of the predictive distribution in the SAEM algorithm.

\begin{figure}[htb!]
	\centering
	\subfigure{\includegraphics[scale=0.29]{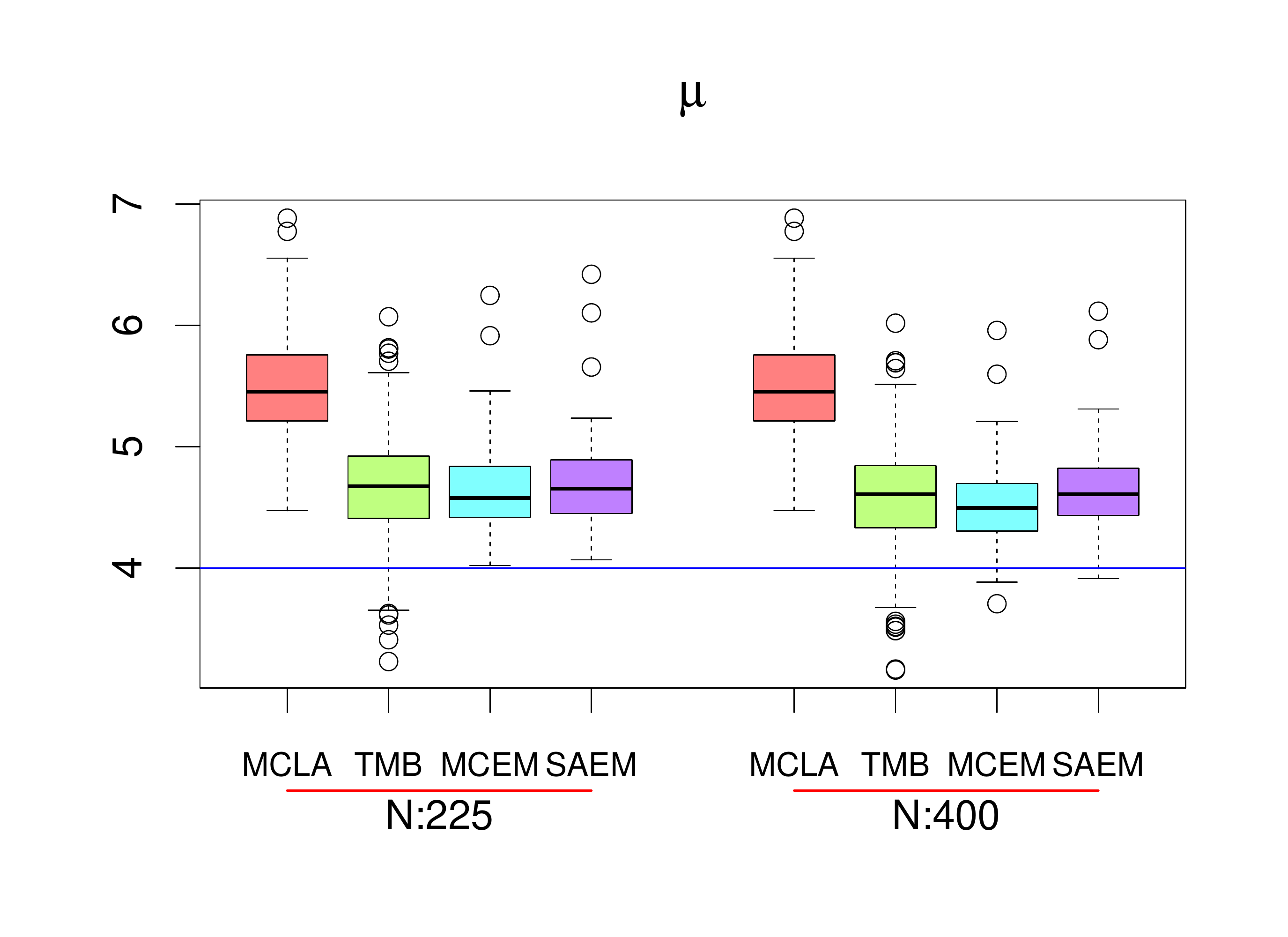}}
	\subfigure{\includegraphics[scale=0.29]{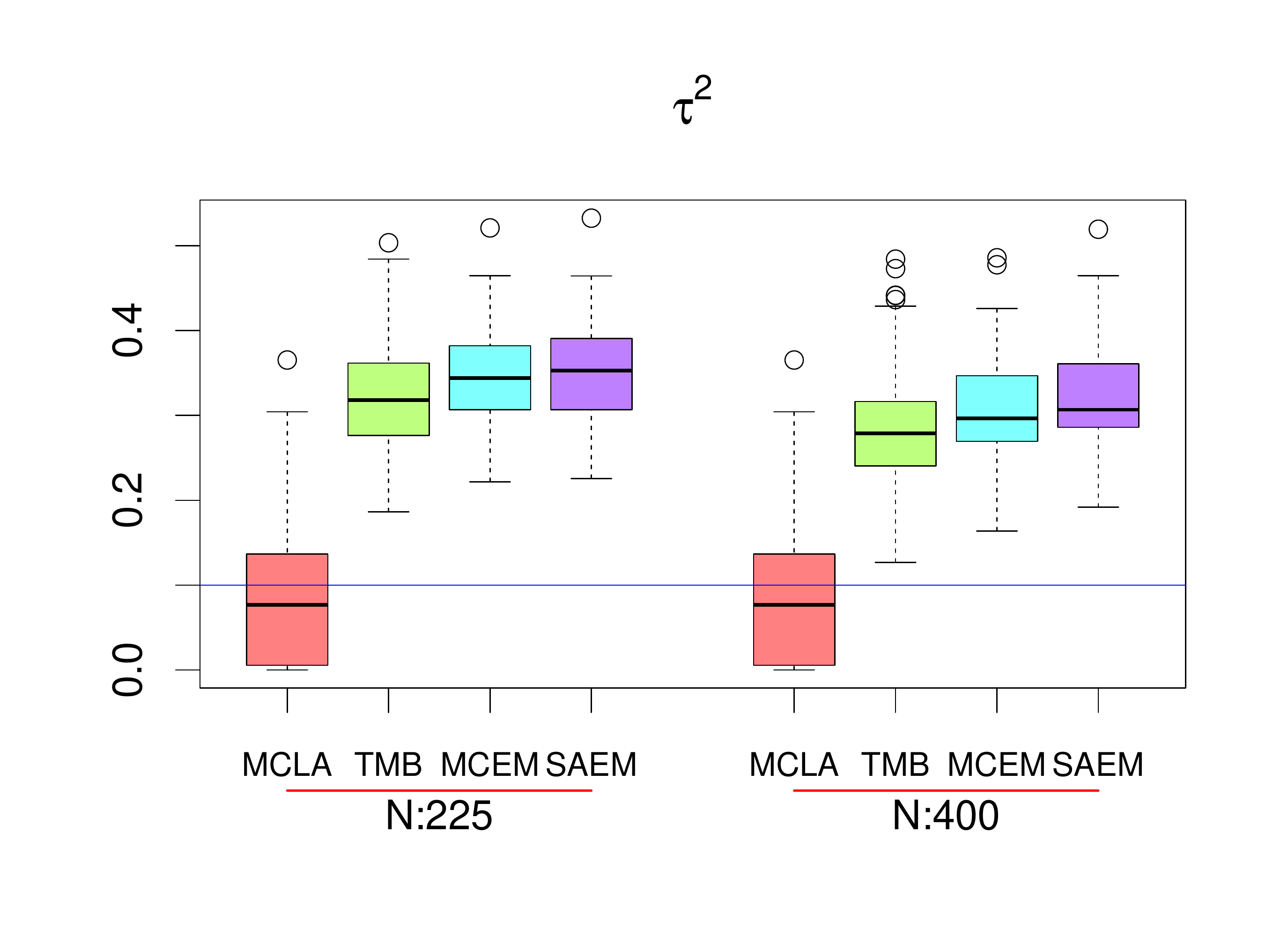}}
	\subfigure{\includegraphics[scale=0.29]{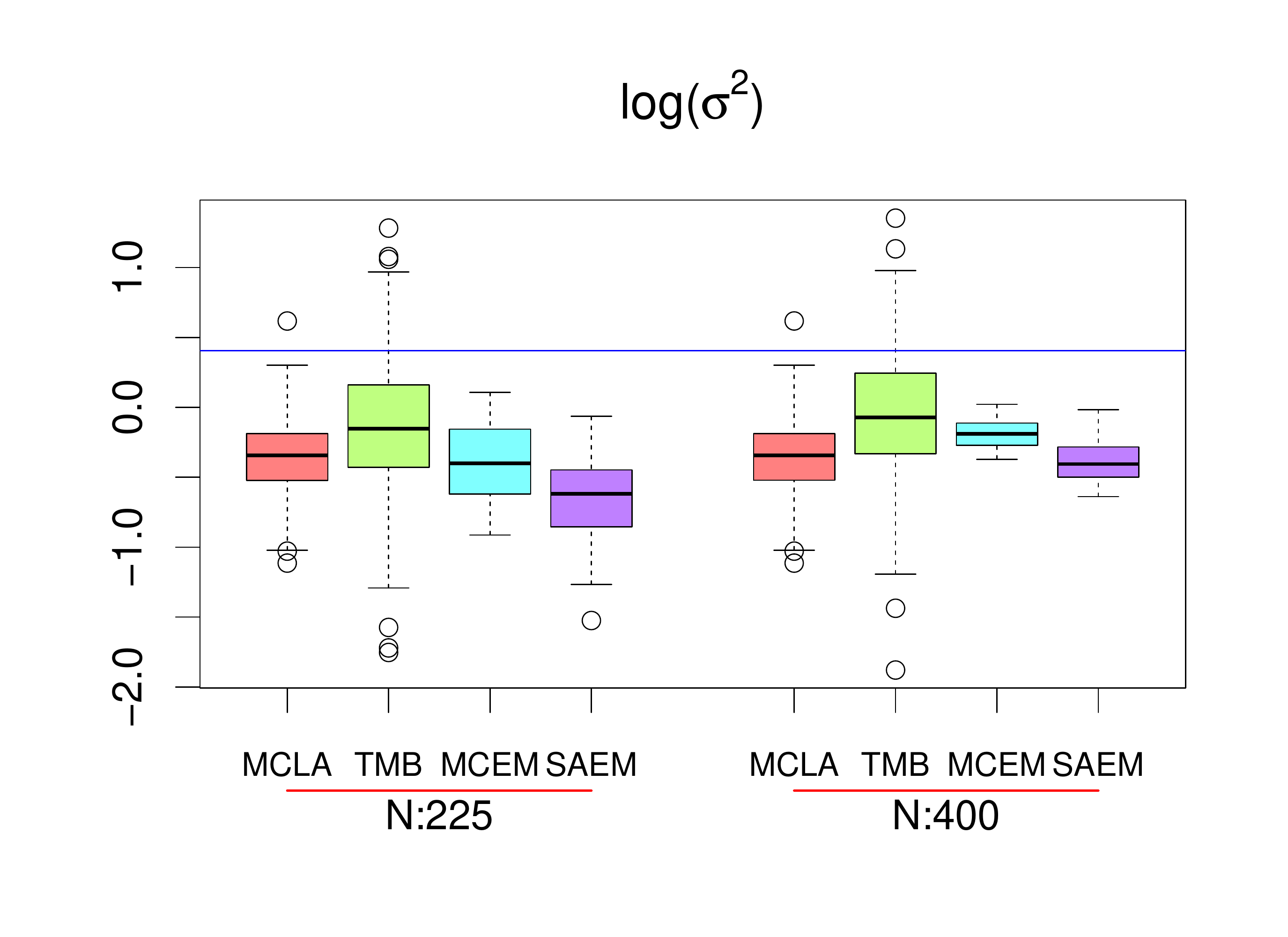}}
	\subfigure{\includegraphics[scale=0.29]{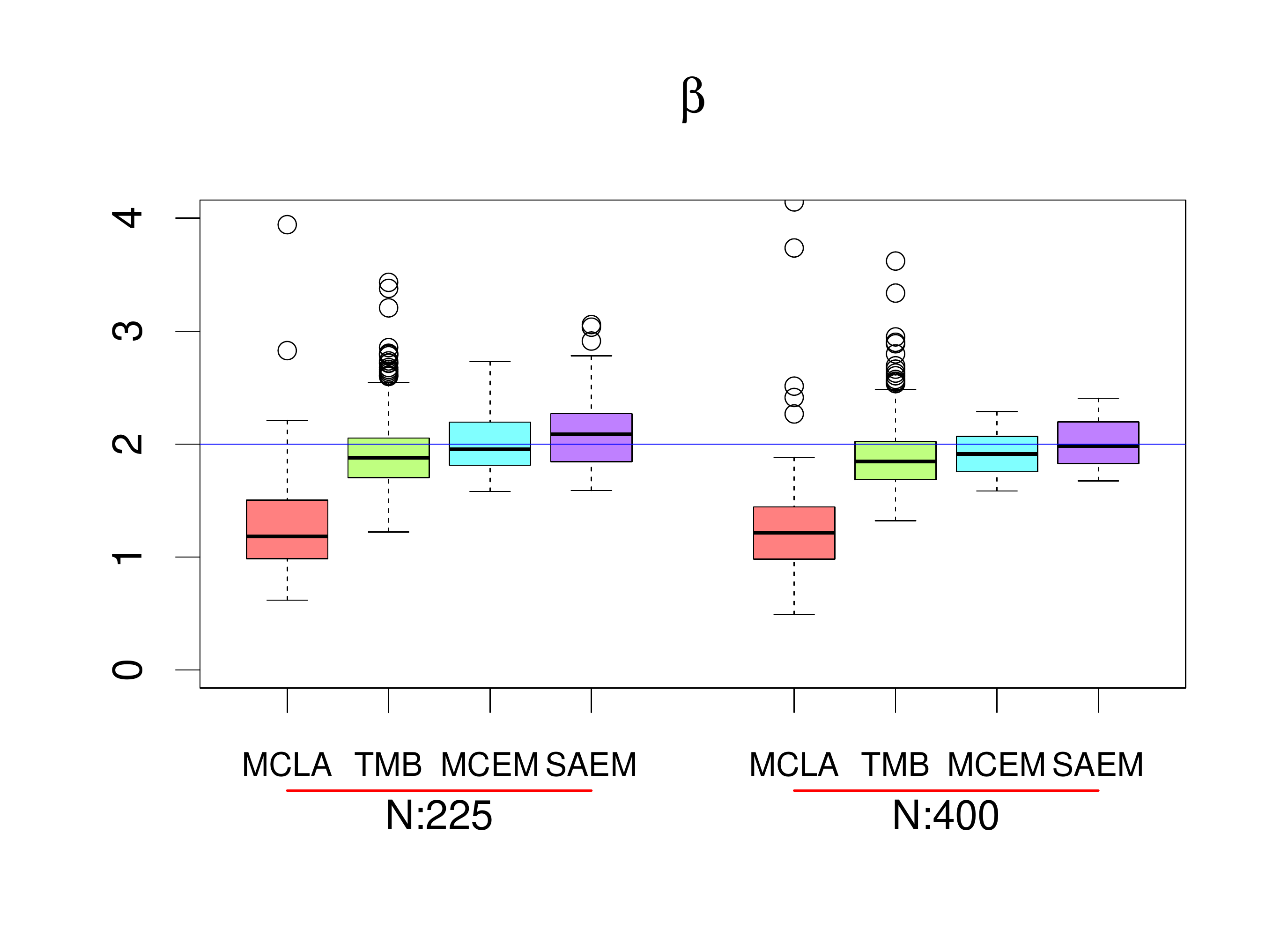}}
	\caption{Boxplot of the parameter estimates of the model by using MCLA, TMB, MCEM and SAEM methods, considering grid sizes of 225 and 400. For the MCEM and SAEM methods, the value of $\phi$ was fixed in 0.15, which is it's true value. The graph of $\beta$ parameter had your vertical axis limited, omitting 1.5\% of points from the MCLA method. \label{fig:est:comparacao}}
\end{figure}

The results are shown in Figure \ref{fig:est:comparacao}. We can observe that the mean parameter is better estimated by SAEM, MCEM and TMB method. As we expected, MCLA provides very superestimated values for $\mu$, since \cite{diggle2010} considered the non preferential distribution of $Y$ on their construction of the algorithm. Although TMB provides better estimates of $\sigma^2$, we obtained better estimates of $\beta$ parameter by using MCEM and SAEM methods. Table \ref{tab:simul:cme:gauss:tempo} shows the mean time for parameter estimation by each method and we can note that MCEM and SAEM are the methods that has taken more time for estimation. Since MCEM and SAEM provide similar estimation results and SAEM takes more time to execute, we choose to proceed the analysis with MCEM only.

\begin{table}[thb!]
	\caption{Mean time (in min) for parameter estimation of the model.}\label{tab:simul:cme:gauss:tempo}
	\centering \renewcommand\arraystretch{1.2}
	\begin{tabular}{lrrrr}
		\hline
		N$\backslash$Method & MCLA & TMB & MCEM & SAEM  \\ \hline
		225 & 0,02 & 0,14 & 0,46 & 0,83 \\
		400 & 0,03 & 0,63 & 3,75 & 13,38 \\
		\hline
	\end{tabular}
\end{table}

\begin{figure}[htb!]
	\centering
	\subfigure[]{\includegraphics[scale=0.29]{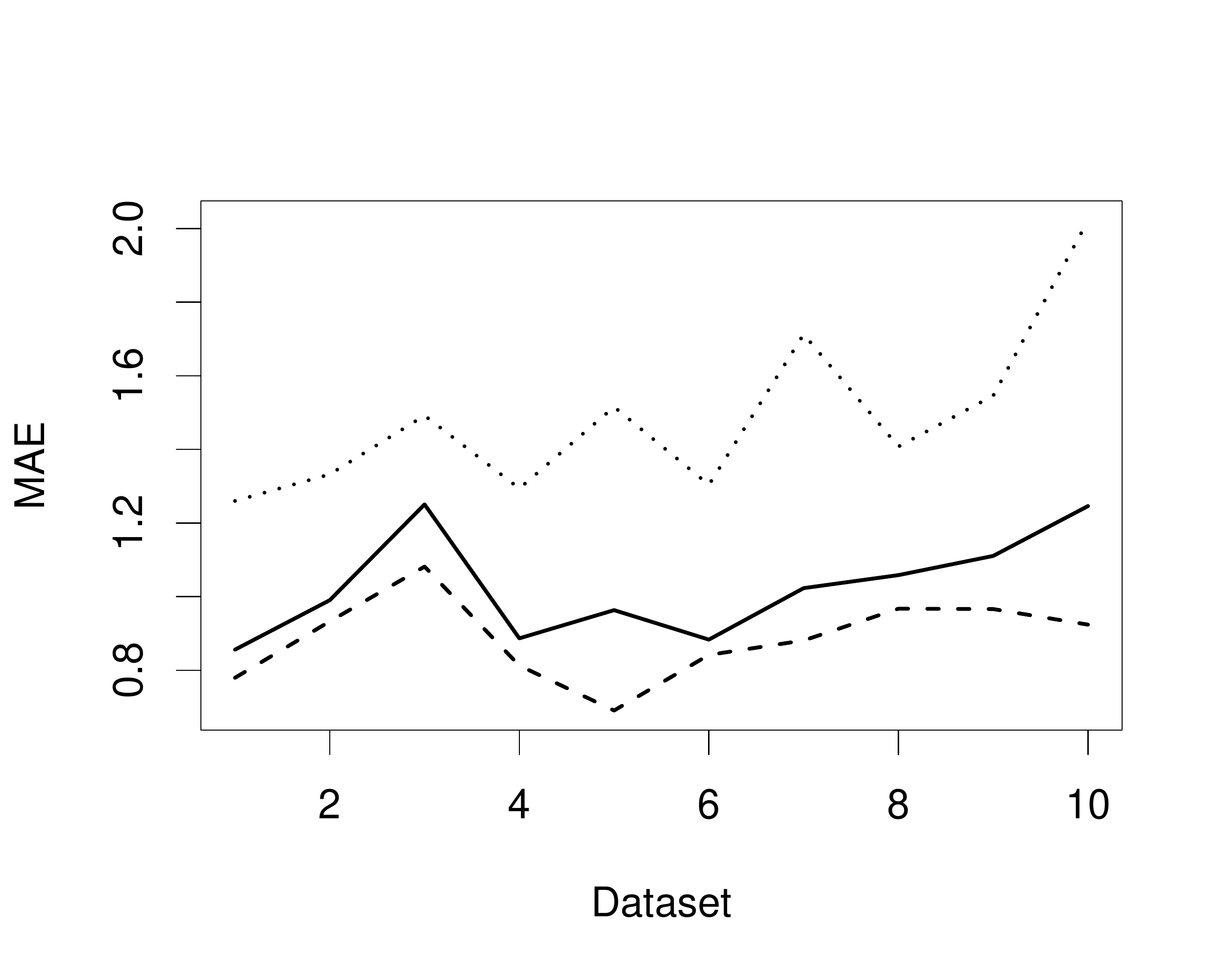}}
	\subfigure[]{\includegraphics[scale=0.29]{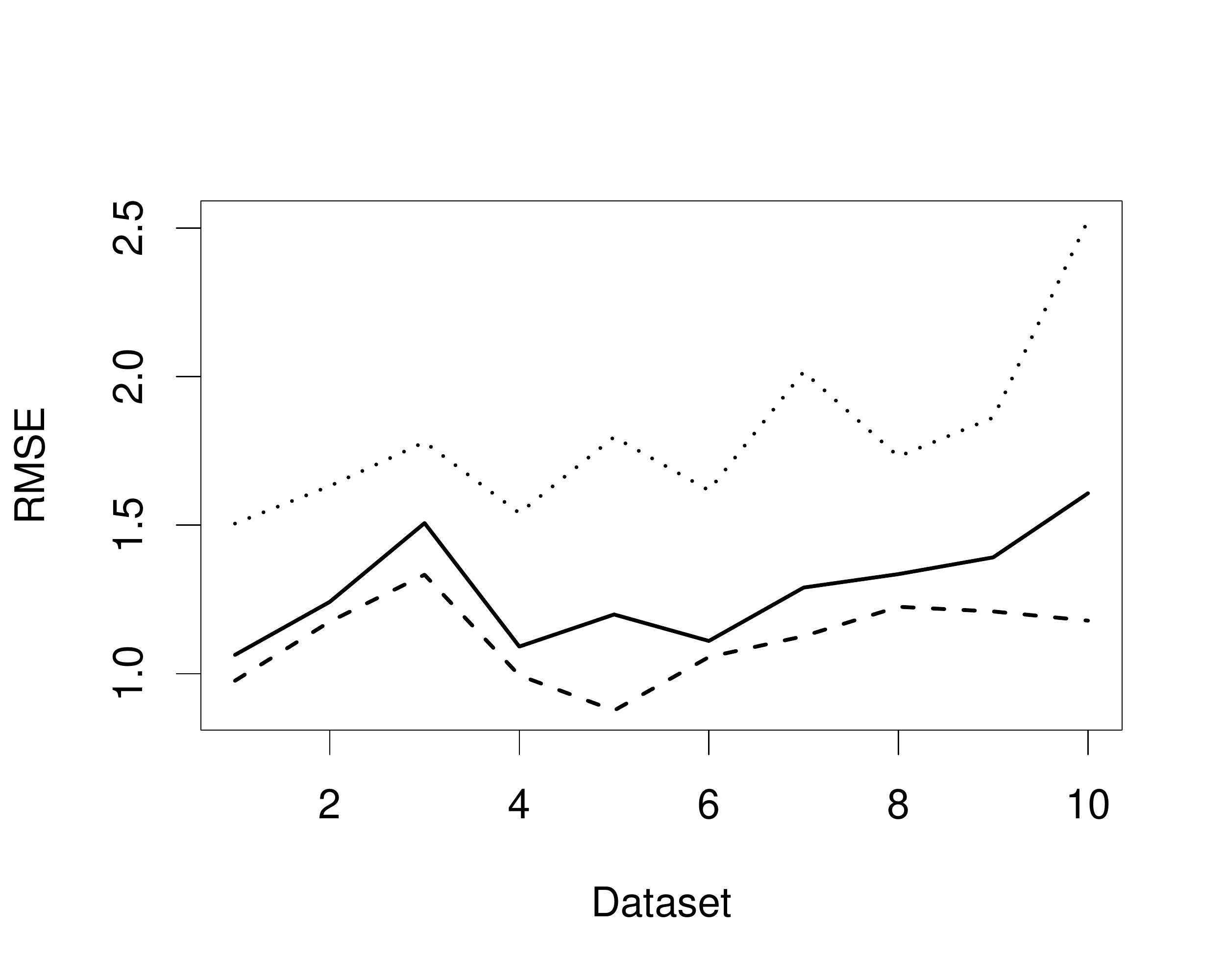}}
	\caption{MAE (a) and RMSE (b) of predicted $Y$ process for ten simulated data sets considering MCEM (continuous line), TMB (traced line) and MCLA (dotted line). \label{fig:est:comparacao:pred}}
\end{figure}

\begin{figure}[htb!]
	\centering
	\subfigure[]{\includegraphics[scale=0.29]{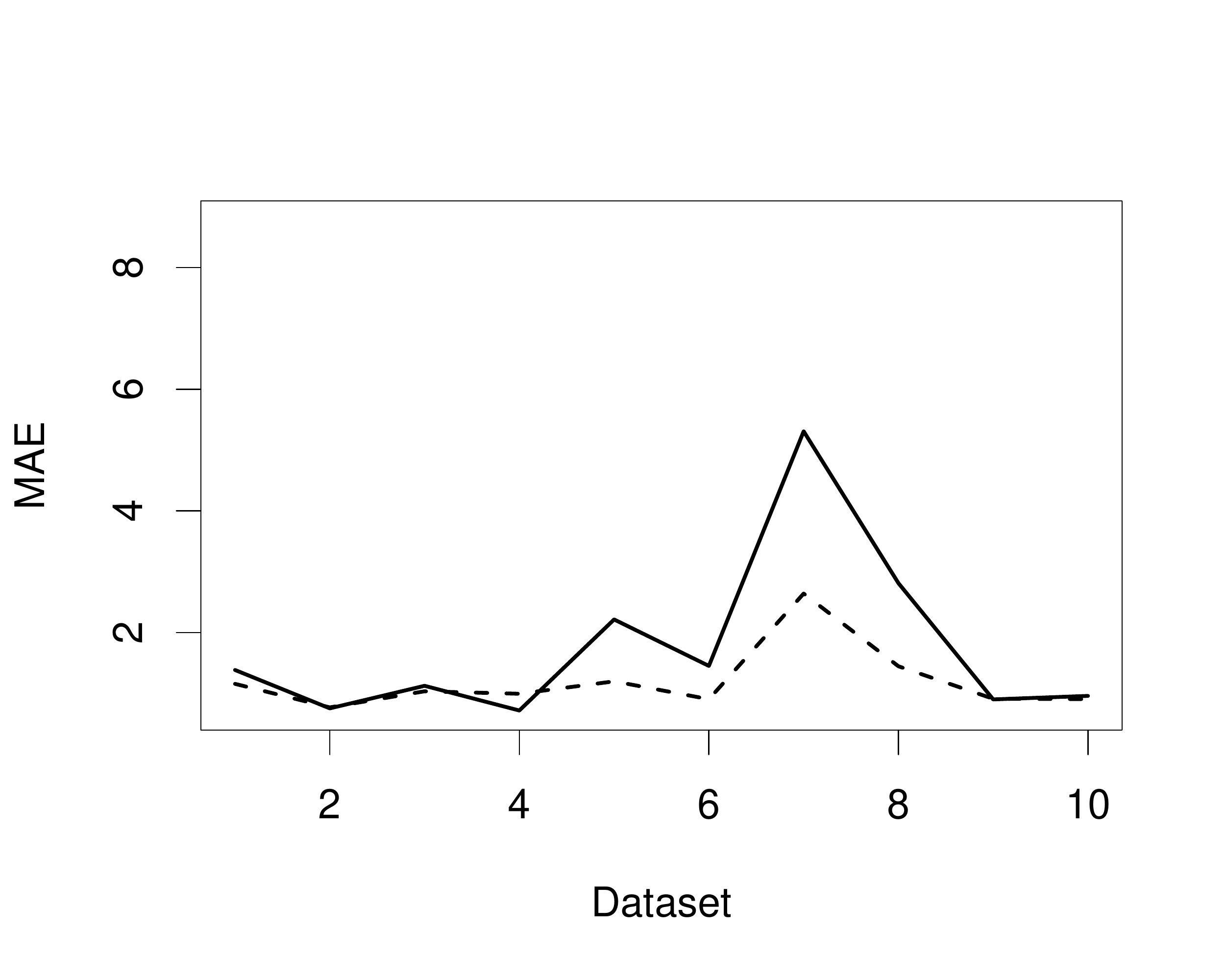}}
	\subfigure[]{\includegraphics[scale=0.29]{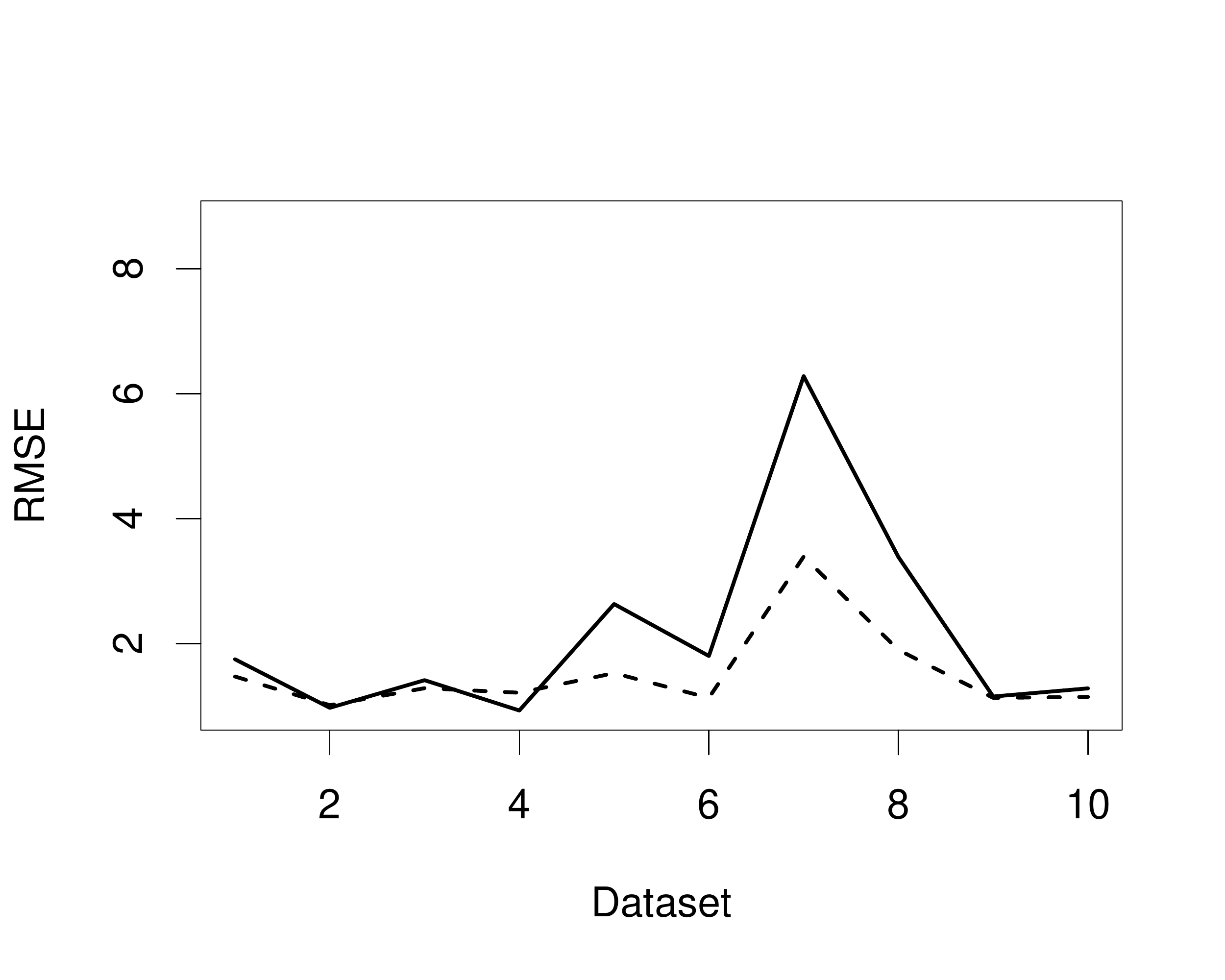}}\\
	\subfigure[]{\includegraphics[scale=0.29]{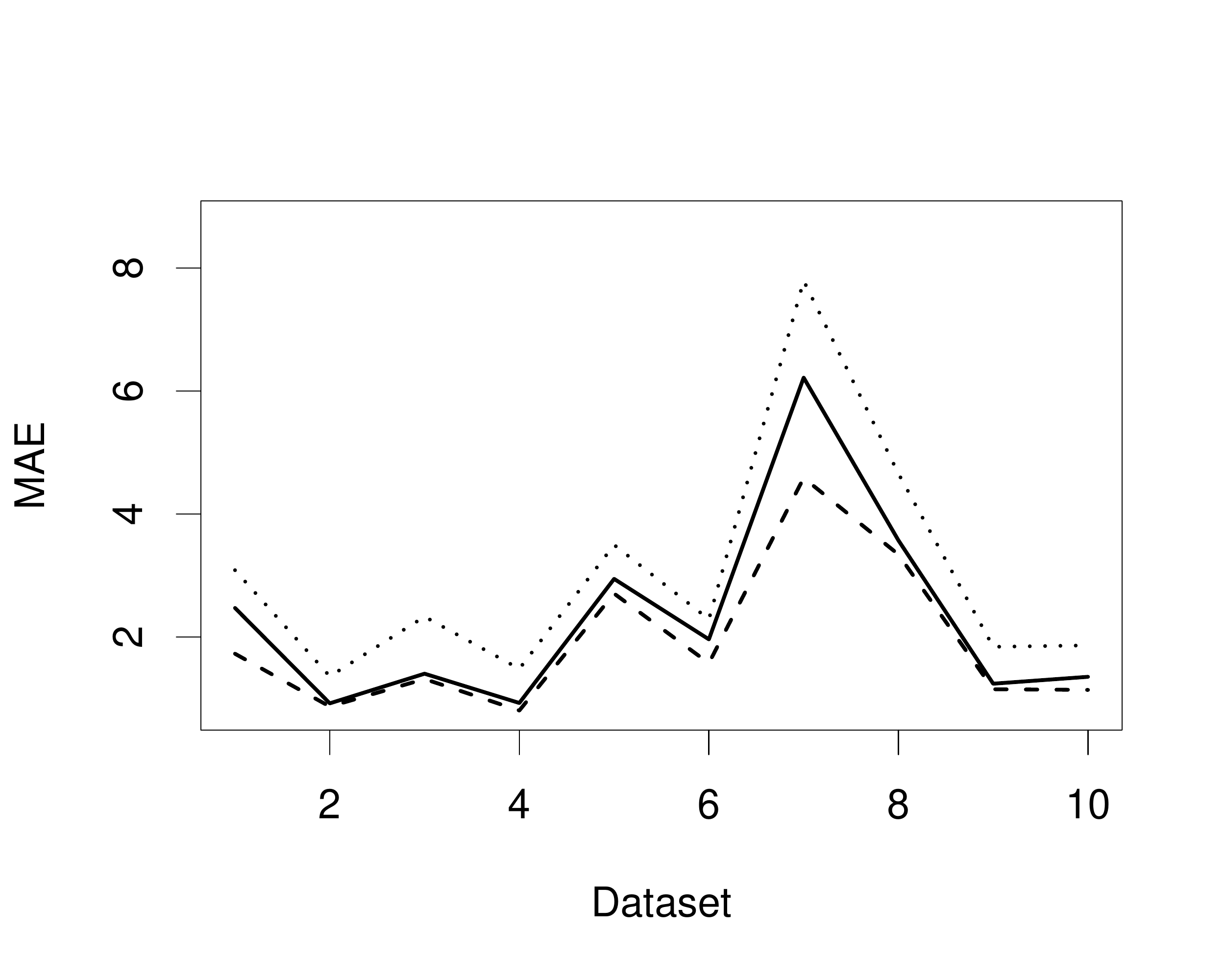}}
	\subfigure[]{\includegraphics[scale=0.29]{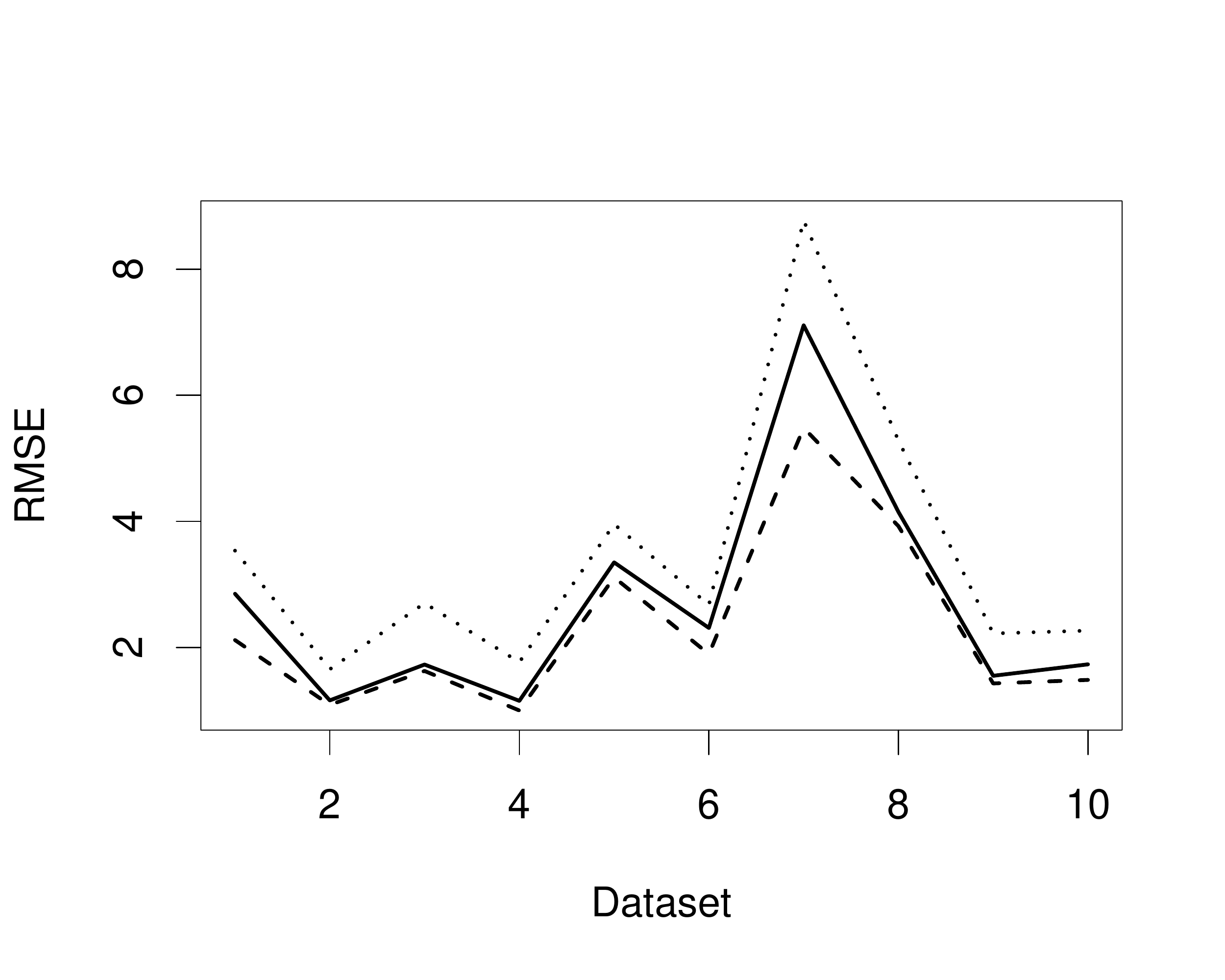}}\\
	\caption{MAE (a) and RMSE (b) of predicted $Y$ process for ten simulated data sets considering MCEM (continuous line), TMB (traced line) and MCLA (dotted line). \label{fig:est:comparacao:predt}}
\end{figure}

\section{Application}
\begin{figure}[htb!]
	\centering
	\subfigure{\includegraphics[scale=0.4]{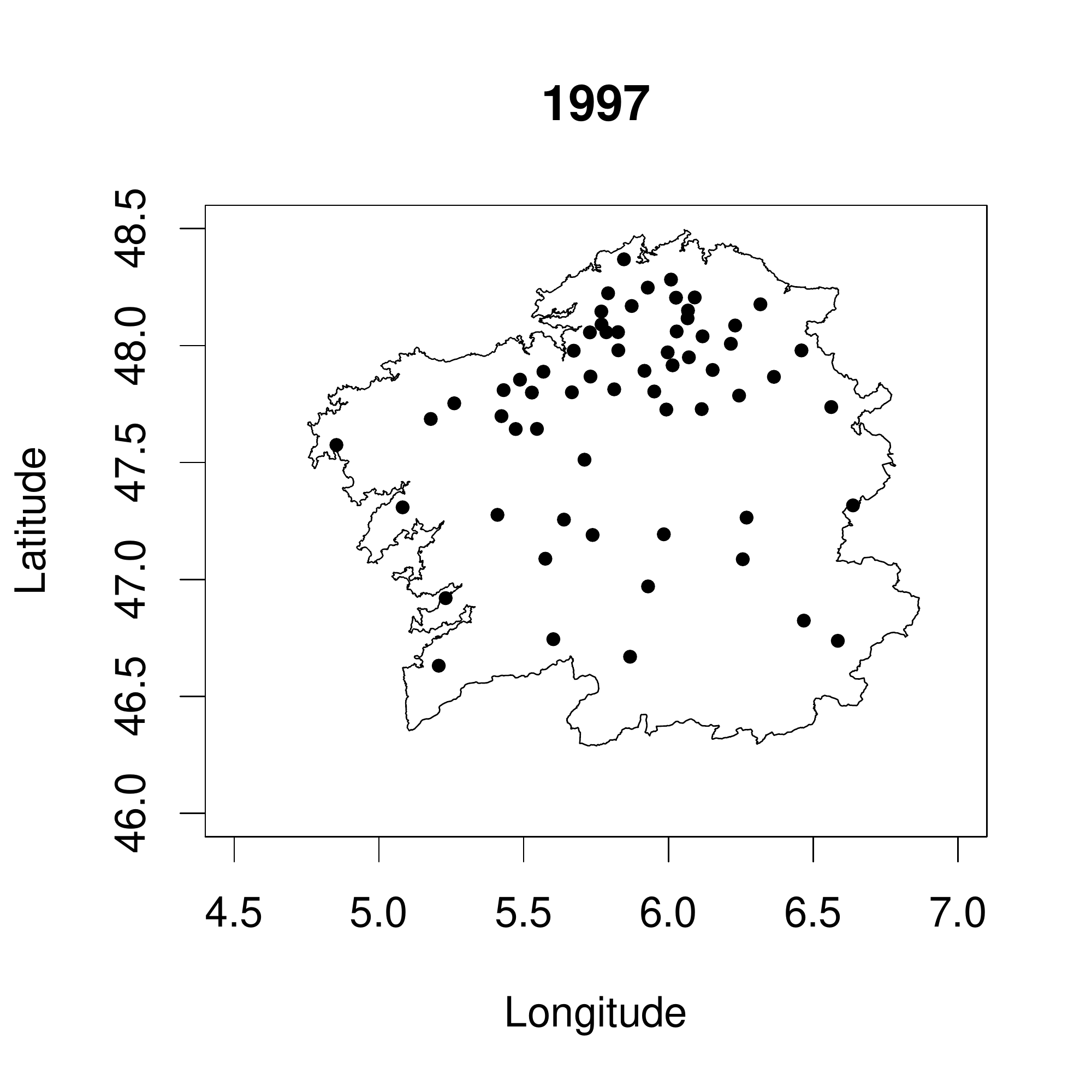}}
	\subfigure{\includegraphics[scale=0.4]{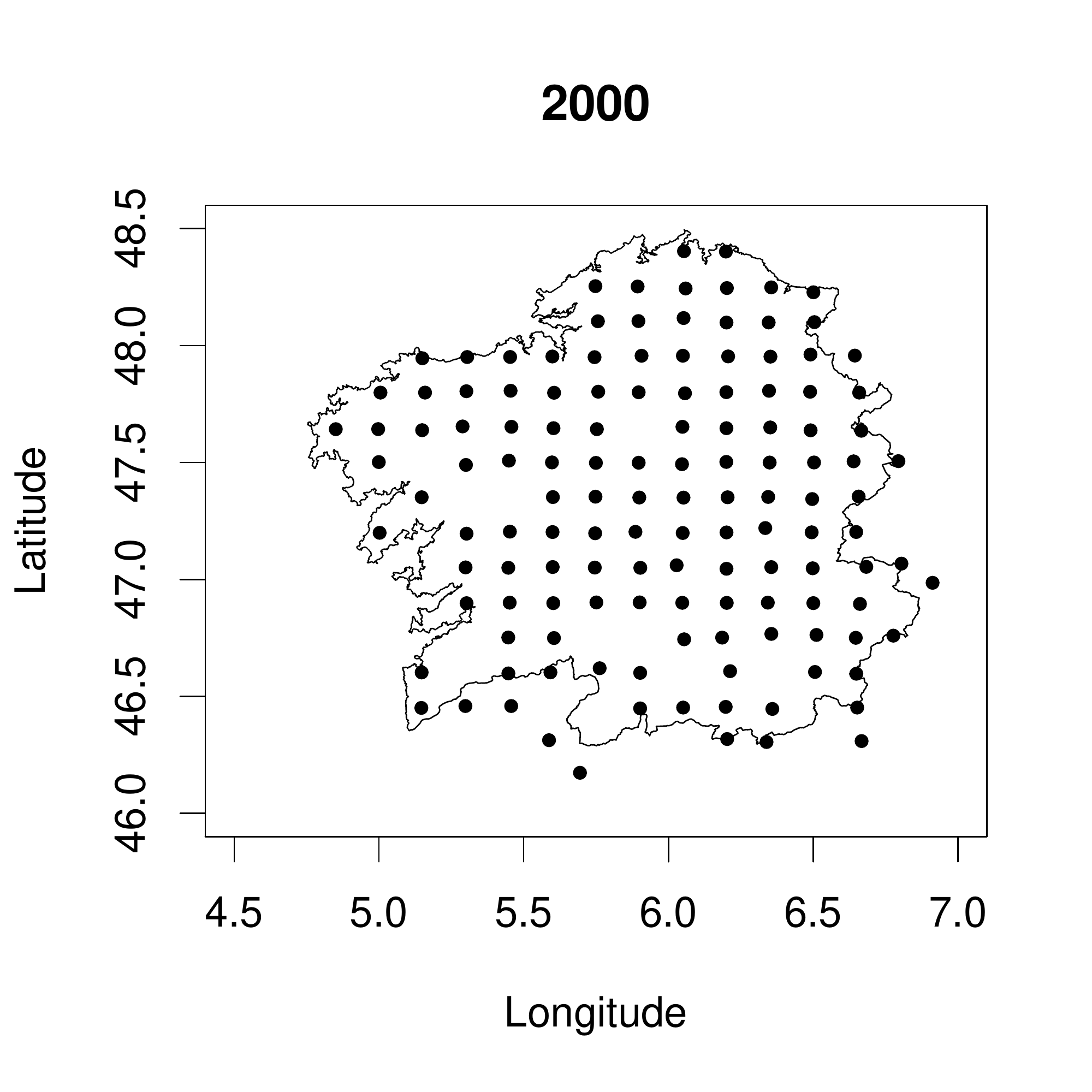}}
	\caption{Sampling locations for 1997 and 2000 of moss data. \label{fig:aplic:moss:mapa}}
\end{figure}

In this section we illustrate our approach on a real data set. 
The moss data is well known in literature, see for example \cite{fernandez2000}, \cite{diggle2010} and \cite{dinsdale2019}. The data consists in measures of lead concentrations in samples of moss in Galicia, nothern Spain. The uptake of heavy metals in mosses occurs mainly from atmospheric deposition, which turns mosses as a biomonitoring of pollution. The study was conducted in October 1997 and July 2000 and Figure \ref{fig:aplic:moss:mapa} shows the sampling design of each year. The choice of sampling locations in 1997 was conducted more intensively in subregions where high lead concentration was expected, turning this sampling design potentially preferential. The second survey was conducted using an approximately regular grid over the Galicia region and, thus, the sampling design of this survey is considered non preferential.

\begin{figure}[hbt!]
	\centering
	\subfigure{\includegraphics[scale=0.41]{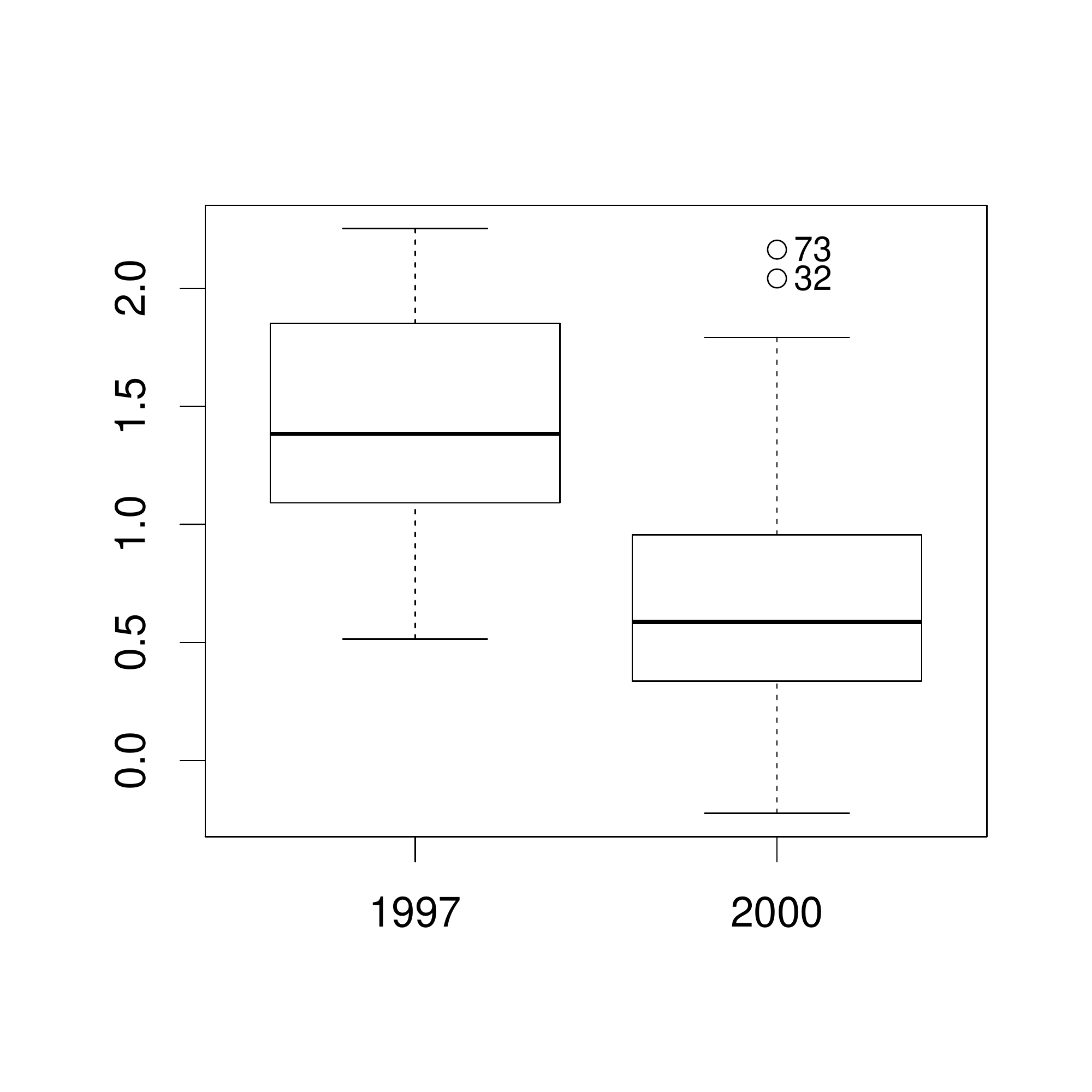}}
	\caption{Boxplot of log-lead concentration in moss in 1997 and 2000. \label{fig:aplic:moss:boxplot_loglead}}
\end{figure}

Table \ref{tab:aplic:moss:est_desc} gives summary statistics for log-lead concentration of moss in both years and Figure \ref{fig:aplic:moss:boxplot_loglead} shows the boxplot of moss data. The log scale was applied to make data distribution symmetric. We observe two outliers in the data set of 2000, the observations 32 and 73. \cite{diggle2010} replaced those outliers by the average of the remaining values from that year's survey. To assess the effect of this replacement, we do not replace or exclude those outliers. In this way, we proceed to the analisys.
\begin{table}[hbt!]
	\caption{Summary statistics of log-lead concentration of moss measured in 1997 and 2000.}\label{tab:aplic:moss:est_desc}
	\centering \renewcommand\arraystretch{1.2}
	\begin{tabular}{lcccccccc}
		\hline
		Year&n&Min&1st Quartile&Median&Mean&3rd Quartile&Max&Stand.dev.\\ \hline
		1997 & 63 &0.515&1.091&1.385&1.440&1.851&2.163&0.480 \\
		2000 & 132&-0.223&0.336&0.586&0.661&0.955&2.163&0.432\\
		\hline
	\end{tabular}
\end{table}

To compare the estimatioin obtained by using our approach with MCEM and the others methods, we fitted a model for data of each year and estimate the parameter's model by using MCLA, TMB and MCEM. We also estimate the parameters of the non preferential geostatistical model (NPG). We considered a $20 \times 20$ grid and exponential correlation function. Results can be viewed in Table \ref{tab:aplic:moss:estimates}. Note that NPG and MCLA gives the same estimates of parameters for both years, except for $\beta$, but different standard erros for 2000. This occurs because MCLA uses the non preferential distribution of $Y$ and, in it's formulation, it does not depend on simulations of $S$. Analysing the results for year 1997, we can observe the value of $\hat{\mu}$ is higher for TMB and MCEM, the methods that consider the correct preditive distribution of $S|X,Y$. Since we have a potentially preferential sampling in this year, it means that those methods give corrected estimates for $\mu$. In the other hand, NPG and MCLA gives biased estimates since those methods do not incorporate the preferability in the estimate process of $\mu$. Comparing TMB with MCEM, we can note the value of $\hat{\beta}$ is higher for TMB, with higher standard error. With MCEM, we got estimates for $\beta$ with more precision.
\begin{table}[hbt!]
	\caption{Parameter estimates (and standard erros) for log-lead concentration in moss in 1997 and 2000.}\label{tab:aplic:moss:estimates}
	\centering \renewcommand\arraystretch{1.2}
	\footnotesize
	\begin{tabular}{lccccccccc}
		\hline
		&\multicolumn{4}{c}{1997} &\phantom{}& \multicolumn{4}{c}{2000}\\ \cmidrule{2-5} \cmidrule{7-10}
		& NPG & MCLA & TMB & MCEM && NPG & MCLA & TMB & MCEM \\ \hline
		\multirow{2}{*}{$\mu$} & 1.542 & 1.542 & 1.608 & 1.985  & &0.724& 0.724 & 0.923 & 0.702   \\
		&(0.113)& (0.113) & (0.159) & (0.050)  & &(0.100)& (0.100) & (0.191) & (0.026) \\ \hline
		\multirow{2}{*}{$\tau^2$} & 0.083 & 0.083 & 0.171 & 0.138  && 0.000 & 0.000 & 0.057 & 0.056  \\
		&(0.042)& (0.042) & (0.039) & (0.027) &&(0.047)& (0.001) & (0.026) & (0.010)  \\ \hline
		\multirow{2}{*}{$\sigma^2$} & 0.147 & 0.146 & 0.059 & 0.194  && 0.192 & 0.192 & 0.174 & 0.144  \\
		&(0.062)& (0.062) & (0.062) & (0.033)  &&(0.058)& (0.037) & (0.073) & (0.015)  \\ \hline
		\multirow{2}{*}{$\phi$} & 0.193 & 0.193 & 0.521 & 0.268  && 0.206 & 0.206 & 0.425 & 0.137  \\
		&(0.120)& (0.120) & (0.836) & (0.046)  &&(0.082)& (0.052) & (0.260) & (0.018)  \\ \hline
		\multirow{2}{*}{$\beta$} & - & -1.514 & -3.896 & -2.754  && - & -0.481 & -0.847 & -0.269  \\
		&& (0.346) & (2.785) & (0.371) &&& (0.254) & (0.308) & (0.246)  \\
		\hline
	\end{tabular}
\end{table}

\begin{figure}[hbt!]
	\centering
	\subfigure{\includegraphics[scale=0.25]{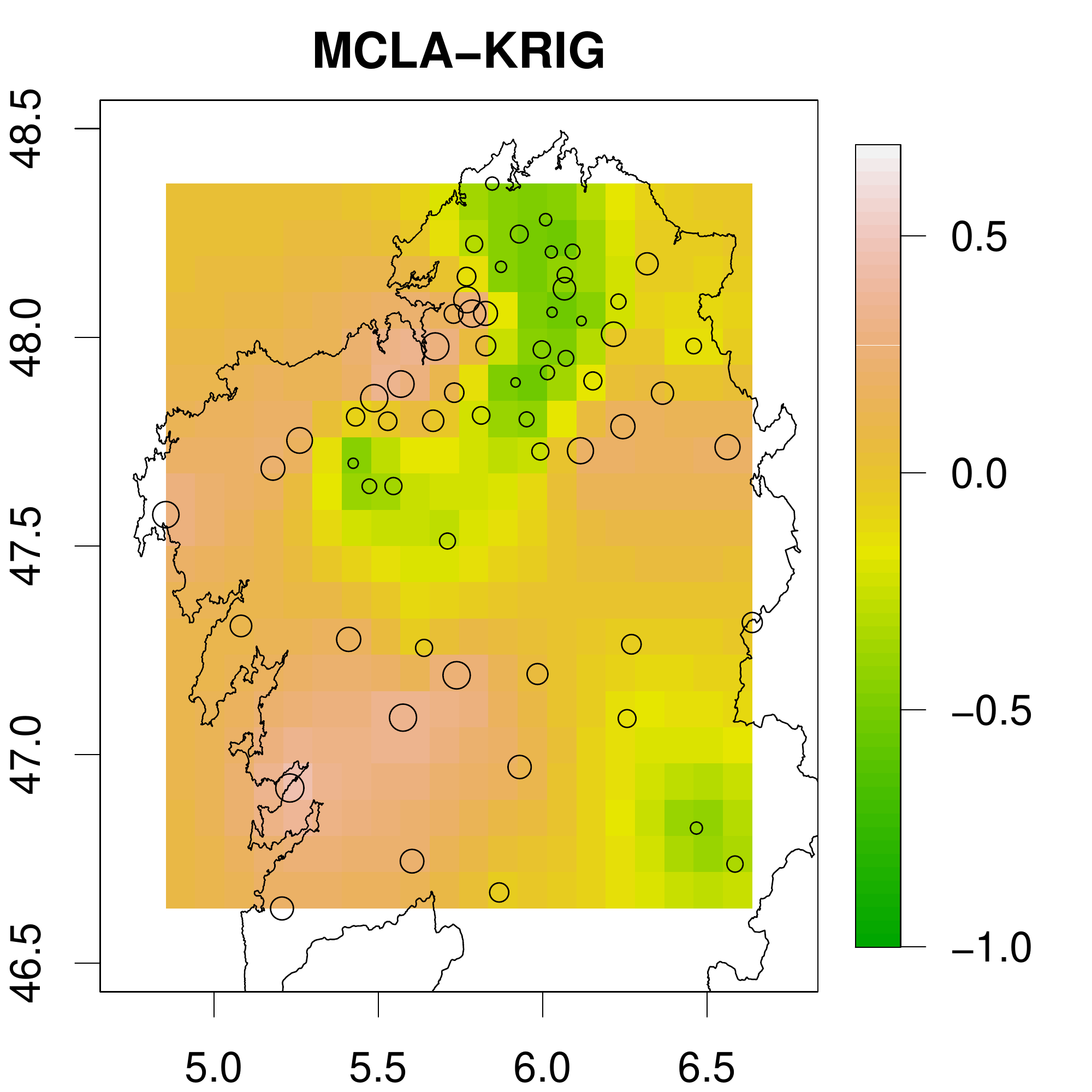}}
	\subfigure{\includegraphics[scale=0.25]{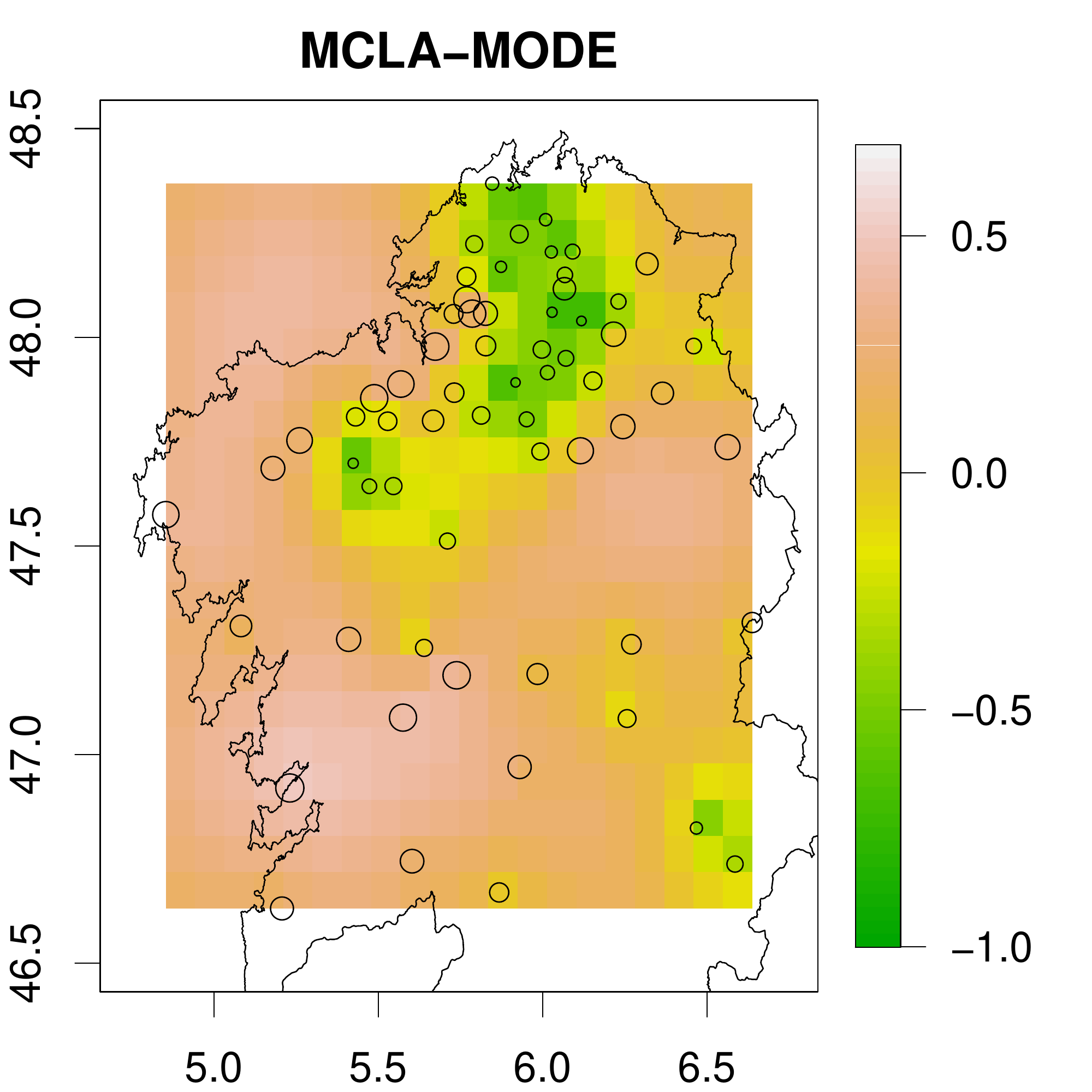}}
	\subfigure{\includegraphics[scale=0.25]{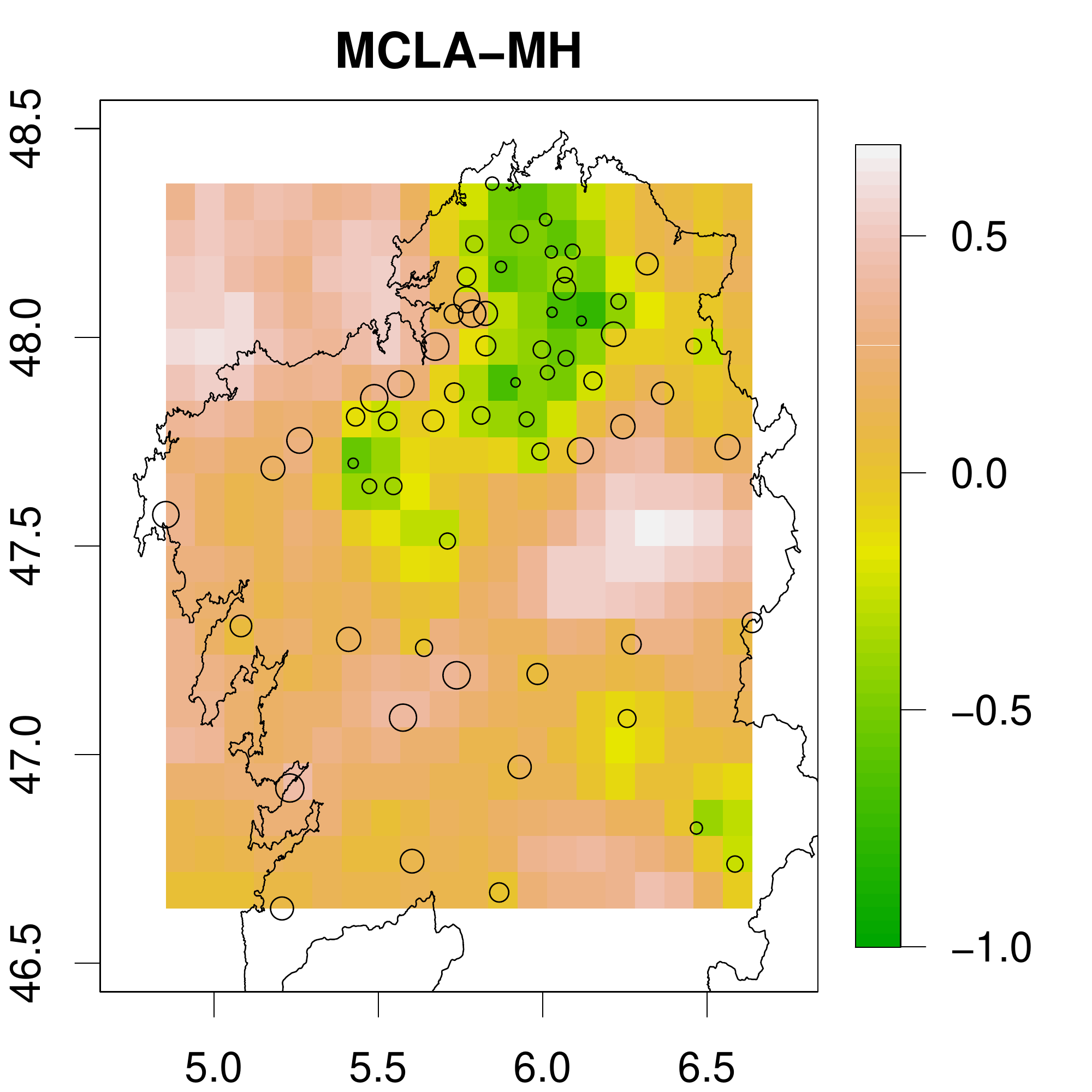}}\\
	\subfigure{\includegraphics[scale=0.25]{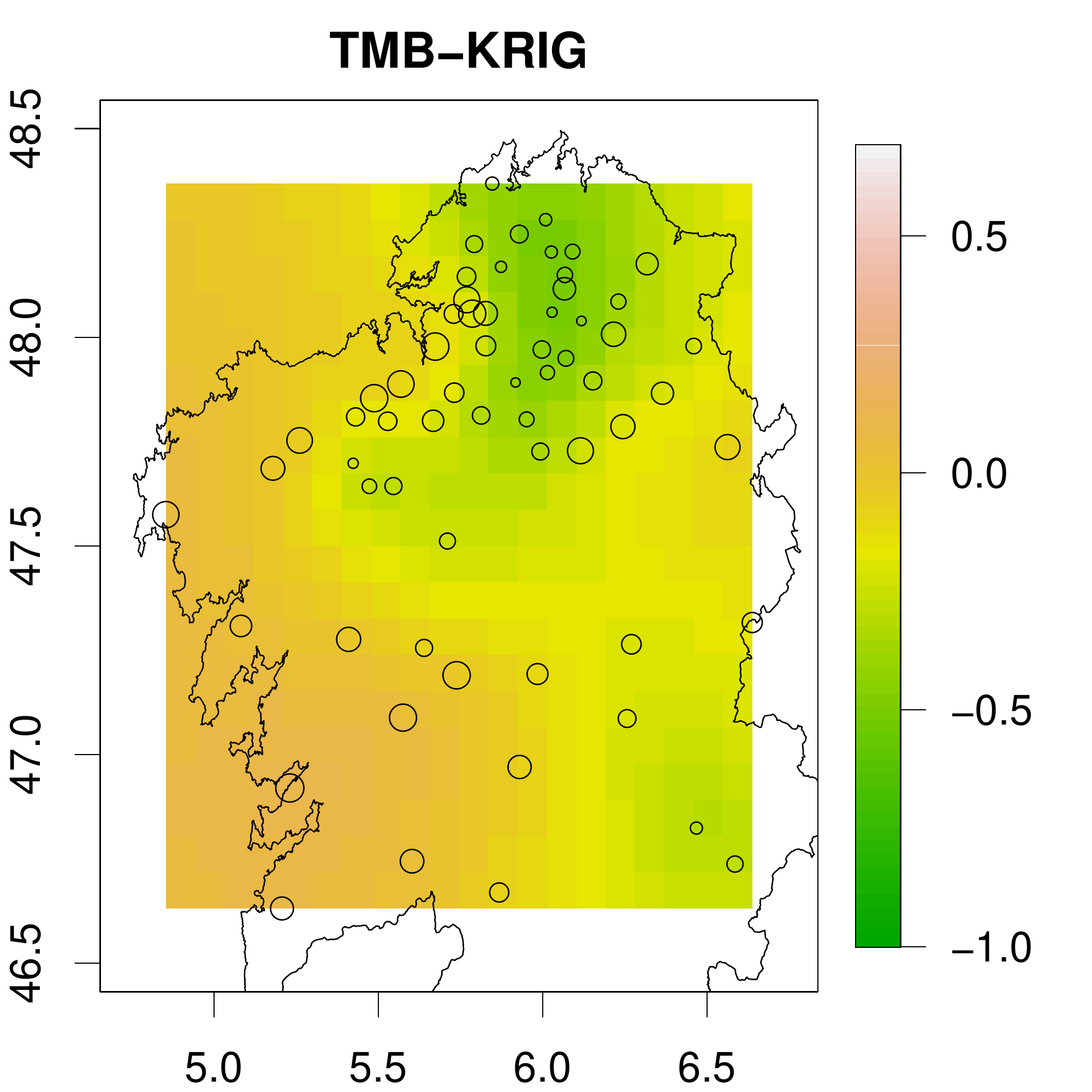}}
	\subfigure{\includegraphics[scale=0.25]{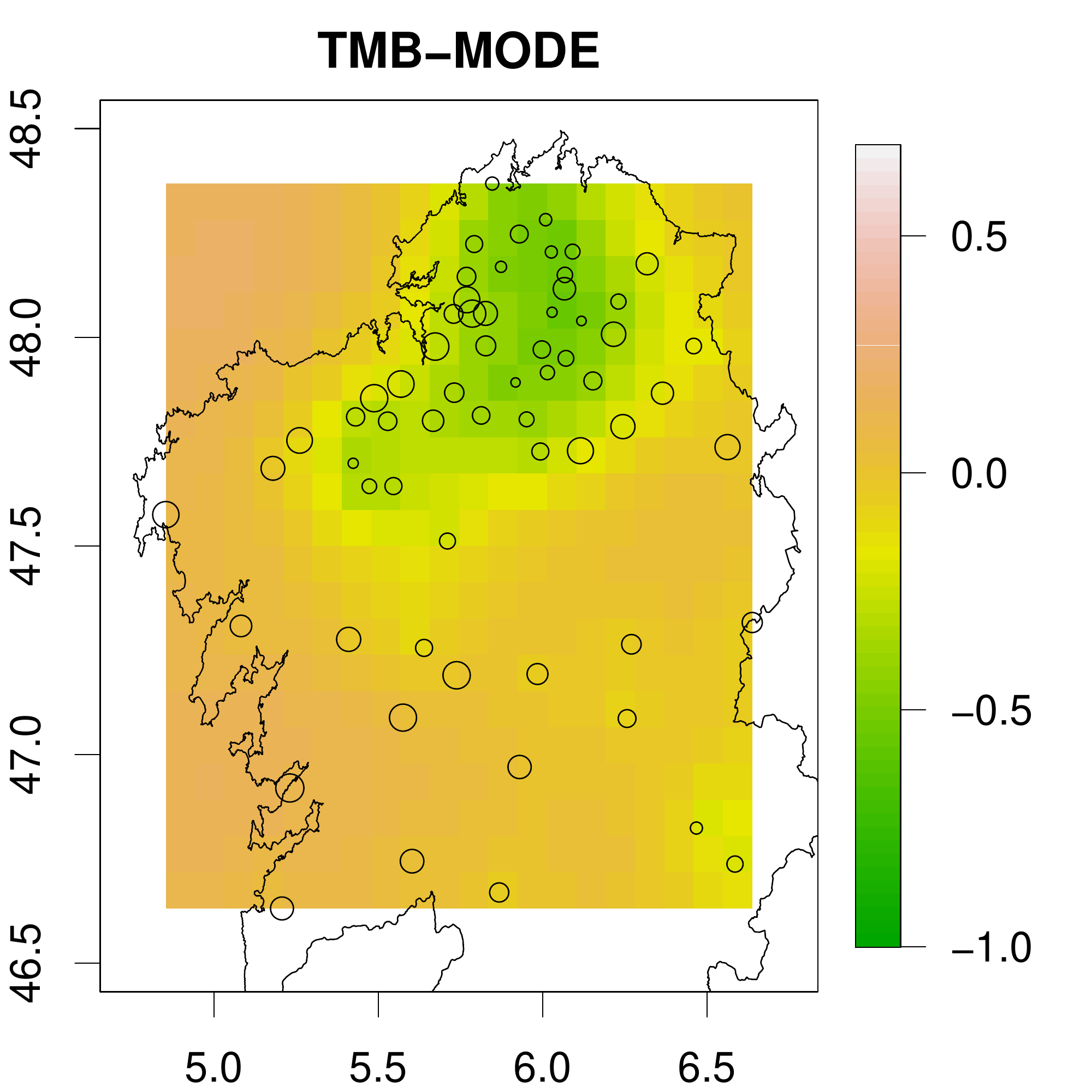}}
	\subfigure{\includegraphics[scale=0.25]{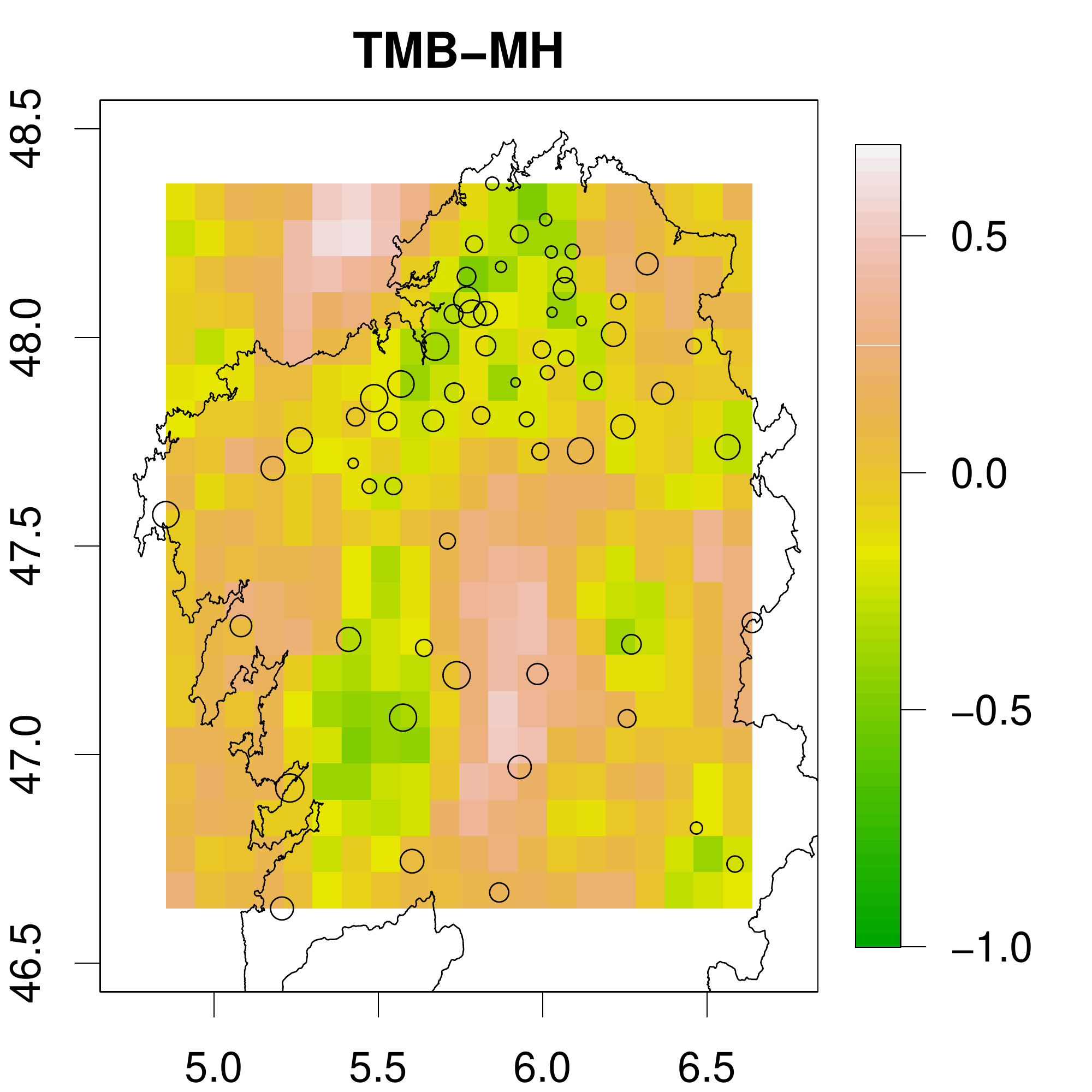}}\\
	\subfigure{\includegraphics[scale=0.25]{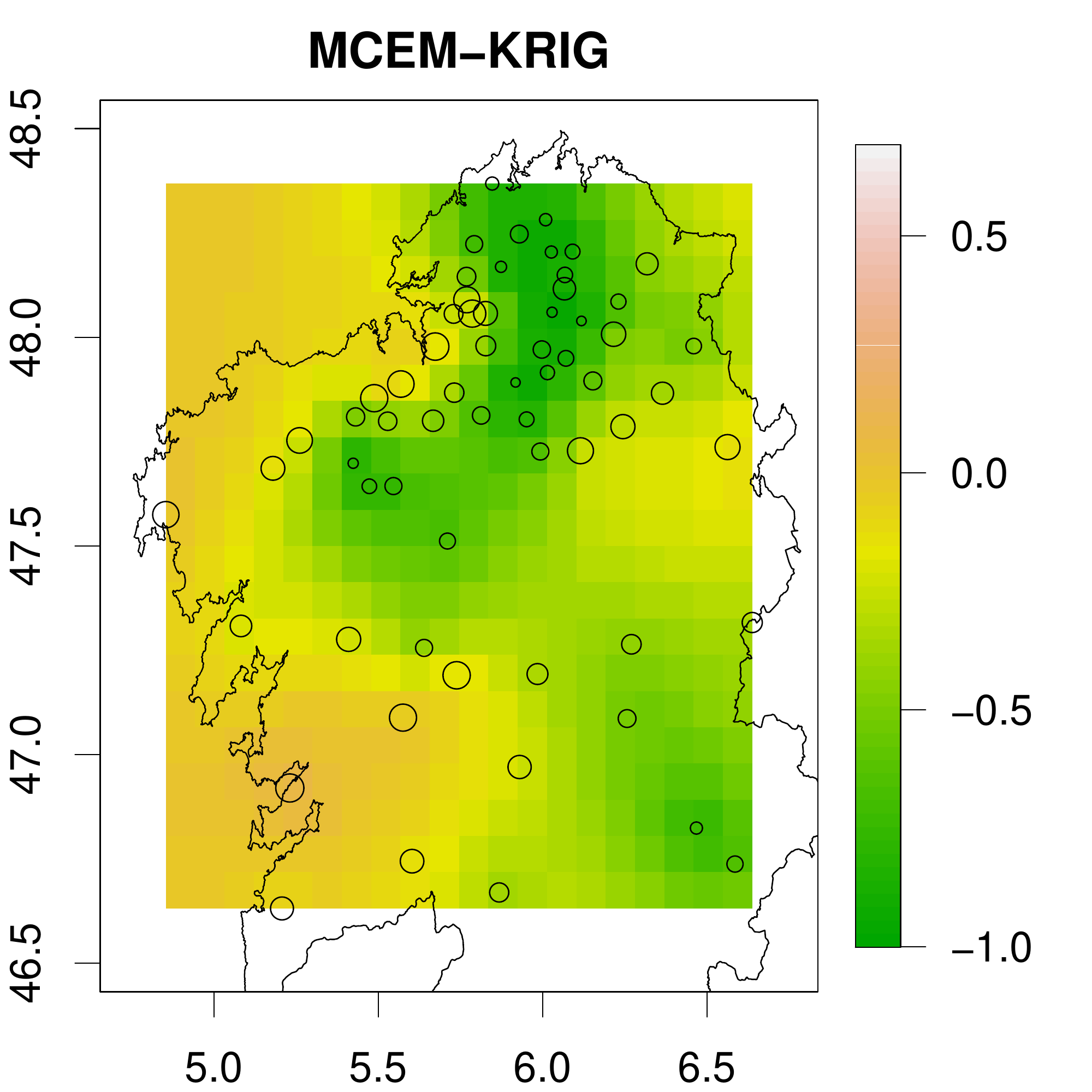}}
	\subfigure{\includegraphics[scale=0.25]{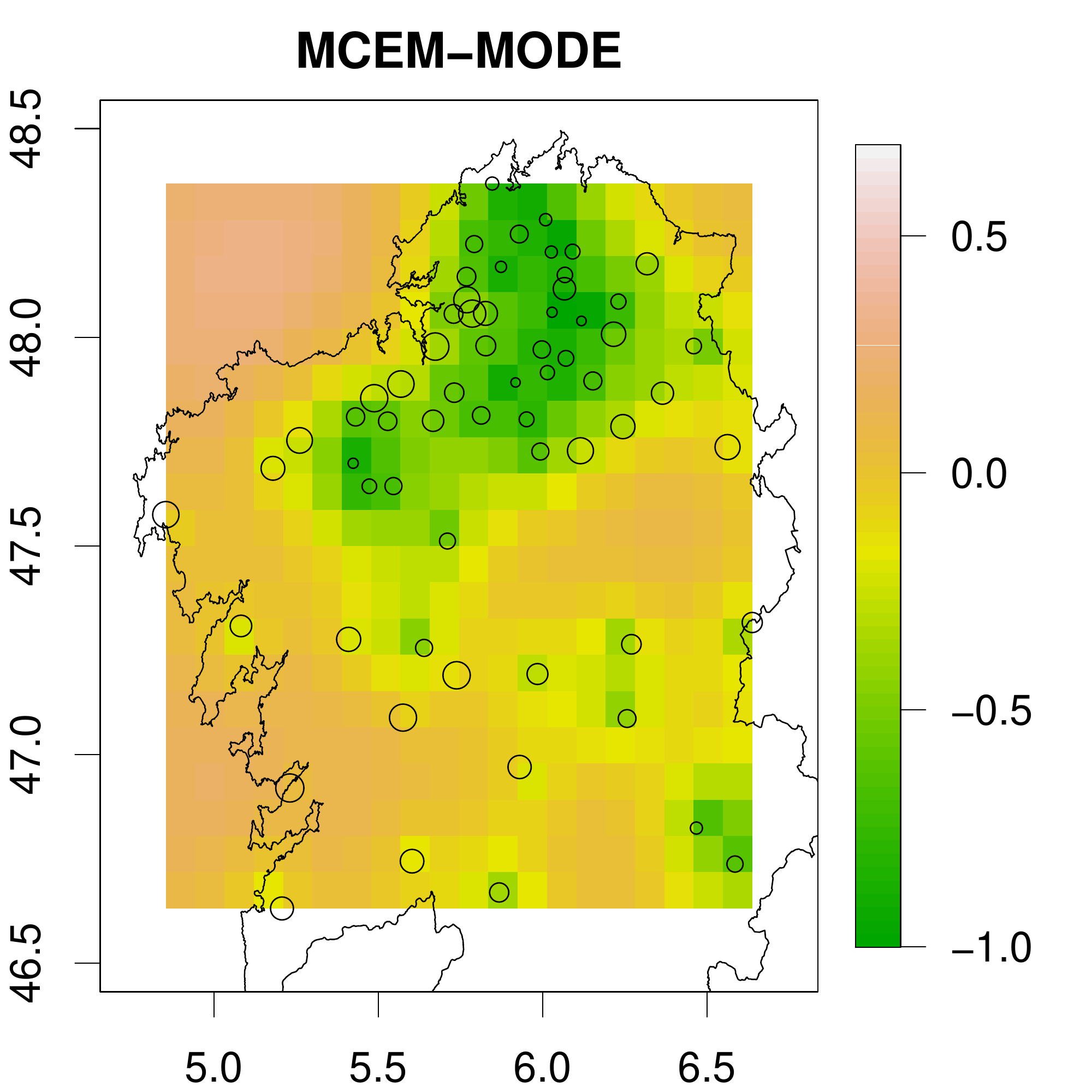}}
	\subfigure{\includegraphics[scale=0.25]{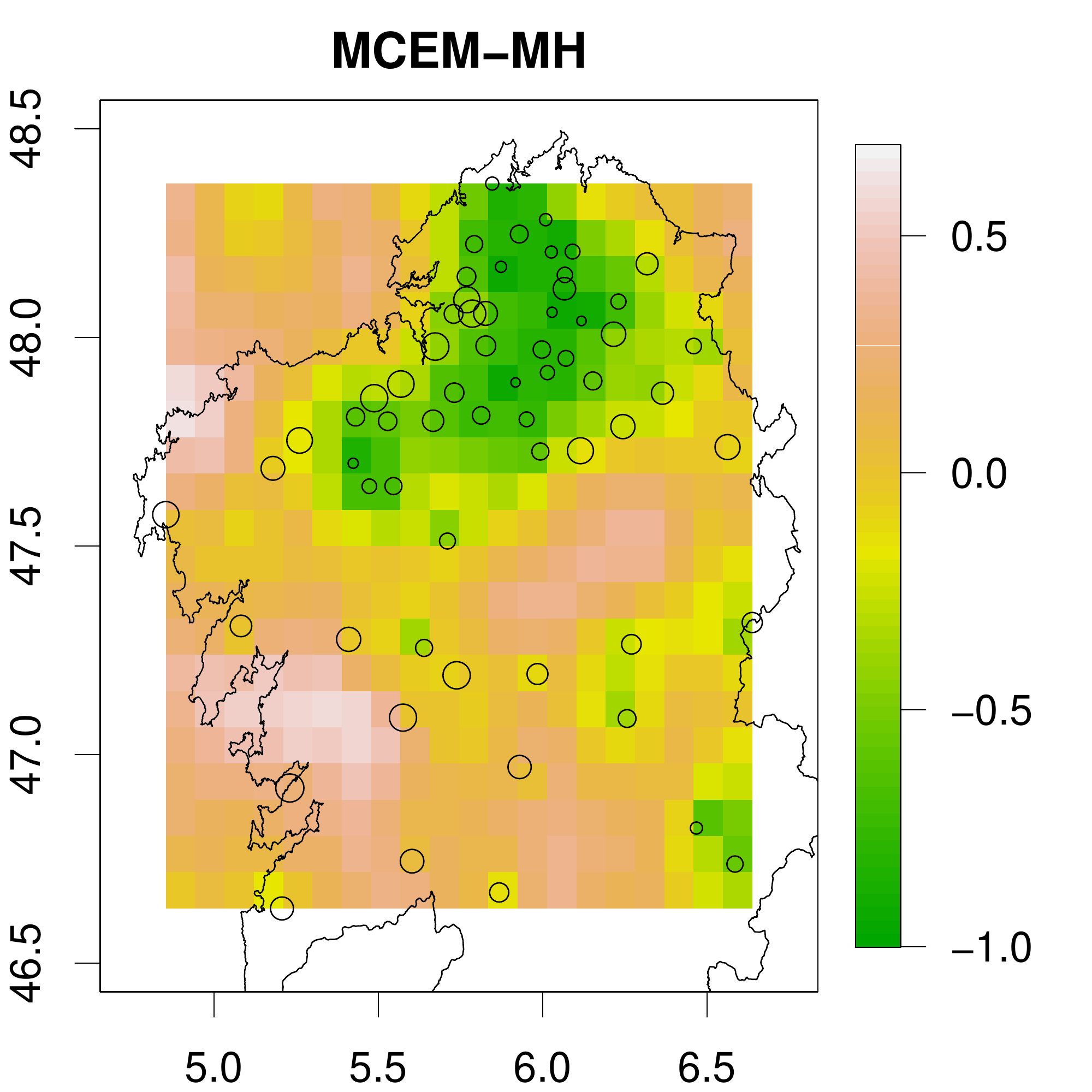}}
	\caption{Predicted maps of $S$ of log-lead concentration in moss in 1997 for combinations of parameter estimation methods and prediction methods. \label{fig:aplic:moss:predS_loglead97}}
\end{figure}

We can observe in Figure \ref{fig:aplic:moss:predS_loglead97} the predicted surface of process $S$ over Galicia in 1997. For construction of the maps, we use the parameters estimates as the true values and plug in them in the predictive distribution considered. So, we combined the prediction methods kriging, the mode of $\boldsymbol{S}$ and the predictive distribution $S|X,Y$ by MH with the estimation methods considered. In relation to the prediction methods, we note the ones that consider the pontual process provide wider ranges of $S$. In relation to the estimation methods, the estimate of correlation parameter has great impact in the range of prediction. MCLA provided lower estimation of $\phi$, which implies that the values of $S$ can vary more in a specific range of distance if we compare to the other methods that provided higher estimate of this parameter. This justifies the fact that MCLA provided a wider range of predicted values of $S$.

\begin{figure}[hbt!]
	\centering
	\subfigure{\includegraphics[scale=0.25]{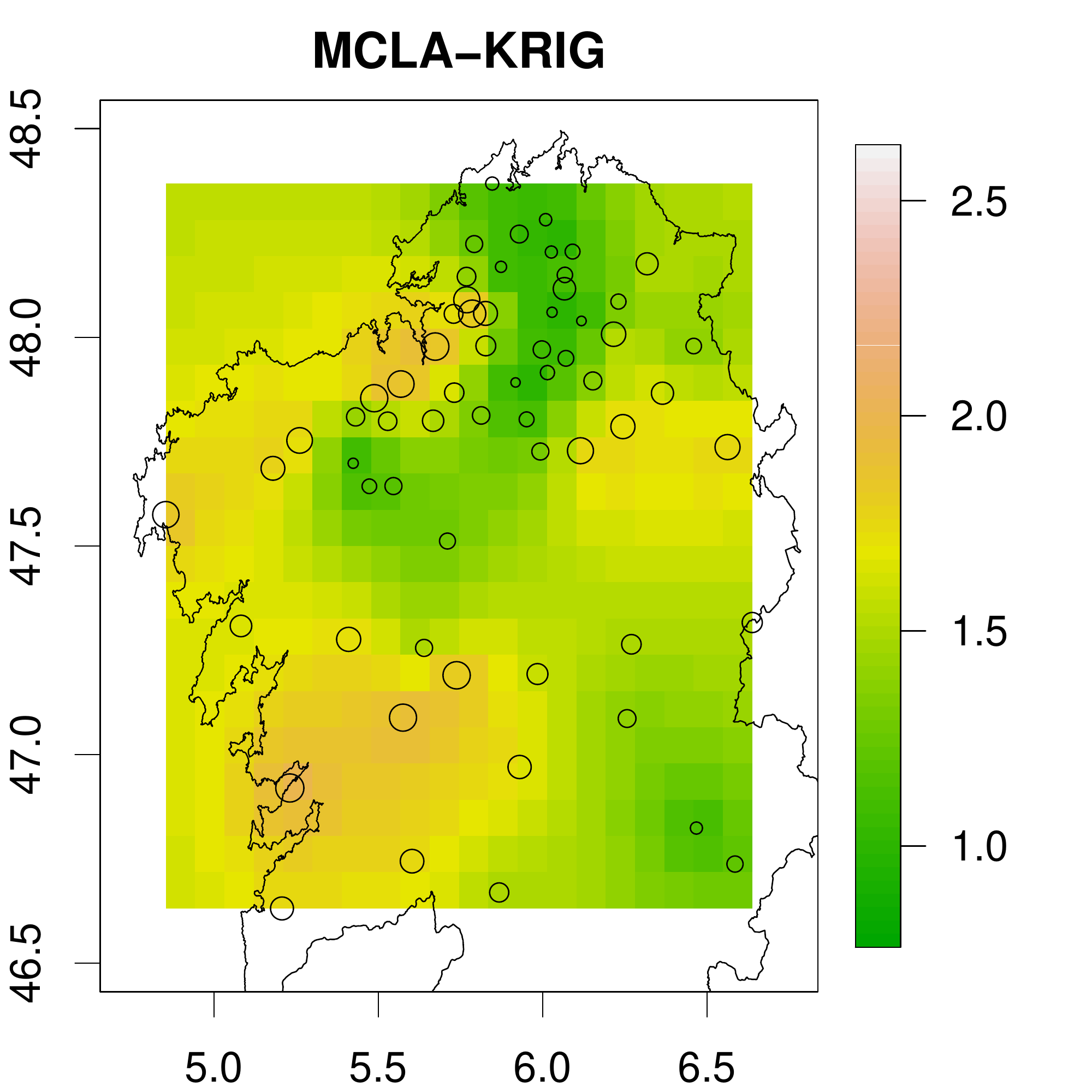}}
	\subfigure{\includegraphics[scale=0.25]{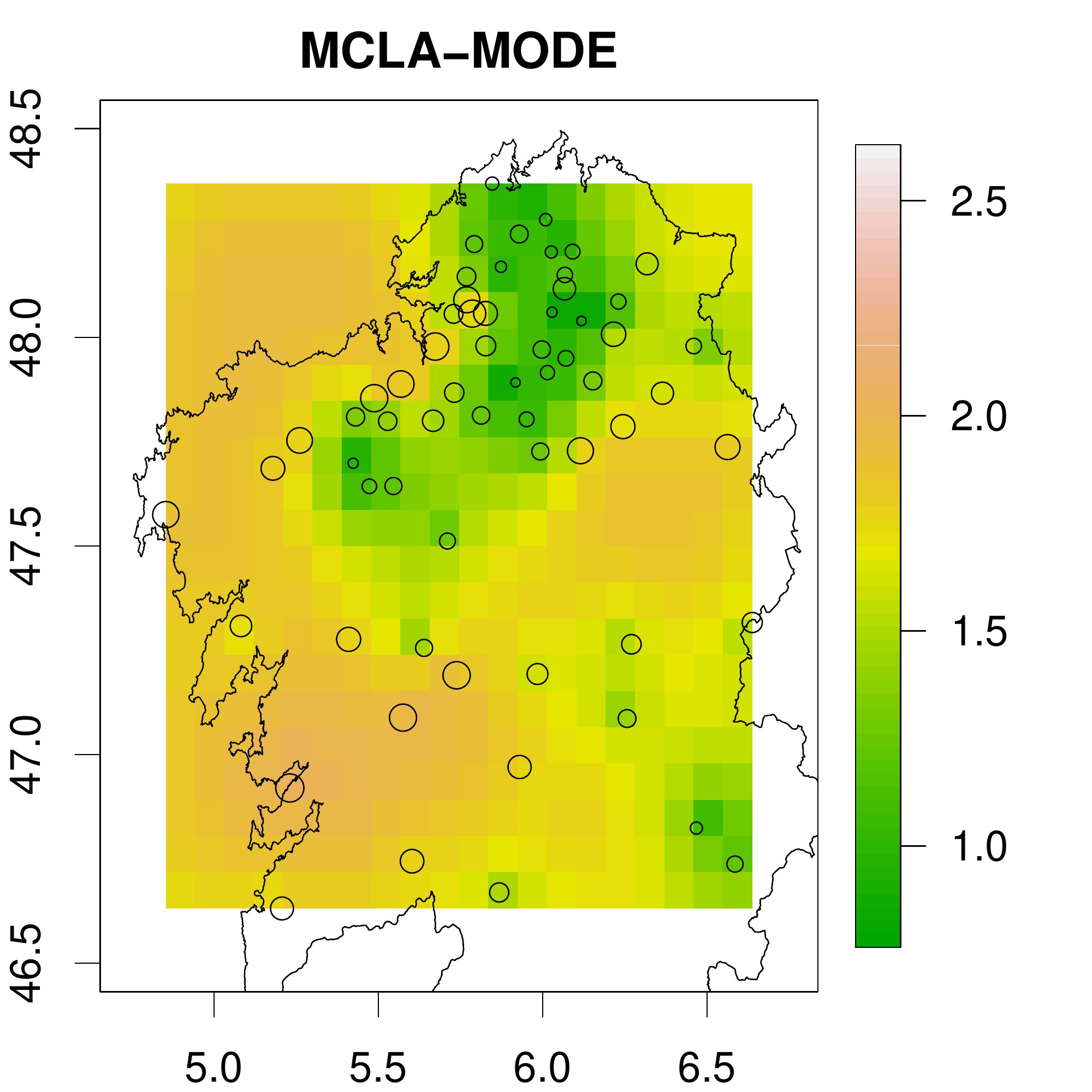}}
	\subfigure{\includegraphics[scale=0.25]{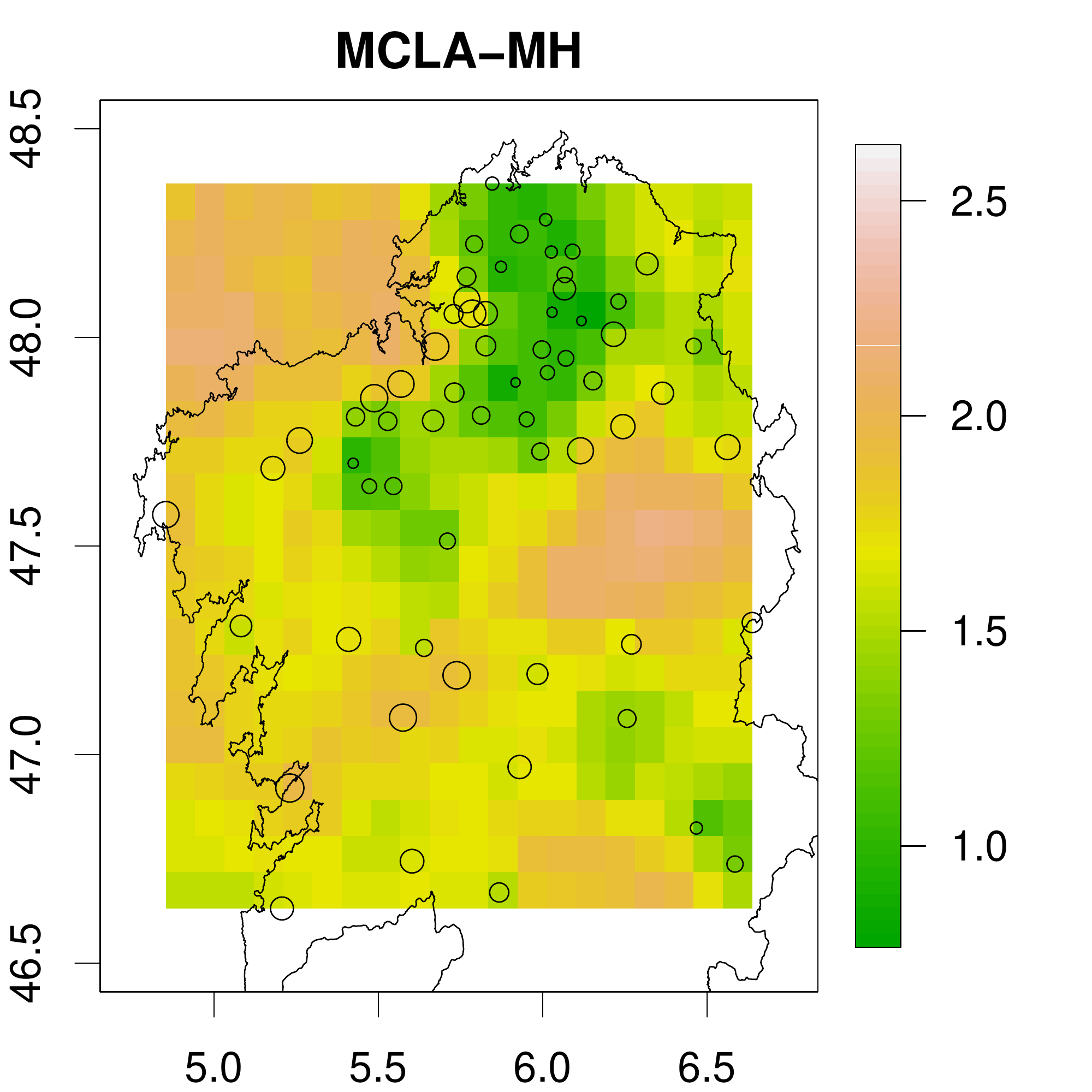}}\\
	\subfigure{\includegraphics[scale=0.25]{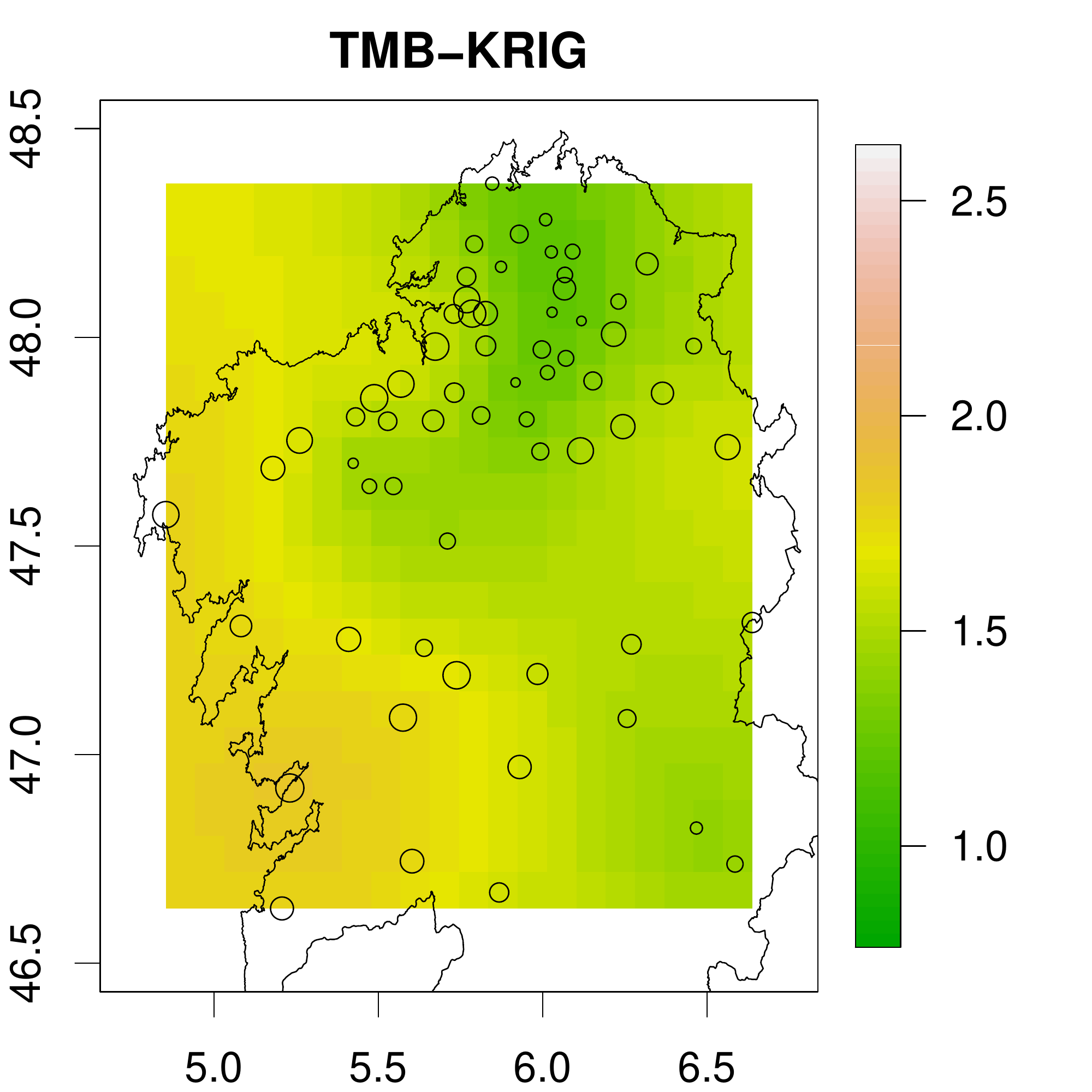}}
	\subfigure{\includegraphics[scale=0.25]{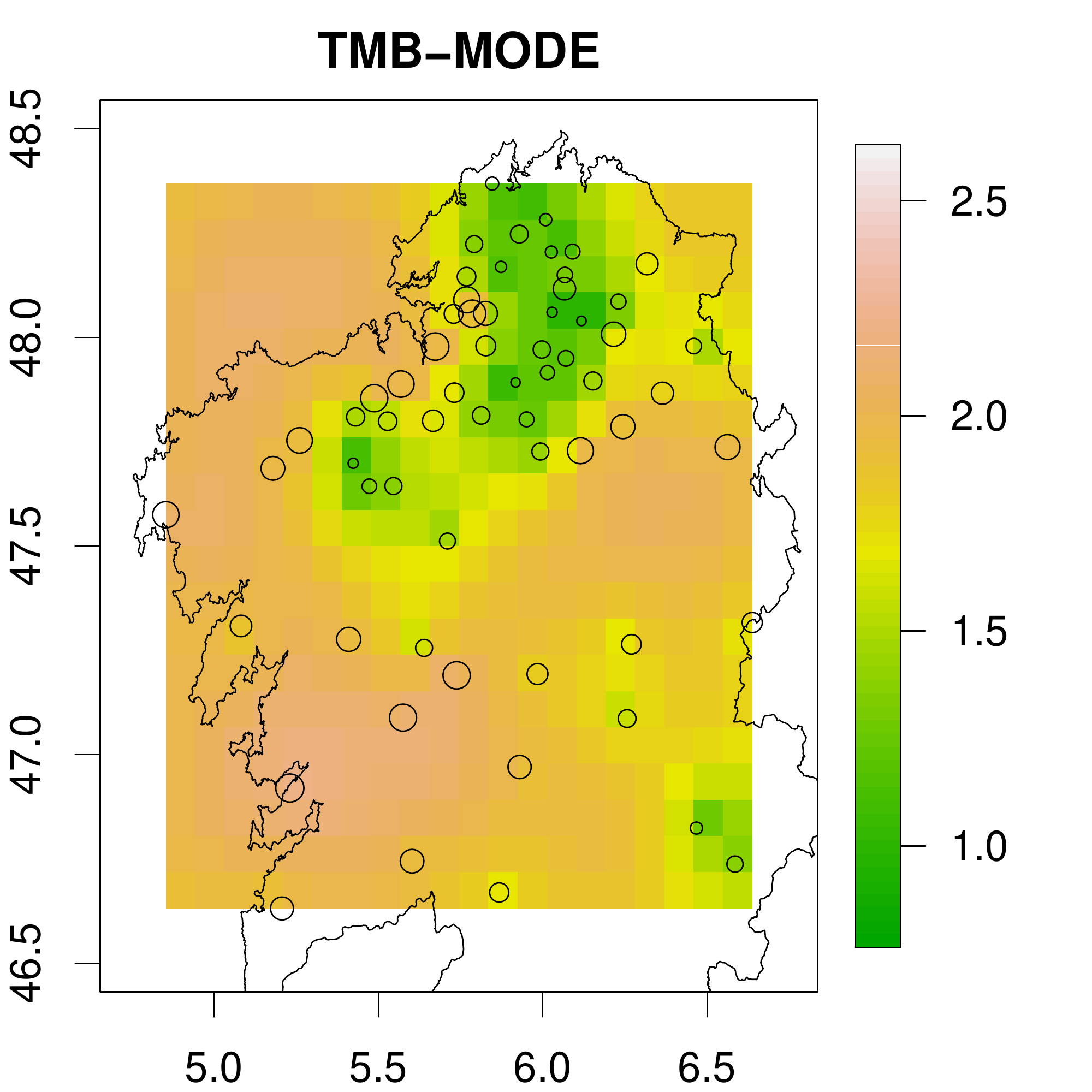}}
	\subfigure{\includegraphics[scale=0.25]{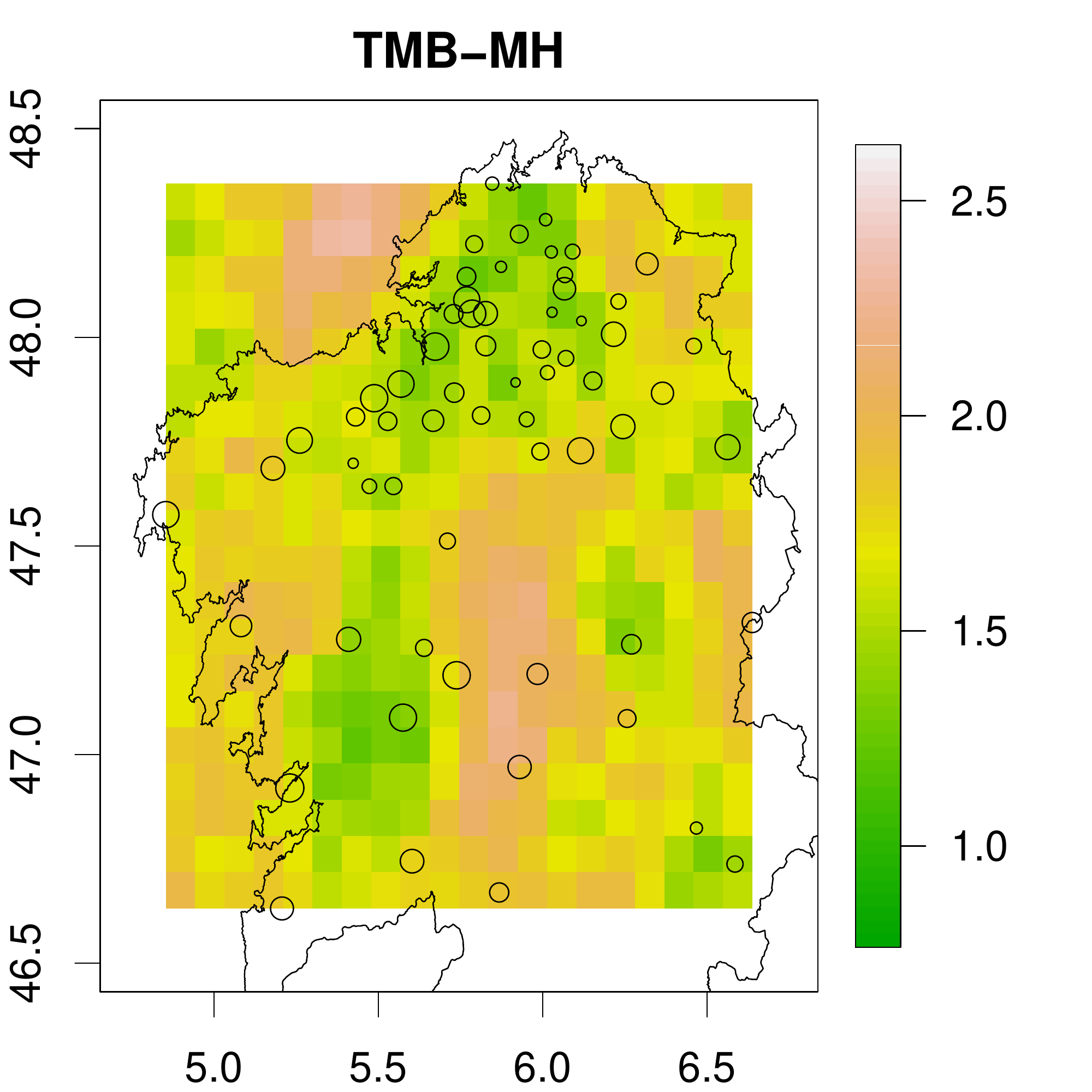}}\\
	\subfigure{\includegraphics[scale=0.25]{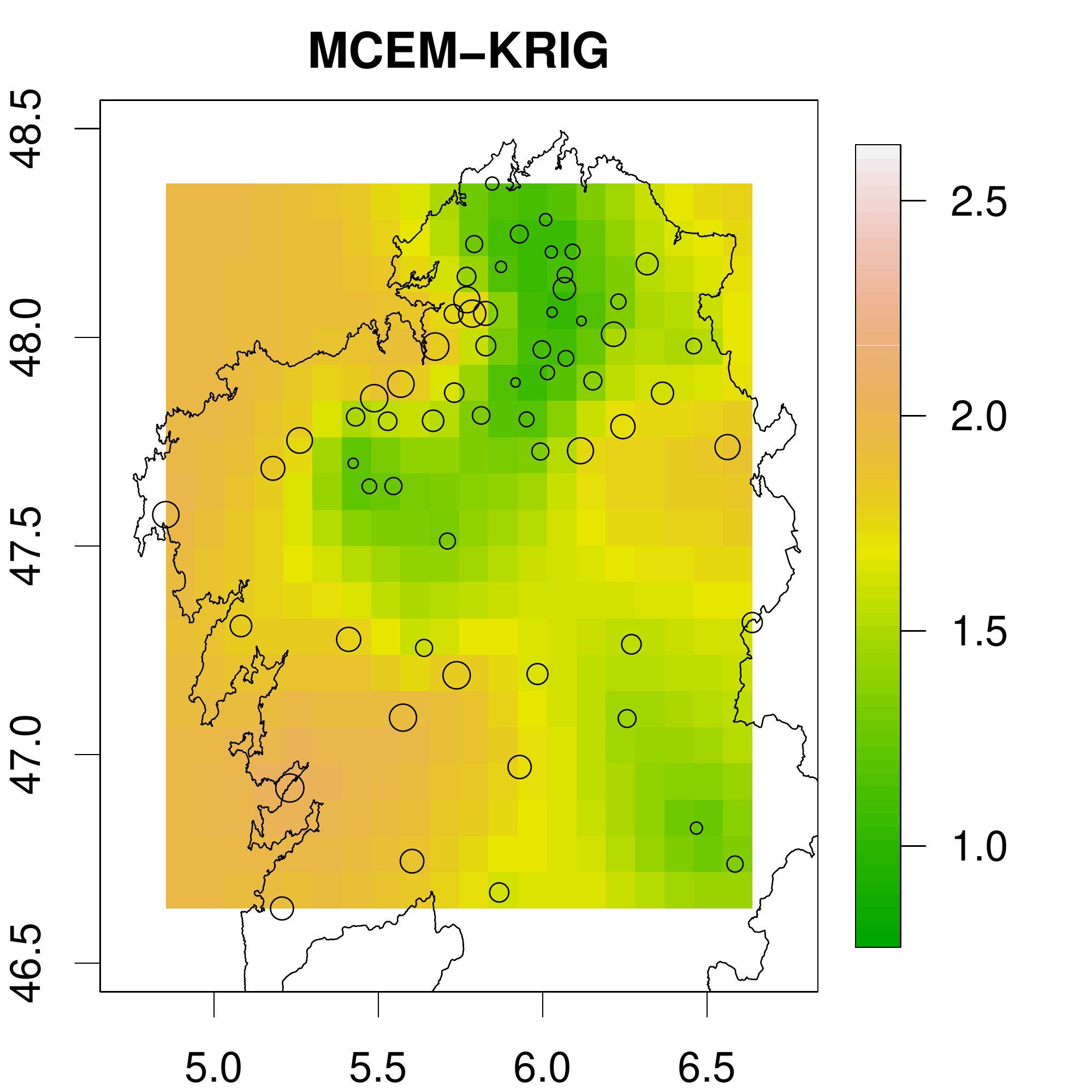}}
	\subfigure{\includegraphics[scale=0.25]{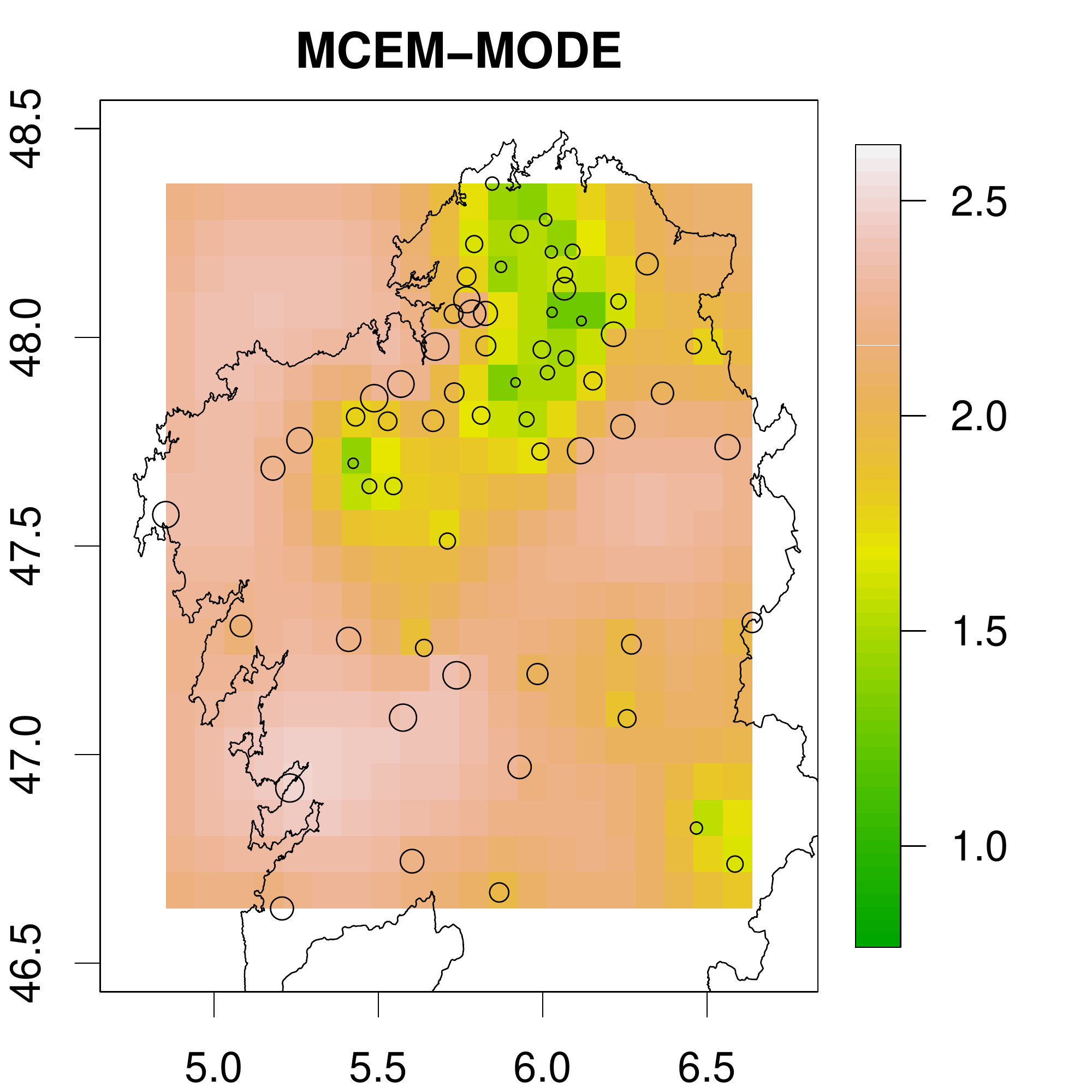}}
	\subfigure{\includegraphics[scale=0.25]{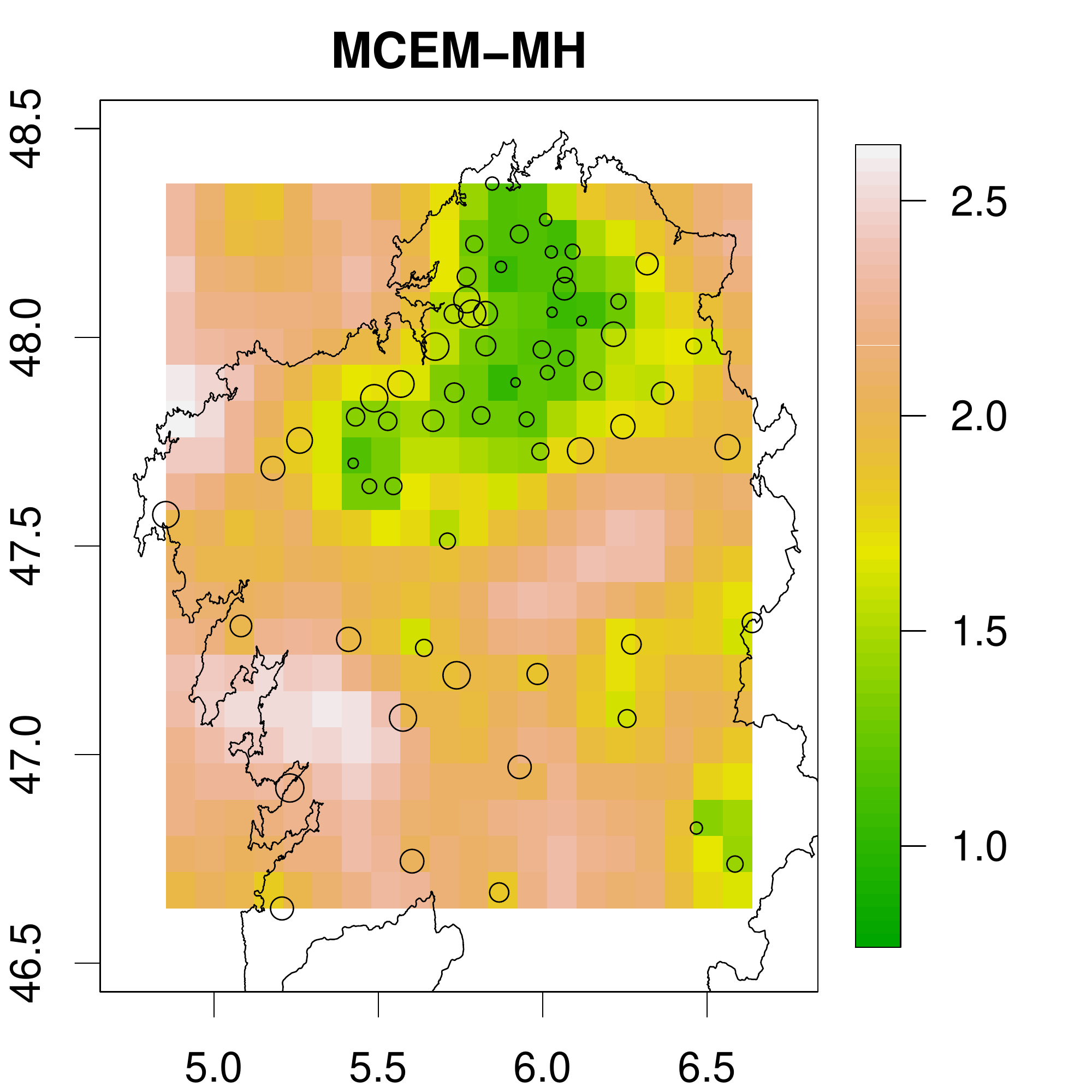}}
	\caption{Predicted maps of $Y$ of log-lead concentration in moss in 1997 for combinations of parameter estimation methods and prediction methods. \label{fig:aplic:moss:predY_loglead97}}
\end{figure}

The aim of geostatistics is to predict the variable in study ($Y$) at unobserved locations and Figure \ref{fig:aplic:moss:predY_loglead97} shows the predicted surface of log-lead concentrations over Galicia's region in 1997. 
We can observe that predictions methods that considered the information of the pontual process gave higher values at locations with no information compared to kriging. This is desired, since we obtained a negative value for $\beta$. 
Note that, since MCLA provided lower estimate of mean parameter, prediction of $Y$ will also be lower. In this way, methods that considere the information of the preferential sampling will give better prediction of $Y$ because they provide a corrected estimate of $\mu$.
We observe a wider range of the predicted values associated with the parameters estimates of MCEM and a less smoother map associated with the MH algorithm for prediction. The higher predicted values of MCEM-MODE and MCEM-MH is justified by the higher estimate of $\mu$ in comparision to the other methods. In this way, our methodology provided good results of parameter estimation and prediction. 




Since we know that the sampling design is not preferential in 2000 survey, the better model to be used is the NPG model. We use our model on this data to show that, even in the non preferential context, our model gives goods estimates. Indeed, the parameters estimates for MCEM are similar to the ones obtained by using NPG model. We note the estimate of $\beta$ is very near to zero, indicating the data is not preferential. 

\section{Discussion}
We have shown that inference can be incorrect if we use traditional geostatiscal model in a preferential sampling context. Both parameter estimation and predictions are biased, since this model do not consider the characteristic of the sampling design. The geostatistical model under preferential sampling, proposed by \cite{diggle2010}, can handle data with preferential sampling, but parameter estimation is difficult, since the values of underlying Gaussian process is needed for the entire region of study. Thus, approximated methods are necessary.

To the best of the authors' knowledge, there is no EM algorithm for parameter estimation for this model presented in literature before. In this way, we proposed a MCEM and a SAEM algorithms and showed that our method gives good parameter estimation, better than the other methods considered. For sampling from the correct predictive distribution of $S$ given data, \cite{ferreiragam} uses a MH algorithm, sampling element by element of the approximated Gaussian process. We proposed to sample from this distribution by blocks, showing that convergence of the Markov chains is achieved for a large number of iterations but in a very smaller time, turning the algorithm more efficient. Finally, we showed that our method gives predictions with wider range than the other methods, which means that we can predict values far from the mean parameters and obtain a better correction of predicted values. 

Although the estimates from SAEM converge to the MLE, this method presented much more time to execute than the MCEM algorithm and both gives similar estimates. In this way, MCEM seems to be more attractive for parameter estimation. The computations that were reported in the paper were run on a 3.60 GHz Intel i7 processor with 28 GB of random-access memory, using the R software enviroment \citep{r2020}. The data are available on the R package \cite{pack_prevmap}.

\clearpage
\noindent ACKNOWLEDGEMENTS

\noindent The authors acknowledge the partial financial support from CAPES-Brazil. \vskip 3mm

{\bibliographystyle{jasa}
	\renewcommand{\refname}{\normalsize References}
	\setlength{\bibsep}{0.2pt}
\bibliography{referencias}}

\begin{thebibliography}{}
\newcommand{\enquote}[1]{``#1''}

\bibitem[Delyon et~al.(1999)Delyon, Lavielle, e Moulines]{delyon1999}
Delyon, B., Lavielle, M., e Moulines, E. (1999), \enquote{Convergence of a
  stochastic approximation version of the EM algorithm,} \emph{The Annals of
  Statistics}, 27, 94--128.

\bibitem[Dempster et~al.(1977)Dempster, Laird, e Rubin]{dempster1977}
Dempster, A.~P., Laird, N.~M., e Rubin, D.~B. (1977), \enquote{Maximum
  likelihood from incomplete data via the EM algorithm,} \emph{Journal of the
  Royal Statistical Society, Series B}, {39}, 1--22.

\bibitem[Diggle et~al.(1998)Diggle, Tawn, e Moyeed]{diggle1998}
Diggle, P., Tawn, J., e Moyeed, R. (1998), \enquote{Model-based geostatistcs,}
  \emph{Journal of the Royal Statistical Society, Series C}, 47, 299--350.

\bibitem[Diggle et~al.(2010)Diggle, Menezes, e Su]{diggle2010}
Diggle, P.~J., Menezes, R., e Su, T. (2010), \enquote{Geostatistical inference
  under preferential sampling,} \emph{Journal of the Royal Statistical Society,
  Series C}, 59, 191--232.

\bibitem[Dinsdale e Salibian-Barrera(2019)Dinsdale e
  Salibian-Barrera]{dinsdale2019}
Dinsdale, D. e Salibian-Barrera, M. (2019), \enquote{Methods for preferential
  sampling in geoestatistics,} \emph{Journal of the Royal Statistical Society,
  Series C}, 68, 181--198.

\bibitem[Fernández et~al.(2000)Fernández, Rey, e Carballeira]{fernandez2000}
Fernández, J., Rey, A., e Carballeira, A. (2000), \enquote{An extended study of
  heavy metal deposition in Galicia (NW Spain) based on moss analysis,}
  \emph{The Science of the Total Enviroment}, 254, 31--44.

\bibitem[Ferreira e Gamerman(2015)Ferreira e Gamerman]{ferreiragam}
Ferreira, G. e Gamerman, D. (2015), \enquote{{Optimal design in geostatistics
  under preferential sampling (with discussion)},} \emph{Bayesian Analysis},
  10, 711 -- 735.

\bibitem[Galarza et~al.(2017)Galarza, Bandyopadhyay, e Lachos]{galarza2017}
Galarza, C.~E., Bandyopadhyay, D., e Lachos, V.~H. (2017), \enquote{Quantile
  regression in linear mixed models: A stochastic approximation EM approach,}
  \emph{Statistics and Its Interface}, 10, 471--482.

\bibitem[Gelfand et~al.(2012)Gelfand, Sahu, e Holland]{gelfand2012}
Gelfand, A.~E., Sahu, S.~K., e Holland, D.~M. (2012), \enquote{On the effect of
  preferential sampling in spatial prediction,} \emph{Environmetrics}, 23,
  565--578.

\bibitem[Giorgi e Diggle(2017)Giorgi e Diggle]{pack_prevmap}
Giorgi, E. e Diggle, P. (2017), \enquote{{PrevMap}: An R Package {for}
  Prevalence Mapping,} \emph{Journal of Statistical Software}, 78.

\bibitem[Kristensen et~al.(2016)Kristensen, Nielsen, Berg, Skaug, e
  Bell]{kristensen2016}
Kristensen, K., Nielsen, A., Berg, C.~W., Skaug, H., e Bell, B.~M. (2016),
  \enquote{TMB: Automatic Differentiation and Laplace Approximation,}
  \emph{Journal of Statistical Software}, 70, 1--21.

\bibitem[Møller e Waagepetersen(2003)Møller e Waagepetersen]{mollerw}
Møller, J. e Waagepetersen, R.~P. (2003), \emph{Statistical Inference and
  Simulation for Spatial Point Processes}, Chapman \& Hall, Taylor \& Francis.

\bibitem[Pati et~al.(2011)Pati, Reich, e Dunson]{pati2011}
Pati, D., Reich, B.~J., e Dunson, D.~B. (2011), \enquote{Bayesian
  geostatistical modelling with informative sampling locations,}
  \emph{Biometrika}, 98, 35--48.

\bibitem[{R Core Team}(2020){R Core Team}]{r2020}
{R Core Team} (2020), \emph{R: A Language and Environment for Statistical
  Computing}, R Foundation for Statistical Computing, Vienna, Austria.

\bibitem[Rue e Held(2005)Rue e Held]{rueheld2005}
Rue, H. e Held, L. (2005), \emph{Gaussian Markov Random Fields: Thoery and
  Applications}, Chapman \& Hall, Taylor \& Francis.

\bibitem[Wei e Tanner(1990)Wei e Tanner]{weitanner1990}
Wei, G. C.~G. e Tanner, M.~A. (1990), \enquote{{A Monte Carlo implementation of
  the EM algorithm and the poor man's data augmentation algorithms},}
  \emph{Journal of the American Statistical Association}, 85, 699--704.

\end{thebibliography}

\end{document}